\documentclass[a4paper,fleqn]{cas-dc}

\usepackage{graphicx,subcaption}
\usepackage{rotating}
\usepackage{lscape}
\usepackage{multicol}
\usepackage[justification=centering]{caption}
\usepackage{layout}
\usepackage{caption}
\usepackage[normalem]{ulem}
\usepackage[utf8]{inputenc}
\usepackage{aas_macros}
\usepackage{threeparttable}
\usepackage[authoryear]{natbib}
\usepackage{epstopdf}
\newcommand{\gm}{$\gamma$}
\newcommand{\src}{1H~0323+342\,\,}

\newcommand{\gp}{\gamma^{\prime}}
\newcommand{\e}{\epsilon}
\newcommand{\ep}{\epsilon^\prime}
\def\tsc#1{\csdef{#1}{\textsc{\lowercase{#1}}\xspace}}
\tsc{WGM}
\tsc{QE}

\begin{document}
\UseRawInputEncoding
\let\WriteBookmarks\relax
\def\floatpagepagefraction{1}
\def\textpagefraction{.001}

\shorttitle{1H 0323+342}    

\shortauthors{Aminabi, Paliya, Kharb}  

\title [mode = title]{A Rare Gamma-ray Flaring episode of the Narrow-Line Seyfert 1 Galaxy 1H~0323+342}  



%

\author[1]{Aminabi Thekkoth}[orcid=0000-0003-3729-0198]



\ead{athekkoth@ncra.tifr.res.in}



\affiliation[1]{organization={National Centre for Radio Astrophysics},
            city={Pune},
            postcode={411007}, 
            state={Maharashtra},
            country={India}} 
\author[2]{Vaidehi S. Paliya}[]
\ead{vaidehi.s.paliya@gmail.com}

\affiliation[2]{organization={Inter-University Centre for Astronomy and Astrophysics (IUCAA),SPPU Campus},
            city={Pune},
            postcode={411007}, 
            state={Maharashtra},
            country={India}}            
\author[1]{Preeti Kharb}
\ead{kharb@ncra.tifr.res.in}


\nonumnote{}

\begin{abstract}
Gamma-ray-emitting narrow-line Seyfert 1 (\gm-NLSy1) galaxies represent an enigmatic class of active galactic nuclei (AGNs) bridging populations of radio-quiet and radio-loud AGNs. Here we report the multi-wavelength investigation of a rare $\gamma$-ray flaring episode of an NLSy1 galaxy, 1H~0323+342 ($z=0.063$), using data from \emph{Fermi}-Large Area Telescope, \emph{Swift}, and \emph{Nuclear Spectroscopic Telescopic Array}. The source exhibited a significant enhancement in $\gamma$-ray flux along with rapid variability on $\sim$ sub-hour timescales in both $\gamma$-ray and hard X-ray wavelengths, hinting that the energy dissipation is happening from a compact region close to the central supermassive black-hole. The time-resolved X-ray spectral study revealed a transition between a jet-dominated and a mixed jet+corona emission state. Reproducing the broadband spectral energy distribution with a one-zone leptonic model suggested that the high-energy emission is mainly produced by external Compton scattering of the accretion disc and broad line region (BLR) photons. The inferred jet power reaches values of the order of $10^{46}$ erg s$^{-1}$ comparable to those of powerful flat-spectrum radio quasars, suggesting that even low black-hole mass systems can occasionally produce powerful \gm-ray outbursts. 
\end{abstract}

\maketitle


\section{Introduction}\label{sec:intro}
According to the prevailing unification model of Active Galactic Nuclei (AGNs) \citep[]{1995PASP..107..803U,2006agna.book.....O}, Type~1 AGNs are objects viewed at relatively small angles to the jet axis, allowing the central accretion disc and the broad-line region (BLR) to be directly visible. As a result, their optical spectra display both broad and narrow emission lines. In contrast, Type~2 AGNs are observed through the torus that obscures the central region, and consequently, their optical spectra are dominated by narrow emission lines \citep[]{2011MNRAS.415.1928C,1981PASP...93....5B}.

Within this framework, the Narrow-line Seyfert~1 (NLSy1) galaxies occupy a remarkably interesting position. They exhibit broad optical emission lines similar to Seyfert~1 galaxies, yet their Balmer lines are unusually narrow, with $\mathrm{FWHM}(\mathrm{H}\beta) < 2000~\mathrm{km,s^{-1}}$. NLSy1s are also characterized by weak [O\,III] emission relative to H$\beta$ ($[\mathrm{O\,III}]/\mathrm{H}\beta < 3$) and prominent Fe\,II multiplet features \citep[]{1989ApJ...342..224G,1985ApJ...297..166O}. Multiple studies indicate that NLSy1 galaxies host relatively low-mass supermassive black-holes ($10^{6-8}\,M_\odot$) that accrete at rates close to the Eddington limit \citep[cf.][]{2000MNRAS.314L..17M,2012AJ....143...83X,2024MNRAS.527.7055P}. NLSy1 galaxies are characterized by bright X-ray emission with steeper spectral indices than broad-line AGNs and a distinctive soft X-ray excess observed below 2 keV \citep[]{1996A&A...305...53B,2019MNRAS.490..683P,2016A&A...591A..98B}.

Only about $5-7\%$ of the NLSy1 population is known as radio-loud \citep[e.g.,][]{2018MNRAS.480.1796S}, with a small subset showing strong radio emission with compact, high brightness cores, and flat or inverted radio spectra, indicating the presence of relativistic jets similar to those in blazars \citep[e.g.,][]{2008ApJ...685..801Y}. Within this population, a few NLS1 galaxies exhibit extended radio structures ranging from several tens of kpc up to $\sim 100$~kpc \citep[e.g.,][]{Congiu2017, 2018ApJ...869..173R,2024ApJ...963...32C, 2025ApJ...995..125U}. The discovery of high-energy \gm-ray emission from a few NLSy1 galaxies demonstrated that even low-mass AGN can launch powerful relativistic jets \citep[][]{2009ApJ...707L.142A,2011nlsg.confE..24F,2018ApJ...853L...2P}. 

The \gm-ray loud NLSy1s show many interesting features similar to those of blazars. For example, \cite{2021ApJS..255...10M} reported rapid mid-infrared (MIR) variability in a sample of \gm-ray loud NLSy1 galaxies using the multi-epoch data of the Wide-field Infrared Survey Explorer (WISE). Their MIR emission is dominated by synchrotron emission, exhibiting a ``bluer when brighter'' trend as observed in blazars. The \gm-ray loud NLSy1 galaxies are also variable in the radio band, and a positive correlation between these two emissions is also reported \citep[e.g.,][]{2014ApJ...781...75W,2025MNRAS.536.1344S}. The optical-to-ultraviolet (UV) emission in NLSy1 galaxies is often dominated by the synchrotron radiation and/or thermal radiation from the accretion disc manifested as a ``big blue bump'' \citep[see, e.g.,][]{2011MNRAS.413.1671F,2019ApJ...872..169P}. The high-energy emission from \gm-ray loud NLSy1s is found to be dominated by the jet, even though several \gm-ray detected NLSy1s exhibit a soft excess in the X-ray band \citep[cf.][]{2014MNRAS.440..106B,2020MNRAS.496.2922M}. Majority of NLSy1 galaxies are faint \gm-ray emitters and have occasionally exhibited \gm-ray flaring episodes \citep[e.g.,][]{2011nlsg.confE..24F,2015MNRAS.446.2456D,2015AJ....149...41P,2021A&A...649A..77G}. 

The broadband spectral energy distribution (SED) of \gm-ray emitting NLSy1 galaxies shows a typical double hump structure akin to blazars \citep[][]{2009ApJ...707L.142A}. In particular, the requirement of the external Compton process to explain the high-energy hump indicates their similarity with flat-spectrum radio quasars (FSRQs) than BL Lac objects \citep[][]{2009ApJ...707L.142A,2013ApJ...768...52P}. Considering relatively small black-hole masses, high accretion luminosity, compact radio structures, and significant variability indicate that this class of AGN could be the low-mass tail of FSRQs and are likely to be young or recently triggered jetted systems \citep[][]{2015ANA...575A..13F,2020ApJ...892..133P}.

As the nearest known \gm-ray-emitting NLSy1 galaxy at the redshift of $z=0.063$, 1H~0323+342 serves as a prototype of the class of AGN that bridges Seyfert galaxies and blazars. This source hosts a relatively low-mass supermassive black-hole ($M_{\rm BH} \sim 10^7\, M_\odot$) embedded in a galaxy exhibiting a distorted spiral or ring-like morphology, likely the result of a recent merger event \citep[][]{2008A&A...490..583A,2017MNRAS.464.2565L,2020MNRAS.496.1757D}. 
The radio morphology of 1H~0323+342 resembles a Z-shaped radio galaxy (see Figure in the Appendix) \citep[see also][who classify it instead as an X-shaped radio galaxy with precessing jets caused by the presence of a binary black hole]{2020MNRAS.496.1757D}. The archival VLA C-array 1.42~GHz image shows east-west oriented FRII-like-jets with an extent of 84 arcsec or 104 kpc, in projection. The brighter approaching jet is towards the east. The hotspot on the counterjet side is not clearly detected most likely due to Doppler de-boosting effects. The end-to-end size of the Z-shaped lobes in the northeast-southwest direction is 137 arcsec or 170 kpc in projection. Given its total 6~cm luminosity of $1.3\times10^{24}$~W~Hz$^{-1}$ \citep[using the total 1.42 GHz flux density of 500 mJy and a lobe spectral index of $-1.0$, e.g.,][]{2020MNRAS.496.1757D}, this source falls into the radio-loud NLS1 category \citep[i.e., $L_{6}>10^{23.2}$~W~Hz$^{-1}$;][]{Kellermann2016}. The VLBI image of 1H 0323+342 from the MOJAVE program shows a particularly straight one-sided jet pointed towards the south-east with an inferred magnetic field direction aligned well with the jet flow. The VLBI jet direction is mostly aligned with the brighter VLA jet to the east, and exhibits a large jet speed of $9.05\pm0.32$c \citep{Lister2016}.

Previous multi-wavelength studies have reported a broadband SED featuring a ``big blue bump'' in optical/UV, a prominent soft X-ray excess below 2 keV, and non-thermal jet emission at higher energies \citep{2014ApJ...789..143P,2017MNRAS.464.2565L}. Detailed X-ray analysis has consistently revealed relativistic reflection features, including a broadened Fe-K$\alpha$ emission line and a high black-hole spin value of $a=0.96\pm 0.14$, suggesting that extreme spin may be vital for its jet production \citep{2014ApJ...789..143P,2020MNRAS.496.2922M}. 1H~0323+342 is one of the very few NLSy1s that have exhibited bright \gm-ray flares. In the first 17 years of the \emph{Fermi}-Large Area Telescope (LAT) operation, 1H~0323+342 exhibited only a single \gm-ray outburst in 2013 \citep[][]{2013ATel.5344....1C}. Interestingly, during its 2013 \gm-ray flare, a rapid \gm-ray flux variability with flux doubling timescales $\leq$ 3 hours was observed \citep[][]{2014ApJ...789..143P}. A rapid change in the degree of optical polarization coincident with the \gm-ray flaring event was also reported \citep{2014PASJ...66..108I}. A long-term multiwavelength study spanning 2006 to 2023 reveals three distinct zones of spectral behavior in the X-ray emission: a high-flux hard index (zone 1), a high-flux soft index (zone 2), and a low-flux soft index \citep[zone 3;][]{2025A&A...698A.160R}. Furthermore, the observed decline in the \gm-ray activity after 2017 is attributed to the transition of the source's oscillation from zones 1 and 2 to zones  2 and 3. These transitions were explained within the framework of intermittent jet activity driven by radiation-pressure instabilities in the accretion disc \citep[][]{2025A&A...698A.160R}. 

After the 2013 \gm-ray flaring event, 1H~0323+342 exhibited increased activity in the GeV band in September 2025, which was covered by \emph{Fermi}-Large Area Telescope (LAT), Nuclear Spectroscopic Telescopic Array (NuSTAR), and \emph{Swift} telescopes. The flux in the LAT energy range was enhanced up to 20 times its average value on 2025 September 20 \citep[][]{2025ATel17407....1L}. Near-simultaneous \emph{Swift} observations revealed a high level of X-ray flux \citep[][]{2025ATel17411....1D}. Over 17 years of \emph{Fermi}-LAT observations, only a handful of radio-loud NLSy1 (1H~0323+342, SBS~0846+513, PMN~J0948+0022, PKS~1502+036, PKS~2004-447) have been reported to exhibit bright \gm-ray flares \citep[][]{2013MNRAS.436..191D,2014ApJ...789..143P,2015MNRAS.446.2456D,2016ApJ...820...52P,2021A&A...649A..77G}. 

Since NLSy1s possess blazar-like jets but host low black-hole masses with high accretion rates, a multi-wavelength investigation of the rare \gm-ray flaring events detected from an NLSy1 galaxy provides the unique opportunity to explore the jet physics in this relatively poorly understood class of AGN. With this goal in mind, we have carried out a detailed investigation of the physical properties of 1H~0323+342 using all publicly available multi-band datasets taken with the \emph{Fermi}-LAT, NuSTAR, and \emph{Swift} telescopes. 

We adopted the following cosmology parameters: Hubble constant $H_0=67.8$ km s$^{-1}$ Mpc$^{-1}$, $\Omega_m=0.308$, and $\Omega_{\Lambda}=0.692$ \citep[][]{2016A&A...594A..13P}.

\section{Multi-wavelength Data Reduction}
\label{sec:data}
\subsection{\emph{Fermi}-Large Area Telescope}
Since its launch in June 2008, the \emph{Fermi-LAT} has been performing almost uninterrupted monitoring of the whole \gm-ray sky. In this study, we analyzed the LAT observation of \src spanning MJD~60933--60948 (September 15-30, 2025). The LAT data was analyzed using \emph{fermipy} software package \citep[][]{2017ICRC...35..824W}. The events were extracted from a circular region with a $12^\circ$ radius centered at the source position. We adopted the standard criteria recommended by \emph{Fermi}-LAT documentation for the event filtering ($\tt{evclass=128}$ \& $\tt{evtype=3}$,  $\tt{DATA\_QUAL>0}$ \& $\tt{LAT\_CONFIG==1}$). A zenith angle cut of $z_{\rm max}<90^{\circ}$ was applied to remove contamination from Earth limb \gm~rays. Using the fourth data release of the fourth \emph{Fermi}-LAT \gm-ray source catalog \citep[4FGL-DR4;][]{2022ApJS..260...53A,2023arXiv230712546B}, all \gm-ray sources lying within 20$^{\circ}$ from 1H~0323+342 were included in the sky model. The latest templates were also used to model the diffuse background \gm-ray emissions ({${\tt{gll\_iem\_V07}}$} and ${\tt{iso\_P8R3\_SOURCE\_V2\_v1}}$). 

We first optimized the sky model and froze the spectral parameters of all background sources having the test statistic or TS$<$9. A binned likelihood fitting was employed to determine the detection significance and spectral parameters of source of interest. In order to probe the flux variability, we generated 12- and 3-hour binned light curves in the selected period using an unbinned likelihood fitting algorithm. From the light curve, we selected two epochs for broadband SED modeling, depending on the source activity and the availability of multi-wavelength information.

\begin{table*}[ht]
\setlength{\tabcolsep}{3pt}
\setlength\extrarowheight{5pt}
\caption{Table showing the results of spectral analysis of \gm-ray spectrum of \src~. The average \gm-ray spectrum corresponds to the total time period of this study: MJD 60933-60948. }
\label{tab:lat_spec}
\begin{tabular}{llccr}

\hline
& & Average \gm-ray spectrum & &\\
\hline
Model& Index & Curvature & Flux & TS\\
& $\Gamma_{\rm \gamma}$/$\alpha$ & $\beta$ & (10$^{-6}$ ph cm$^{-2}$ s$^{-1}$)& \\

\hline
Power-law & $2.74\pm 0.01$ & - & $3.05\pm 0.01 $ & 120 \\
Log-parabola & $2.66\pm 0.03$ & $0.11\pm 0.02$ & $2.96\pm 0.01$ & 120 \\
\hline
& &Individual \gm -ray spectra (Power-law)&& \\
& &(SED analysis)&&\\
\hline
State & Time & Flux & Index$\Gamma_{\rm \gamma}$ &  TS \\
     &MJD & (10$^{-7}$ ph cm$^{-2}$ s$^{-1}$) &  $\Gamma_{\gamma}$ &    \\
\hline
SED1 & 60939.40 - 60942 & 2.47 $\pm$ 0.80 & 2.50 $\pm$ 0.30 & 30 \\
SED2 & 60944.20 - 60945.50 & 2.65 $\pm$ 1.21  & 2.42 $\pm$ 0.39 & 11 \\
\hline
\end{tabular}
\end{table*}
\subsection{Nuclear Spectroscopic Telescopic Array}
The NuSTAR observation of 1H~0323+342 was conducted on 2025 September 25 (obsid: 91101637002). The events were calibrated, screened, and filtered using the \emph{nupipeline} tool applying standard criteria. The products were extracted for both focal plane modules FPMA and FPMB. A circular region of 60 arcseconds centered at 1H~0323+342 was used as the source region, while the background region was set at a circular region of the same radius from a source-free position near the target. After this, the \emph{nuproducts} tool was used to obtain the spectra and light curves for both modules. The light curves were generated with 1000-sec,ond binning and the background subtraction was done using the package \emph{lcmath}. We chose {\tt saacalc=1}, {\tt saamode=OPTIMIZED}, and {\tt tentacle=NO} to filter out the high background flaring epochs caused by the South Atlantic Anomaly and to ensure that the observed flux variability is intrinsic to the source.  The spectra were rebinned to a minimum of 30 counts in each energy bin using the \emph{grppha} tool. The spectral fitting was carried out using XSPEC \citep{1996ASPC..101...17A}. 

\begin{table*}[]
    \centering
    \setlength\extrarowheight{5pt}
	\setlength{\tabcolsep}{12pt}
    \caption{The results of spectral analysis of X-ray and optical/UV observations. The NuSTAR and Swift-XRT fluxes were estimated covering the energy ranges of 3-70 keV and 0.3-10 keV, respectively. The \emph{Swift}-UVOT fluxes in the filters $U$, $B$, $V$ are in units of $10^{-11}$ erg cm$^{-2}$ s$^{-1}$ and those in $UVW1$, $UVM2$, $UVW2$ are in $10^{-10}$ erg cm$^{-2}$ s$^{-1}$ units. The \emph{Swift} observation with superscript 1 is used in SED1, while the observation with superscript 2 is used in SED2 and in the combined spectral fitting of XRT+NuSTAR observations as well.}
    \label{tab:nu}
    \begin{tabular}{llllr}
        \hline
     & & NuSTAR - FPMA + FPMB  & & \\
     \hline
     Model & Index/alpha($\Gamma_X$) & Curvature ($\beta$) & Un-absorbed flux & $\chi^2$(DOF) \\
     &&& (erg cm$^{-2}$ s$^{-1}$) &\\
     \hline
       Power-law &  $1.81^{+0.01}_{-0.01}$ & - &  $7.12^{+0.09}_{-0.08} \times 10^{-11}$
       & 442.75 (421) \\
       Logparabola & $1.93^{+0.04}_{-0.04}$ & $-0.12^{+0.04}_{-0.04}$ & $7.48^{+0.10}_{-0.09} \times 10^{-11}$ & 419.20 (420) \\
       \hline
       Obs.Id && \emph{Swift}-XRT (Power-law fitting) && \\
       \hline
        00035372003$^1$ & $1.60^{+0.10}_{-0.09}$ & -- & $8.40^{+0.58}_{-0.48}\times 10^{-11}$ & 64 (71)\\
     00035372004 & $2.12^{+0.13}_{-0.13}$ & -- & $4.08^{+0.41}_{-0.58}\times 10^{-11}$ & 78 (60)\\
         00035372005$^2$ & $2.17^{+0.09}_{-0.09}$ & -- & $6.40^{+0.67}_{-0.61}\times 10^{-11}$ & 55 (62)\\
         \hline
        Obs.Id & & \emph{Swift}-UVOT Fluxes &&\\
         \hline
         &  U & B & V \\
         \hline
         00035372003$^1$ & $3.51\pm 0.10$ & $2.40\pm 0.06$ & $1.02\pm 0.04$\\
         00035372004 &  $3.17\pm 0.09$ & $2.09\pm 0.06$ & $0.90\pm 0.04$ \\

          00035372005$^2$ & $3.47\pm 0.10$ & $2.27\pm 0.06$ & $0.86\pm 0.03$ \\
          \hline
          &  UVW1 & UVM2 & UVW2 \\
          \hline
          
         00035372003$^1$& $0.81\pm 0.02$ & $2.13\pm 0.06$ & $1.90\pm 0.04$  \\
         
         00035372004& $0.79\pm 0.02$ & $2.05\pm 0.07$ & $1.91\pm 0.05 $  \\
         00035372005$^2$& $0.85\pm 0.02$ & $2.32\pm 0.08$ & $2.12\pm 0.05 $  \\
              \hline
       & &\emph{Swift}-XRT + NuSTAR && \\
       \hline
       &Index1 ($\Gamma_1$) & Break energy ($E_b$) & Index1 ($\Gamma_2$) & $\chi^2$(DOF) \\
     &$2.3^{+0.4}_{-0.2}$ & $1.6^{+0.9}_{-0.5}$ & $1.81^{+0.01}_{-0.01}$ & 506.48 (483) \\

       \hline
  \end{tabular}
\end{table*}

\subsection{\emph{Swift} Observations}
The \emph{Swift}-X-Ray Telescope (XRT) observed the source thrice during September 15-30, 2025. The XRT spectra of these observations were obtained from the automated online tool $\tt{Swift-XRT \ data\ product's\ generator}$ \citep{2009MNRAS.397.1177E}. The spectral fitting was done in XSPEC using an absorbed power-law model with Galactic column density fixed to $N_{\rm H}=1.27\times10^{21}$\ cm$^{-2}$ \citep[][]{Kalberla_2005}. The X-ray spectra were rebinned to include at least 20 counts per bin.
The obtained spectra were fitted with a power-law, including Galactic absorption.

The \emph{Swift}-UVOT datasets were downloaded from the High Energy Astrophysics Science Archive Research Center and analyzed using standard procedures\footnote{\url{https://swift.gsfc.nasa.gov/analysis/threads}}. The tool \mbox{\emph{uvotimsum}} was used to sum the images over extensions in each filter. A circular region of radius 5 arcsec around the source was used for extracting the source counts, while a circle with a 20 arcsec radius was chosen from a nearby source-free region for background estimation. The aperture photometry task \emph{uvotsource} was considered to measure magnitudes in each filter, which were corrected for Galactic extinction by fixing \mbox{E(B-V)=\,0.561 for R$_V$=A$_V$/E(B-V)=3.1} \citep{Schlafly_2011}. We converted the source magnitudes to the energy flux units using the standard zero-points provided by \citet[][]{2011AIPC.1358..373B}.
\section{Results}
\label{sec:res}
\subsection{Gamma-ray Flux Variations}
\label{sec:res1}
\begin{figure*}
    \centering
    \includegraphics[scale=0.5]{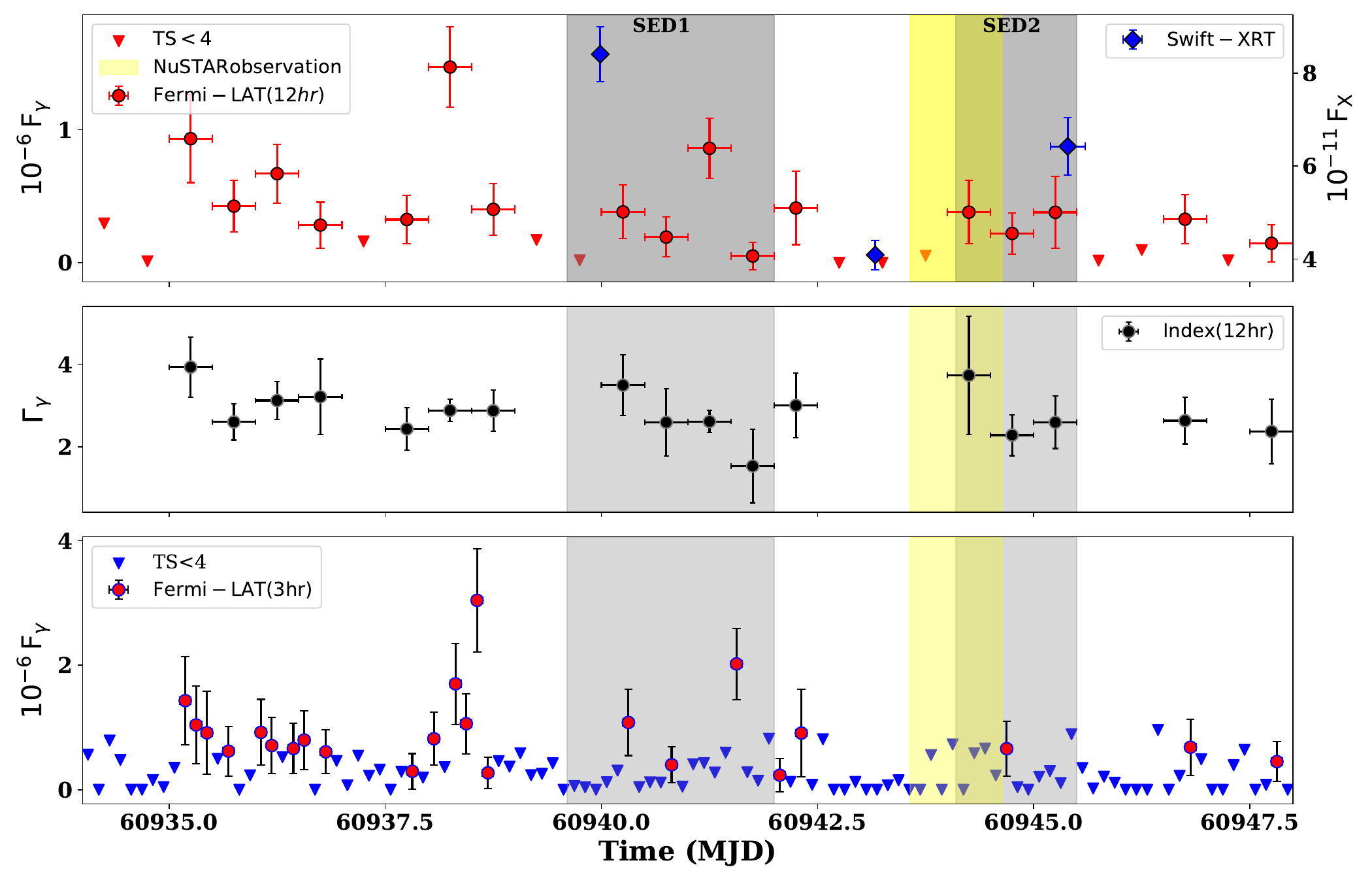}
    \caption{Top panel: \emph{Fermi}-LAT \gm-ray flux light curve. Detections with TS $\geq 4$ are shown by red circles, and inverted triangles are given for TS $<4$. The LAT flux is shown on the left axis in units of $ 10^{-6}\mathrm{ph\,cm^{-2}\,s^{-1}}$. The X-ray flux (\emph{Swift}-XRT) in the energy range 0.3-10 keV is shown in blue upper triangles. The X-ray flux (right axis) is shown in units of $ 10^{-11}\mathrm{erg\,cm^{-2}\,s^{-1}}$.
Middle panel: Corresponding LAT photon index ($\Gamma_{\gamma}$) is plotted across time.
Bottom Panel: \emph{Fermi}-LAT flux lightcurve with 3-hour binning.
Grey shaded regions highlight the SED epochs, and the yellow shaded region shows the NuSTAR observing time period.}
    \label{fig:lc}
\end{figure*}

To understand the \gm-ray flux variations, we examined the 12-hour binned lightcurve (figure \ref{fig:lc}). This binning yielded the highest flux of $(1.47\pm 1.19) \times 10^{-6}$ ph \ cm$^{-2}$\ s$^{-1}$ with a detection significance, TS of 65 on MJD 60938.5. 
The variability timescale of \gm-ray emission from jetted-AGNs often scales down to a few hours or even minutes \citep[]{2016ApJ...824L..20A}. Therefore, to better probe the possible rapid flux variations, we generated a 3-hour binned \gm-ray lightcurve, which showed a peak flux of $(3.04\pm 0.82)\times 10^{-6}$ ph \ cm$^{-2}$\ s$^{-1}$ on MJD~60938.5 (TS= 40). Additionally, we observed the fastest flux decay time of $0.84\pm 0.30$ hour in the time bin MJD~60938.5 - 60983.62 with 3.2 $\sigma$ significant change in flux\footnote{We used the equation $F(t)=F(t_0)2^{-(t-t_0)/t_{\rm var}}$ for searching rapid variability. Here, $F(t)$ and $F(t_0)$ are the fluxes at consecutive times $t$ and $t_0$ respectively, with a minimum of 3$\sigma$ significance for the change in flux. $t_{\rm var}$ is the flux doubling or halving time.}.

The observed peak flux (3-hour binned) corresponds to an isotropic \gm-ray luminosity of $\sim1\times \,10^{46}$ erg \ s$^{-1}$, which is relatively lower than the same observed for the previous \gm-ray flare ($4.7\times \,10^{46}$ erg \ s$^{-1}$).  However, this is still $\sim$ 100 times greater than the average 4FGL-DR4 catalog value of $\sim 2\times \,10^{44}$ erg\ s$^{-1}$.

\begin{figure}
\centering
\includegraphics[scale=0.45]{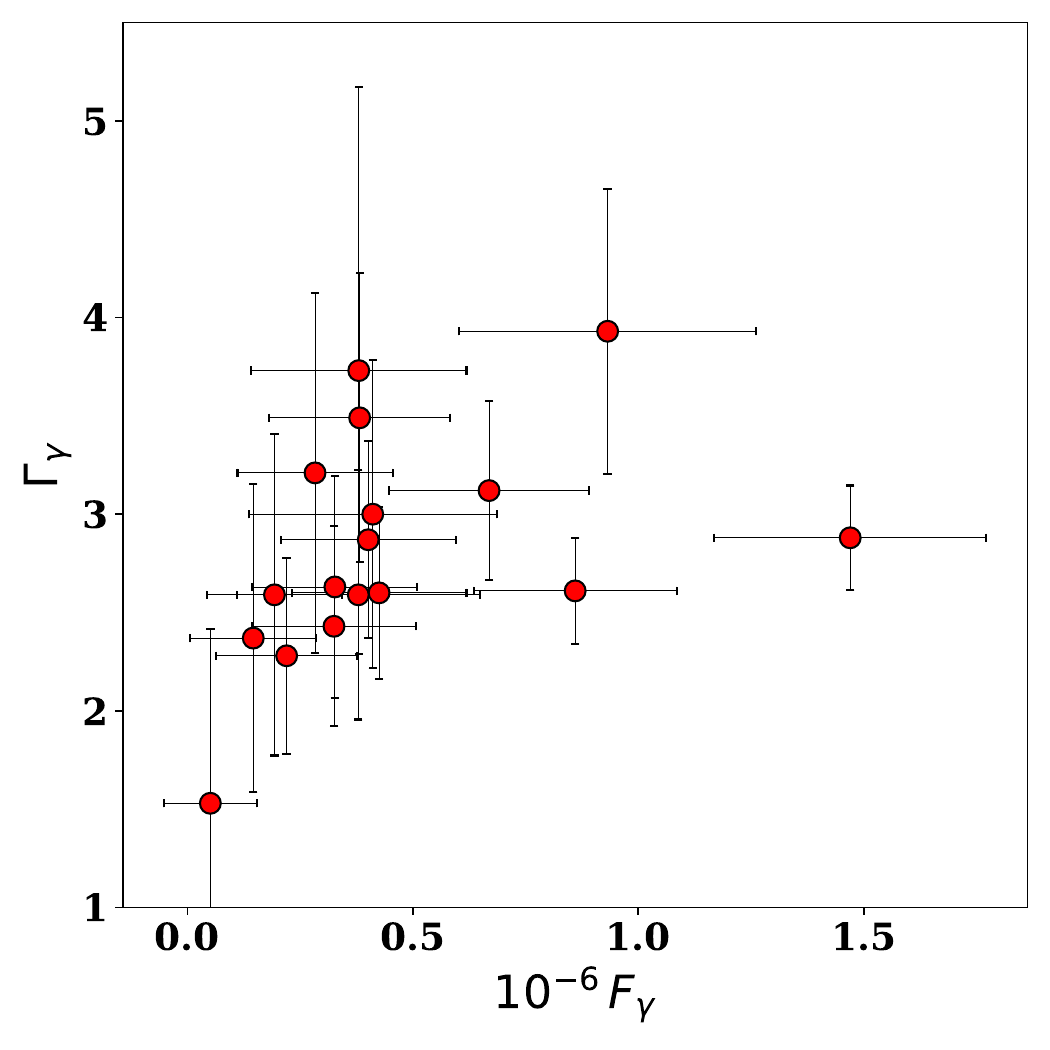}
\caption{The scatter plot between \gm-ray index and flux in the energy range 0.1-300 GeV. Fluxes are in the units of $10^{-6}\,\mathrm{ph\,cm^{-2}\,s^{-1}}$. }
\label{fig:gvf}
\end{figure}

The average \gm-ray spectrum during the period MJD 60932-60948 is fitted with both power-law and log-parabola models. However, we cannot claim significant curvature in the \gm-ray spectrum, as both fits yield similar statistics (Table \ref{tab:lat_spec}). The variation of \gm-ray photon index is displayed in the middle panel of figure \ref{fig:lc}. 
Figure \ref{fig:gvf} shows an apparent ``softer-when-brighter'' behavior.
The strength of the correlation was quantified using a Monte-Carlo assessment incorporating the uncertainties in the flux and index values. This exercise revealed that the correlation is marginal: the resulting distribution yields a Spearman rank-correlation coefficient $\rho$ of $0.28\pm0.18$. 

Under the opacity arguments of \gm-ray photons against pair absorption, the highest energy found for the detected photons can constrain the location of \gm-ray production site in the jet \citep[]{2006ApJ...653.1089L, Costamante_2018}. The highest energy photon with $>$95\% association probability was detected on MJD~60938.5 with the energy of 1.3 GeV. Interestingly, no $>$10 GeV photon was also detected during 2013 September \gm-ray outburst of \src. Together with the observation of a soft \gm-ray spectrum, rapid flux variability, and a low Doppler factor (section~\ref{sec:sed_r}), these results indicate the possible absorption of higher energy photons, possibly due to the compactness of the emission region \citep[cf.][]{2008ApJ...686..181F}, or due to the lack of sufficient energetic electrons to radiate high-energy \gm-ray photons. Alternatively, the non-detection of the higher energy \gm-ray photons could also be due to the emission region being located close to the central engine, where $\gamma \gamma$ pair production with the corona photon field can effectively absorb $>$10 GeV photons \citep[e.g.,][]{2009MNRAS.397..985G}.  

\subsection{X-ray Flux Variations}
\label{sec:xres}
\begin{figure*}
\centering
\includegraphics[scale=0.5]{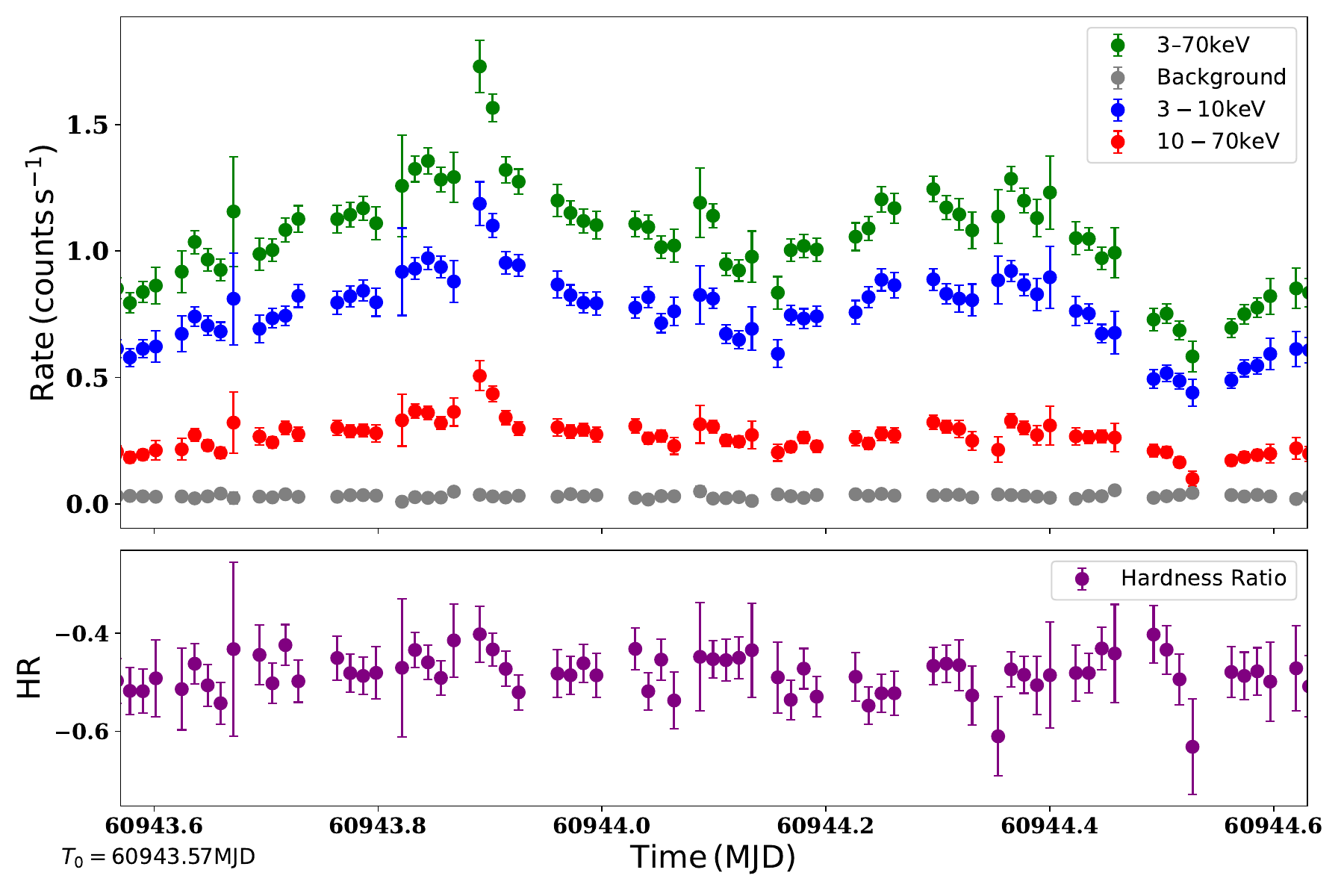}
\caption{
Top panel: Source count rate (green circles) and background noise (gray) in the full energy band (3--70 keV), together with the background-subtracted low and high-energy light curves (blue and red circles, respectively). All lightcurves are binned with the same temporal resolution of 1000s. Bottom panel: The temporal variation of the hardness ratio (HR).}
\label{fig:nulc}
\end{figure*}

The \gm-ray flare peaked on MJD~60938.5 was not covered at other frequencies. However, the source exhibited a bright X-ray flare with a significant spectral hardening ($\Gamma_x=1.60^{+0.10}_{-0.09}$) on MJD~60940. Interestingly, the source was moderately active in the \gm-ray band during the epoch of the \emph{Swift} observation. 
After this, NuSTAR observed \src~ for $\sim$48 ksec during 60943.57-60944.63 MJD, which revealed two distinct flares as can be seen in the 1000-second binned lightcurve (Figure~\ref{fig:nulc}). A rapid flux variation was identified on MJD~60943.8 with the shortest flux doubling time of $0.95\pm0.24$ hours at 3.3$\sigma$ significance level. 

We also looked for the energy-dependent variability by splitting the NuSTAR lightcurve into low-energy (3-10 keV) and high-energy (10-70 keV) bands as shown in the figure \ref{fig:nulc}. As can be seen, the flux variations are more pronounced in the low-energy band (more than twice the magnitude of those in the high-energy band). This suggests that low-energy electrons are more likely to contribute to the upscattering of photons. However, the variability pattern appears similar in both energy bands. We also estimated the hardness ratio which is defined as $HR = \frac{F_{\mathrm{hi}} - F_{\mathrm{low}}}{F_{\mathrm{hi}} + F_{\mathrm{low}}}$, where $F_{\mathrm{hi}}$ and $F_{\mathrm{low}}$ are the background subtracted count rates 
in the high and low energy bands, respectively. Given the large uncertainties, no strong claim can be made regarding the spectral variations (Figure~\ref{fig:nulc}, bottom panel).

\subsection{X-ray spectral study}
The \emph{Swift}-XRT spectra covering 0.3$-$10 keV are fitted with a single power-law model, and the best-fit parameters are listed in Table~\ref{tab:nu}.
We fitted the NuSTAR FPMA and FPMB spectra with both power-law and log-parabola models, and the corresponding best-fit parameters are given in Table~\ref{tab:nu}. The F-test performed on the chi-square statistics of the two models indicates that the log-parabola model provides a significantly better description of the NuSTAR spectrum, with a null-hypothesis probability of $1.7\times10^{-6}$.

We also attempted a joint spectral fit of the simultaneous \emph{Swift}-XRT and NuSTAR observations (obsid: 00035372005 and 91101637002, respectively). A broken power-law model was found to provide a better fit to the combined X-ray spectrum, yielding a break energy of $E_b = 1.6^{+0.9}_{-0.5}$ keV, with photon indices $\Gamma_1 = 2.3^{+0.4}_{-0.2}$ and $\Gamma_2 = 1.81^{+0.01}_{-0.01}$ below and above the break, respectively. The $F$-test comparing the broken and power-law models confirmed this improvement with $\chi^2(\mathrm{DOF})_{\mathrm{BKN}} = 506.48 (483)$, $\chi^2(\mathrm{DOF})_{\mathrm{PL}} = 529.75 (487)$, and the null hypothesis probability $2.2\times 10^{-4}$.
\begin{figure*}
\centering
	\begin{subfigure}{0.5\textwidth}
		\includegraphics[scale=0.32]{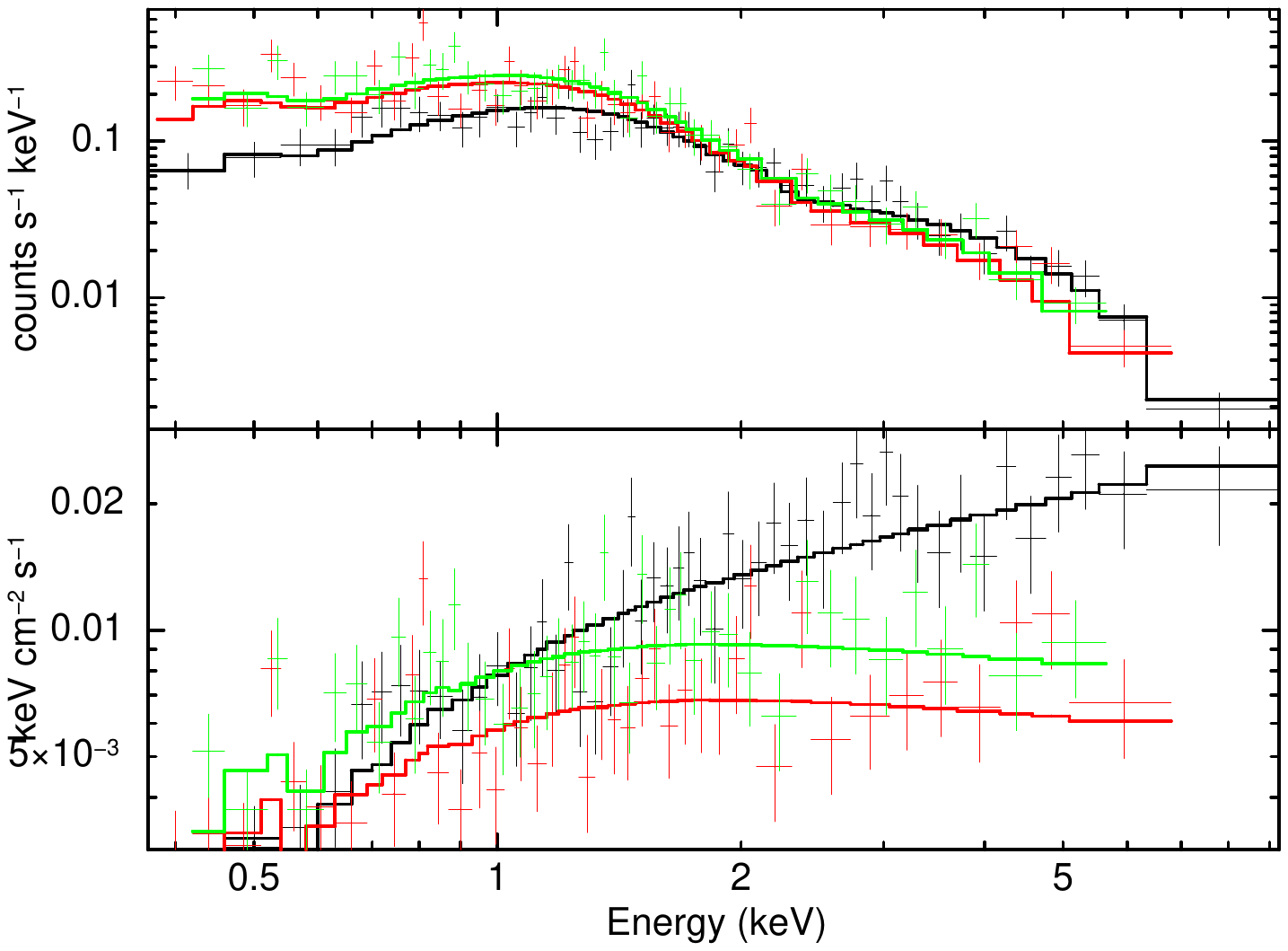}
		\caption{}
		\label{fig:a}
	\end{subfigure}
\hspace{-1cm}
	\begin{subfigure}{0.5\textwidth}
		\includegraphics[scale=0.32]{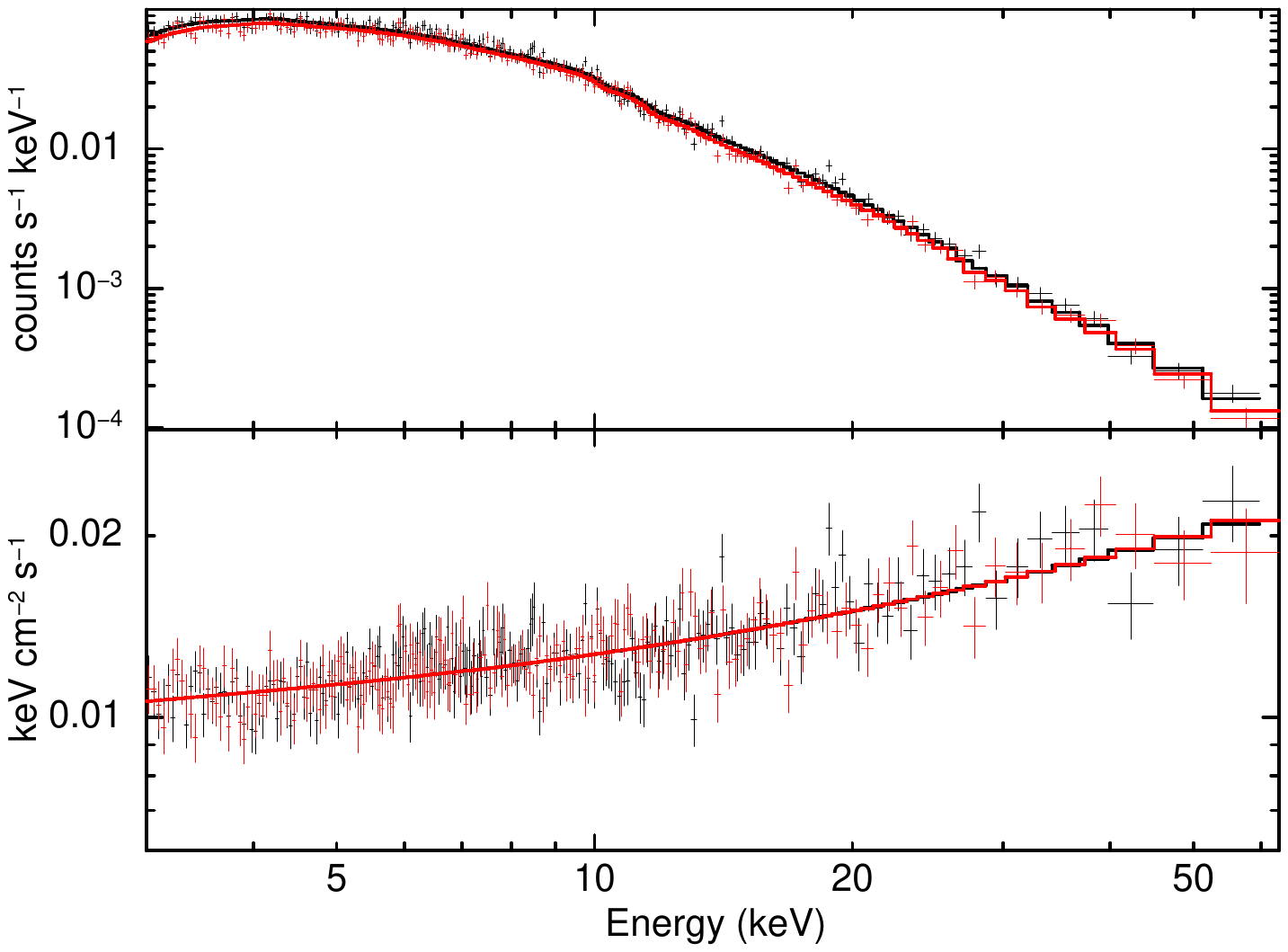}
		\caption{}
		\label{fig:b}
	\end{subfigure}
	\smallskip
	\begin{subfigure}{0.5\textwidth}
		\includegraphics[scale=0.32]{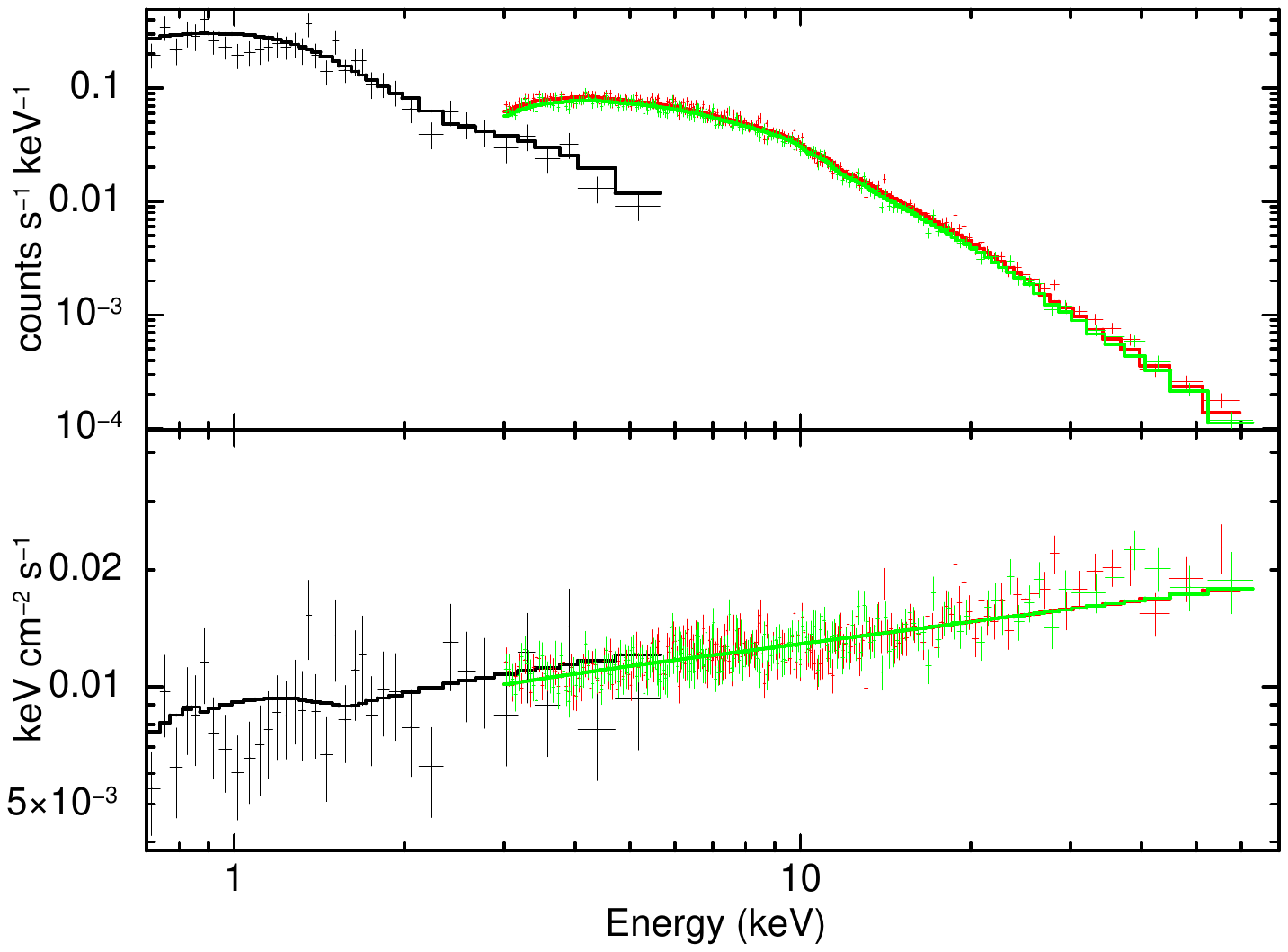}
		\caption{}
		\label{fig:c}
	\end{subfigure}
\caption{The figure shows the X-ray spectra and the corresponding best-fit models generated using XSPEC: 
(a) \emph{Swift}-XRT observations 00035372003 (black), 00035372004 (red), and 00035372005 (green). (b) \emph{NuSTAR} FPMA and FPMB spectra fitted with a log-parabola model. 
(c) The combined \emph{Swift}-XRT observation 00035372005 and \emph{NuSTAR} spectra fitted with a broken power-law model. In all plots: Top panels) Folded spectrum [(\(\mathrm{Counts\,keV^{-1}\,s^{-1}}\)) versus energy (\(\mathrm{keV}\))], and Bottom panel) Unfolded spectrum [(\(\mathrm{keV\,cm^{-2}\,s^{-1}}\)) versus energy (\(\mathrm{keV}\))]. }
\label{fig:XSPEC}

\end{figure*}

\subsection{Broad-band SED analysis}
\label{sec:sed}
\subsubsection{The model}
To get insights about the radiative mechanisms responsible for the recent outburst of \src, we used a one-zone leptonic model, which is commonly adopted to explain the SEDs of jetted AGN \citep[]{2009ApJ...692...32D,2009MNRAS.397..985G}. The model assumes a geometrically thin, optically thick accretion disc \citep[]{1973A&A....24..337S} surrounding the central supermassive black-hole of mass $M_{\rm BH}$. The disc emission is considered to follow a multi-color blackbody with inner and outer radii $3R_{\rm sch}$ and $500R_{\rm sch}$, respectively, where $R_{\rm sch}$ is the Schwarzschild radius. The accretion disc luminosity is denoted by $L_{\rm disc}$. Along with the accretion disk, we also considered emission from the corona, BLR, and dusty torus. The luminosities of these components are parameterized as the fractions of the disk luminosity, such that $L_{\rm j}=f_{\rm j}L_{\rm disc}$, where $f_{\rm j}$ represents the fraction of the accretion-disk radiation reprocessed by the corresponding components \cite{2002ApJ...577...78S, 2009MNRAS.397..985G}. 
The spectrum of the corona radiation is assumed to follow a power law with an exponential cut-off, $L_{\nu}\propto\nu^{-\alpha_{\rm cor}}\exp(-h\nu/\nu_c)$ and is normalized such that its integrated luminosity equals $L_{\rm cor}=f_{\rm cor}L_{\rm disc}$. Following \citet{2009MNRAS.397..985G}, the coronal component is included above the peak frequency of the accretion-disk emission with a cut-off at 150~keV.
Both the BLR and torus are treated as blackbody emitters with their spectra peaking at rest-frame Lyman-$\alpha$ and at a frequency equivalent to the characteristic torus temperature $T_{\rm tor}$, respectively. The spatial extent of BLR and torus are constrained by the accretion disc luminosity following $R_{\rm BLR}=10^{17}L_{\rm disc,45}^{1/2}$ cm and $R_{\rm tor}=2.5\times10^{18}L_{\rm disc,45}^{1/2}$ cm, where $L_{\rm disc,45}$ denotes the disc luminosity in units of $10^{45}$~erg\,s$^{-1}$ \citep[]{2002ApJ...577...78S,2000ApJ...545..107B}.

The spherical emission region is considered to be propagating down the conical jet with a semi-opening angle of 0.1 rad. The emission region size is calculated from the variability timescale $t_{\rm var}$ and adopting the causality relation ($R_{\rm blob}\lesssim ct_{\rm var}\delta/(1+z)$). The emission region is assumed to be filled with energetic electrons whose energy distribution follows a smooth broken power-law shape as given by
\begin{align}
    N(\gamma)\,d(\gamma) = \frac{k_0 \, \gamma_b^{-p}}{\left(\frac{\gamma}{\gamma_b}\right)^{p} + \left(\frac{\gamma}{\gamma_b}\right)^{q}} \,d(\gamma) \quad \textrm{cm}^{-3}
\end{align}\label{eqn:1}
where $\gamma$ is the electron's Lorentz factor or dimensionless energy, $p$ and $q$  are the particle indices below and above the electron's break energy $\gamma_{\rm b}$. 
In the presence of a homogeneous yet tangled magnetic field, $B$, relativistic particles cool via synchrotron and inverse-Compton processes \citep[e.g.,][]{2009herb.book.....D}. For the external Compton (EC) component, we considered the energy densities of the disc, BLR, and torus photon fields in the comoving frame following the prescriptions of \citet[]{2009MNRAS.397..985G}. Finally, we derived the jet power assuming an equal number density of radiating electrons and protons, which are considered to be cold and carry the inertia of the jet \citep[]{2008MNRAS.385..283C}.

\subsubsection{SED Epochs}
The peak of the \gm-ray flare observed on MJD~60938.5 was not covered simultaneously at other wavelengths, as shown in figure~\ref{fig:lc}. Therefore, we selected two epochs for the SED analysis based on the availability of multi-wavelength information and the source activity. The first selected period (MJD~60939.4 - 60942) included the highest X-ray flare observed by \emph{Swift}-XRT (obsid: 00035372003), when the \gm-ray activity is relatively lower. The second SED covers the time interval MJD~60944.2 - 60945.5, which corresponds to the second flare in the NuSTAR lightcurve, which was also supplemented by \emph{Swift}-XRT observations (Figure~\ref{fig:lc}). 
The time periods were determined while also ensuring the source was significantly detected in the \gm-ray band (TS$>$10). \src~was \gm-ray undetected during the first flare observed in the hard X-ray band ($\sim$ MJD 60943-60944.2), and hence the corresponding epoch was not considered for the SED modelling. In the following sections, we refer to the two SED epochs as SED1 (high X-ray flux state) and SED2 (variable hard X-ray state) and highlight them as gray shaded regions in figure~\ref {fig:lc}.

\subsubsection{SED Modeling Constraints}
All \gm-NLSy1 galaxies exhibit bright optical emission lines with rest-frame EWs larger than 5\AA, indicating a radiatively efficient accretion process \citep[e.g.,][]{2017ApJS..229...39R,2019ApJ...872..169P}. In the SED modeling of such broad emission line sources, $M_{\rm BH}$ and $L_{\rm disc}$ are the two key parameters,
since the sizes of the BLR/Torus/corona and corresponding energy densities have a direct dependence on $L_{\rm disc}$ \citep[e.g.,][]{2015MNRAS.448.1060G}.
For \src, these were constrained to be $\sim10^7\, M_\odot$ and $\sim10^{45}$ erg s$^{-1}$, based on single-epoch optical spectroscopy, reverberation mapping, big blue bump modeling, and X-ray variability \citep{2016ApJ...824..149W,2017MNRAS.464.2565L,2018ApJ...866...69P}. The emission region size was constrained from the shortest flux variability time of $\sim$50 minutes observed at hard X-ray and \gm-ray bands. 
The spectral slopes of the particle energy distribution, i.e $p,q$ can be constrained from the photon spectral index $\Gamma_{\rm X/\gamma}$ of the observed X-ray and the \gm-ray spectra.
The observed X-ray spectrum during SED1 lies entirely on the rising edge of the IC component and hence may correspond to the low-energy part of the particle energy distribution. Hence, an initial guess for the value of $p$  is obtained as $2\Gamma_X-1\,=\,2.2$. A fair guess on $q$ is also made in the same way using the \gm-ray photon index as $\approx4$.
The break energy ($\gamma_{\rm b}$) can be constrained by adjusting $p$ and $q$, which shifts $\gamma_{\rm b}$ to the low/high energy part of the distribution. The normalization of the particle energy distribution is computed from the synchrotron emission, which is expected to lie in the radio band. 
For example, the observed synchrotron flux can be expressed as \citep{2008ApJ...686..181F}
\begin{equation}
\label{fesynexact}
f_\e^{syn} = \frac{\sqrt{3}\delta^4 \e^{\prime}e^3 B}{4\pi h d_L^2}
   \int^{\infty}_1 d\gp\ N_e^\prime(\gp)\ R(x)\ .
\end{equation}
where $x = \dfrac{4\pi \ep m_e^2 c^3}{3eB h\gamma^{\prime 2}}$ and primed quantities are measured in the comoving frame. The approximated expression for $R(x)$ is given by
\begin{equation}
R(x) = \left\{ \begin{array}{ll}
	1.80842\ x^{1/3} & x \ll 1 \\
	\frac{\pi}{2}e^{-x}\left[ 1 - \frac{99}{162x}\right] & x \gg 1 \\
	\end{array}
	\right. \ 
\end{equation}\label{eqn:2}
Here, $m_e$ and $e$ are the electron mass and charge, respectively, while $\e$ and $\ep$ are the photon energies in the observed and in the co-moving frame. 
Since there are no radio observations available quasi-simultaneous to the \gm-ray flare, a reasonable approximation would be to consider the synchrotron emission at the level of the archival radio flux, i.e., $f_\e^{syn}$. Inserting $f_\e^{syn}$ and corresponding frequency ($\e$) and $N'_e(\gamma')$ (equation~\ref{eqn:1}) in equation~\ref{eqn:2} and assuming a certain magnetic field, we can solve it to determine $k_0$.

The X- and \gm-ray SEDs can give hints about the target photon field and the bulk Lorentz factor. In our model, the target photon energy densities for the EC process depend on the distance of the emission region from the central black-hole, i.e., R$_{\rm diss}$. The EC peak frequency also depends on the bulk Lorentz factor and the characteristic frequency of the target photon field, thus enabling us to effectively constrain the primary \gm-ray emission mechanism and the location of the emission region along the jet. The level and shape of the X-ray spectrum also allows to constrain the level of the SSC flux, which, in turn, controls the magnetic field strength and the size of the emission region \citep[e.g.,][]{2009ApJ...692...32D,2017ApJ...851...33P}. Unlike blazars, the X-ray spectrum in \gm-NLSy1 can have a contribution from the corona in addition to jetted radiation \citep[e.g.,][]{2019ApJ...872..169P}. Hence, the shape and flux in the X-ray band can bring insights into the possible contributions from different emission components. 
In the modeling of SED2, the soft X-ray spectrum hints at the presence of a coronal component with a spectral index of 1.1 \citep[see, e.g.,][]{2019ApJ...872..169P}.
Finally, we tried to vary the minimum number of SED parameters when modeling both SEDs and attempted to maintain the equipartition between the particle and magnetic energy densities.
\subsubsection{Results}\label{sec:sed_r}
\begin{figure*}
\centering
\includegraphics[width=0.49\textwidth]{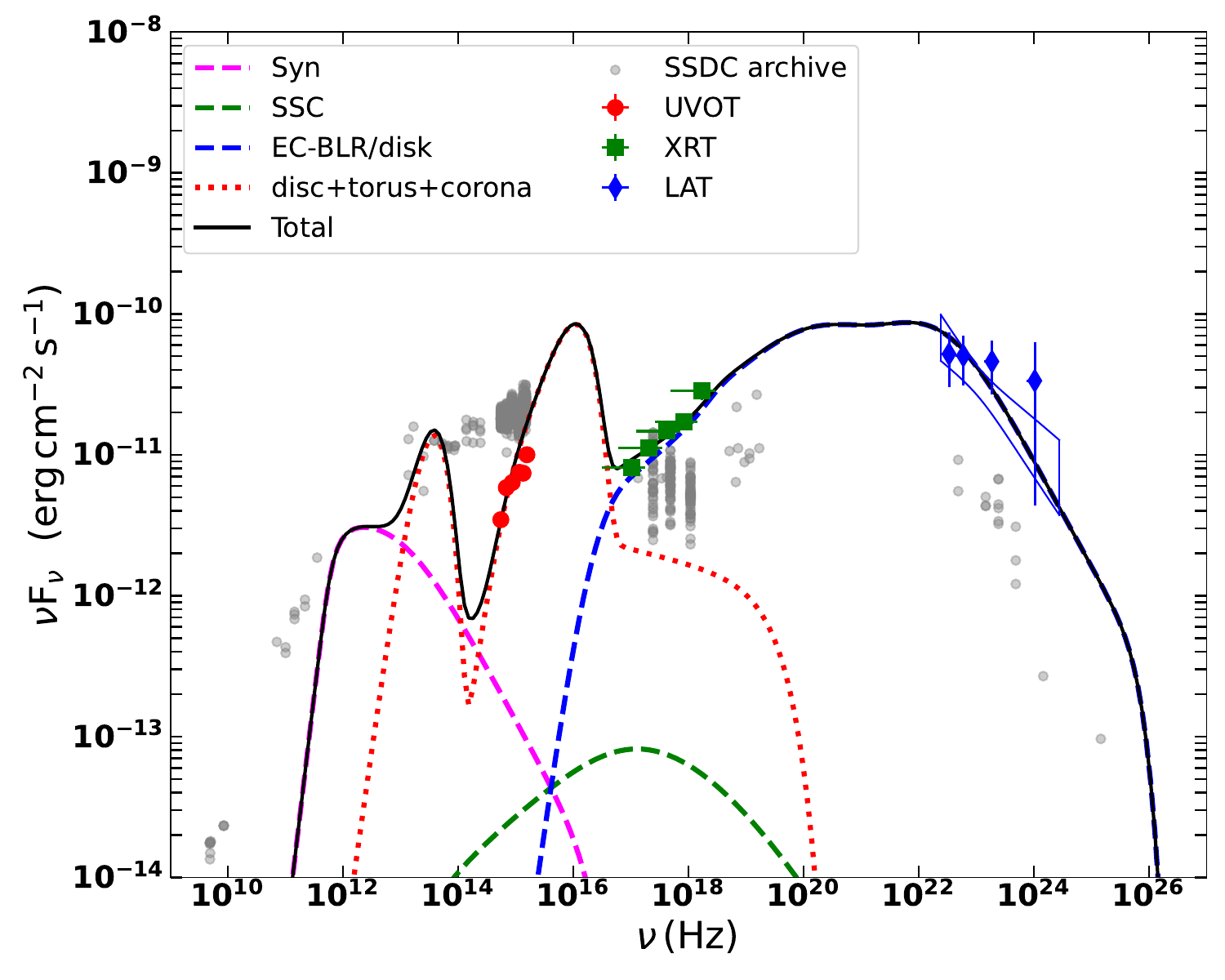}
\includegraphics[width=0.49\textwidth]{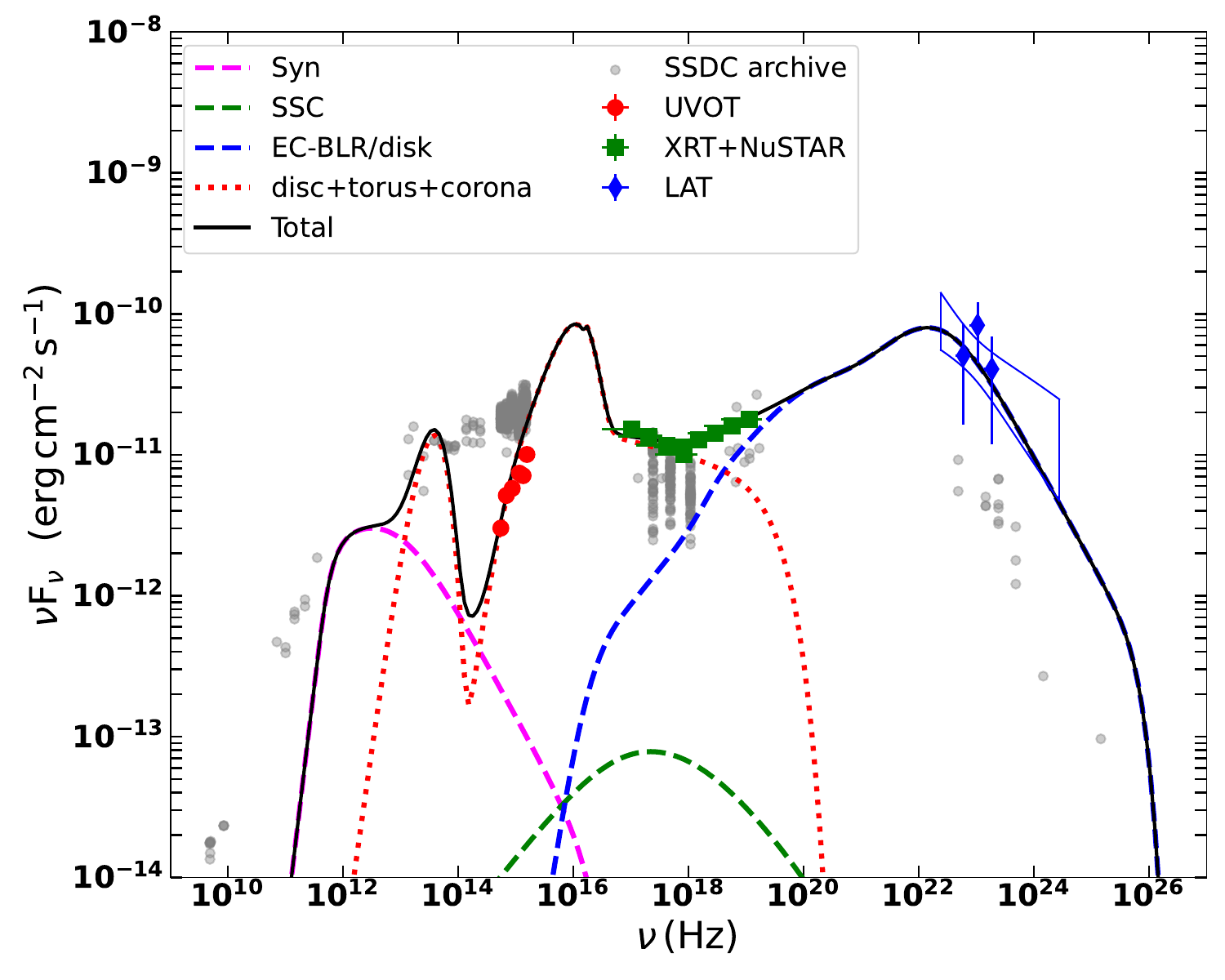}
\caption{The figure shows the observed broadband SEDs corresponding to the time intervals SED1: MJD~60943.6--60944.2 (left) and SED2: 60944.2--60944.6 MJD (right) along with the best-fitted models. In the plots, the dashed curves display the non-thermal components: synchrotron (purple), SSC (green), and EC (blue). The dotted curve gives the sum of the emissions from the torus, disc, and corona.}
\label{fig:sed}
\end{figure*}
The broadband SEDs covering the two epochs of observations were reasonably reproduced by the one-zone leptonic model. 
In figure~\ref{fig:sed}, we show the observed SEDs along with different emission model components. We also plot the archival observations representing the average activity state of 1H 0323+342. The best-fit SED parameters are provided in Table~\ref{tab:sed}. 
\section{Discussion}

\subsection{Flux Variability and Its Implications}
High-energy or $\gamma$-ray flares are far less common in NLSy1 galaxies, making each event valuable for understanding the powerful jets in these systems. \src~is one of few NLSy1 systems that has exhibited $\gamma$-ray outbursts more than once. In its 2025 September \gm-ray flare, the peak $\gamma$-ray flux reached $(3.04\pm 0.82) \times 10^{-6}$~ph~cm$^{-2}$~s$^{-1}$, exhibiting a marginal positive correlation between the $\gamma$-ray flux and the photon index (``softer when brighter''). The positive/negative correlation between fluxes and indices in optical, X-ray, and \gm-ray energies is very often observed in blazars \citep{2017ApJ...841..123P,2018A&A...619A..93B,2020ApJS..247...27K}. 

The detection of rapid flux variability in high-energy bands on a timescale of 
$t_{\rm var} \approx  0.9$~hour imposes a strong causality on the size 
and the location of the emitting region. Using the light-travel time argument, the radius of the emission zone can be expressed as 
\begin{equation}
    R \lesssim \frac{\delta}{1+z} \, c\, t_{\rm var} \,
        \approx \,10^{14}\,\delta_{1}~{\rm cm},
\end{equation}
where $\delta_{1}=\delta/10$. Even for moderate Doppler boosting, this 
corresponds to a compact region, nearly 100 times the gravitational 
radii for a black hole mass of $M_{\rm BH}\simeq 1\times10^{7}\,M_{\odot}$ (with gravitational radius of $r_{g}\approx 1.5\times10^{12}$~cm). Assuming a conical jet scenario, the distance of the emission region from the central black hole with jet opening angle $\theta_{\rm jet}\sim 1/\Gamma$, can be written as 
\begin{equation}
    D \approx \frac{R}{\theta_{\rm jet}} \approx R\,\Gamma 
        \approx 10^{15}~{\rm cm},
\end{equation}
implying that the $\gamma$-ray emitting region lies well within a sub-parsec 
scales at $\approx$ 1000$r_g$. This compactness is notably interesting when compared to the expected radius of the broad-line region,
    $R_{\rm BLR} \simeq 10^{17} 
    \left( \dfrac{L_{\rm disc}}{10^{45}~{\rm erg\,s^{-1}}} \right)^{1/2}
    \approx 1.0\times10^{17}~{\rm cm}$, 
for a disc luminosity of $L_{\rm disc} \approx 10^{45}~{\rm erg\,s^{-1}}$ \citep{2009MNRAS.397..985G}.  
The inferred emission-zone distance is therefore well inside the BLR. 
In this regime, the target photon field is dominated by the dense UV 
radiation from the BLR clouds or photons from the disc/corona, leading to a high external energy density in the comoving frame. Such intense radiation fields can be effectively upscattered to produce high-energy photons; on the other hand, increased $\gamma$--$\gamma$ absorption opacity can strongly influence both the spectral shape and the cooling timescales \citep{2011MNRAS.417L..11S, 2013ApJ...771L...4C, 2021MNRAS.502.5875K}. The observed fastest 
variability, therefore, supports the scenario of high-energy photons originating deep within the BLR, inside a compact and efficient dissipation region. 

\subsection{Broadband Modeling}
\label{sec:sed_dis}
In this study, we performed the broadband SED analysis for two epochs selected on the basis of the source's activity and the availability of information in various energy bands. The first SED involves the high X-ray state shown by the \emph{Swift}-XRT observation. The second selected epoch corresponds again to a high activity state in X-ray, including the second flare in the NuSTAR lightcurve.

For both SEDs, the optical-UV emission was found to be dominated by the accretion disc. Interestingly, the optical-UV flux appeared to be lower compared to the archival observations. The X-ray spectrum during SED1, i.e., the brightest X-ray state, shows a flat, rising shape ($\Gamma_{\rm X}=1.6^{+0.10}_{-0.09}$), which is reproduced by external Compton scattering of the disc photons, thus indicating the X-ray flux enhancement to be related to the jet activity. The X-ray spectrum observed during SED2, on the other hand, revealed a prominent spectral break with a soft spectrum ($\Gamma_{\rm X}>2$) below 2 keV and a rising spectrum ($\Gamma_{\rm X}<2$) at higher energies. We modeled the observed X-ray spectrum with a combination of the corona and EC mechanisms, with the former dominating at soft X-rays ($<$2 keV). Furthermore, the \gm-ray spectrum remained soft during both epochs, which we modeled with the EC process. Similar to previous works, we found that the primary source of seed photons for EC scattering is the accretion disc and BLR \citep[see, e.g.,][]{2018MNRAS.475..404K}. This finding indicates the emission region to be located close to the black hole, within the BLR. 
Based on our SED modeling analysis, the contribution of the SSC mechanism to the observed X-ray spectrum is negligible.

A comparison of the modeled X-ray spectral shapes across the two SED epochs provides clues to the possible coronal flux variations. The X-ray spectrum corresponding to SED1 is flat and shows a rising slope likely due to jet-dominated emission, with coronal flux below the observed flux level. We constrained the corresponding corona luminosity to be $6\times10^{43}$ erg s$^{-1}$, i.e., $\sim$6\% of $L_{\rm disc}$, which is lower than that found in previous work \citep[][]{2018MNRAS.475..404K,2019ApJ...872..169P}. During the SED2 epoch, on the other hand, the 0.3$-$70 keV emission revealed a spectral break ($E_{\rm break}=1.59^{+0.89}_{-0.52}$ keV). We modeled the X-ray spectrum below the break energy, assuming it originates from the corona, which required a higher corona luminosity of $\sim3\times10^{44}$ erg s$^{-1}$. Although coronal flux variations have been observed in radio-quiet Seyfert 1 galaxies \citep[e.g.,][]{2015MNRAS.454.4440W}, such detections are rare in radio-loud NLSy1 galaxies. Before the present study, a comparable X-ray spectral transition attributed to the change in the relative contributions of the jet and Seyfert-like components was also reported for PMN~J0948+0022 by \citet{2015MNRAS.446.2456D}.

One could argue that the soft X-ray spectrum below the break energy can be due to the tail part of the synchrotron spectrum. However, a steep falling synchrotron spectrum is usually detected in the X-ray observations of high-synchrotron peaked BL Lac objects that have a hard \gm-ray spectrum, indicating the inverse Compton peak to be located at very high energies \citep[$>$100 GeV; e.g.,][]{2026ApJS..282...35A}. The \gm-ray spectrum of \src, on the other hand, is soft, thus confirming that the soft X-ray spectrum cannot originate from the synchrotron mechanism. Another possibility could be the SSC origin of the X-ray emission below the break energy. However, it will require injecting more energetic particles, increasing the jet power well above the Eddington limit, which is energetically infeasible. Moreover, since the same electron population is responsible for the synchrotron, SSC, and EC radiations, reproducing the soft X-ray spectrum with SSC demands the synchrotron emission to be peaked at even lower frequencies, hence a lower break Lorentz factor. In turn, the EC mechanism would peak at lower energies, which is unlikely given that the EC model is well constrained from the observed hard X-ray and \gm-ray spectra. Therefore, the corona-origin of the soft X-ray emission is plausible.

\subsection{Comparison with Other Works}
Our SED modeling parameters across the two states (Table \ref{tab:sed}) closely align with a low-power, moderately beamed jet regime, which characterizes $\gamma$-ray loud NLSy1 galaxies in general. The previous study on a sample of 16 $\gamma$-ray loud NLSy1 sources using one-zone leptonic model yields systematically smaller bulk Lorentz factors (typically $\Gamma<15$) and lower radiative jet powers ($\lesssim 10^{45}$ erg s$^{-1}$) than those of classical FSRQs \citep[][]{2015MNRAS.448.1060G,2019ApJ...872..169P}, consistent with our estimations, $\Gamma = 10$ and $P_{\rm rad}<10^{44}$ erg s$^{-1}$. 
However, in a few \gm-ray flaring NLSy1 sources like PKS~1502+036 and PKS~2004-447, the GeV flares have been explained by an enhancement of the bulk Lorentz factor to $>$20 \citep{2016ApJ...820...52P,2021A&A...649A..77G}. 

The previous $\gamma$-ray flare of \src was studied in detail by \citet{2014ApJ...789..143P}, where the \gm-ray emission peak was considered in the SED analysis, due to the availability of the multi-wavelength data. However, our SED epochs match the lower $\gamma$-ray activity state, yet the X-ray flux exceeded previous measurements (see Figure~\ref{fig:sed}), resulting in different SED parameters compared to the earlier analysis. In particular, our study indicates a magnetic field strength of 1.6~G, which is smaller than $\sim5-7$ G reported in \citet{2014ApJ...789..143P}. We found that the slope of the particle spectrum before the break energy is slightly softer ($p=2-2.3$) than the previous flaring period ($p=1.2-1.6$). The inferred jet power\footnote{It is unfortunate indeed that the MOJAVE program stopped monitoring the VLBI jet in 1H 0323+342 after 2019.} during the high X-ray flux state ($\sim 10^{46}\,\mathrm{erg\,s^{-1}}$) is comparatively higher than that reported for previous flare ($\sim 10^{44}$--$10^{45}\,\mathrm{erg\,s^{-1}}$). The distances of the emission region observed for the two SEDs are 1150 and 1550 $R_{\rm sch}$, which is in the order of $10^{15}$ cm. For an emission region at such a distance from a black hole of mass $10^7~M_\odot$, the BLR photons will provide the ambient target field for the EC in the comoving frame \citep{2009herb.book.....D}.
Hence, we can infer that the origin of GeV photons from \src~~is external Compton scattering of BLR photons, similar to earlier studies \citep[][]{2014ApJ...789..143P}. 
Our results also show a significant contribution of coronal emission in X-rays with a notable increase from SED1 to SED2.

A recent multi-wavelength study focused on the long-term flux variations of \src~by \citet[][]{2025A&A...698A.160R} proposed an intermittent jet scenario, driven by radiation-pressure instability in the accretion disc, resulting in three different activity zones in the X-ray emission, 
\begin{enumerate}
    \item Dominated by jet, $F_{0.3-10\,\mathrm{keV}} \gtrsim 10^{-11}\,\mathrm{erg\,cm^{-2}\,s^{-1}}$ and $\Gamma \lesssim 1.9$.
    
    \item $F_{0.3-10\,\mathrm{keV}} \gtrsim 10^{-11}\,\mathrm{erg\,cm^{-2}\,s^{-1}}$ and $\Gamma \gtrsim 1.9$.
    
    \item Dominated by corona, $F_{0.3-10\,\mathrm{keV}} \lesssim 10^{-11}\,\mathrm{erg\,cm^{-2}\,s^{-1}}$ and $\Gamma \lesssim 1.9$.
\end{enumerate}
Furthermore, the source transitioned to an oscillation between zones 2 and 3 after 2017 \citep[]{2025A&A...698A.160R}. Our results provide tentative indications that \src~may be transitioning back to a regime of oscillation between zones 1 and 2, causing the observed variability.

\section{Summary}
The recent high-energy flaring episode exhibited by 1H~0323+342 in 2025 September was only the second observed \gm-ray flare since its discovery in the \gm-ray band, offering a unique opportunity to understand the jet properties from a low mass black hole system. We have presented a dedicated study of this recent active phase in $\gamma$-ray and X-ray bands, combining temporal, spectral, and broadband SED modeling analysis. The main findings are summarized below.

\begin{enumerate}
    \item A strong $\gamma$-ray flare was detected with a peak flux of $3.04\times10^{-6}$ ph cm$^{-2}$ s$^{-1}$, corresponding to an isotropic luminosity of $10^{46}$ erg s$^{-1}$, which is nearly two orders of magnitude above the long-term average \gm-ray luminosity. A rapid \gm-ray flux variation was observed with the shortest flux halving time of $0.84\pm0.3$ hours.

    \item Available \emph{Swift}-XRT observations indicated a bright X-ray flare with modest spectral hardening after the peak level in the \gm-ray band. 
    \item \src~exhibited a rapid hard X-ray flare with the shortest flux doubling time of $0.95\pm0.24$ hours. This is probably the first-ever report of finding sub-hour-scale hard X-ray and \gm-ray flux variability in the \gm-NLSy1 population.
    
    \item The joint \emph{Swift}-XRT and NuSTAR spectral analysis indicated the presence of a break at $\sim1.6$~keV, which could be explained by the presence of multiple emission components.

     \item The broadband SED modeling performed using a conventional one-zone leptonic radiative model reasonably explained the observed SEDs. In SED1, the X-ray spectrum had a flat rising shape, which we explained by EC emission from the jet. On the other hand, the X-ray spectrum covering the epoch of SED2 required a combination of corona emission below $\sim2$~keV and EC at higher energies.
    
    \item The results of broadband SED modeling also suggest that the dissipation region lies inside the BLR.
\end{enumerate}

\begin{table*}[ht]
\centering
\caption{Optimized values of the physical parameters obtained from the broadband SED modeling. The parameters with an asterisk were kept frozen for modeling of the SED1 and SED2 epochs.}
 	\label{tab:sed}
 	\setlength{\tabcolsep}{12pt}
 	\setlength\extrarowheight{4pt}
 	\begin{tabular}{llll}
 		\hline
 		Name of parameter& Symbol & SED1 & SED2  \\
 		 		\hline
        Disc luminosity$^*$ (erg s$^{-1}$) & $L_{\rm disc}$ & $10^{45}$ &  $10^{45}$ \\
        Black-hole mass$^*$ ($M_\odot$) & $M_{\rm BH}$ & $10^{7}$ & $10^{7}$  \\
 	Particle spectral slope before break energy     &  p      &    2.3 &   2.0     \\
 	Particle spectral slope after break energy     &  q &   4.5    &    4.5    \\
 	Bulk Lorentz factor$^*$ & $\Gamma$ & 10 & 10 \\
 	Magnetic Field$^*$ (G)  & B & 1.6 & 1.6 \\
 	Break Lorentz factor$^*$    &     $\gamma_b$  &   198 &   198 \\
 	Emission region size$^*$ (cm)&R & $1.54\times10^{15}$ & $1.54\times10^{15}$ \\
    Distance of the emission region (cm) & R$_{\rm diss}$ & $3.40\times10^{15}$ (1150) & $4.57\times10^{15}$ (1550)\\
    Corona luminosity (erg s$^{-1}$) & L$_{\rm corona}$ & $6.00\times10^{43}$ & $3.50\times10^{44}$ \\
 	Total jet power (erg s$^{-1}$) & P$_{\rm jet}$ & $1.19\times10^{46}$ & $4.15\times10^{45}$ \\
 	Total radiated power (erg s$^{-1}$) & P$_{\rm rad}$ & $8.51\times10^{43}$ & $5.38\times10^{43}$  \\
 	\hline
\end{tabular}
\end{table*}

\section{Appendix}
\begin{figure*}
\centering
\includegraphics[width=7cm,trim=140 140 140 140]{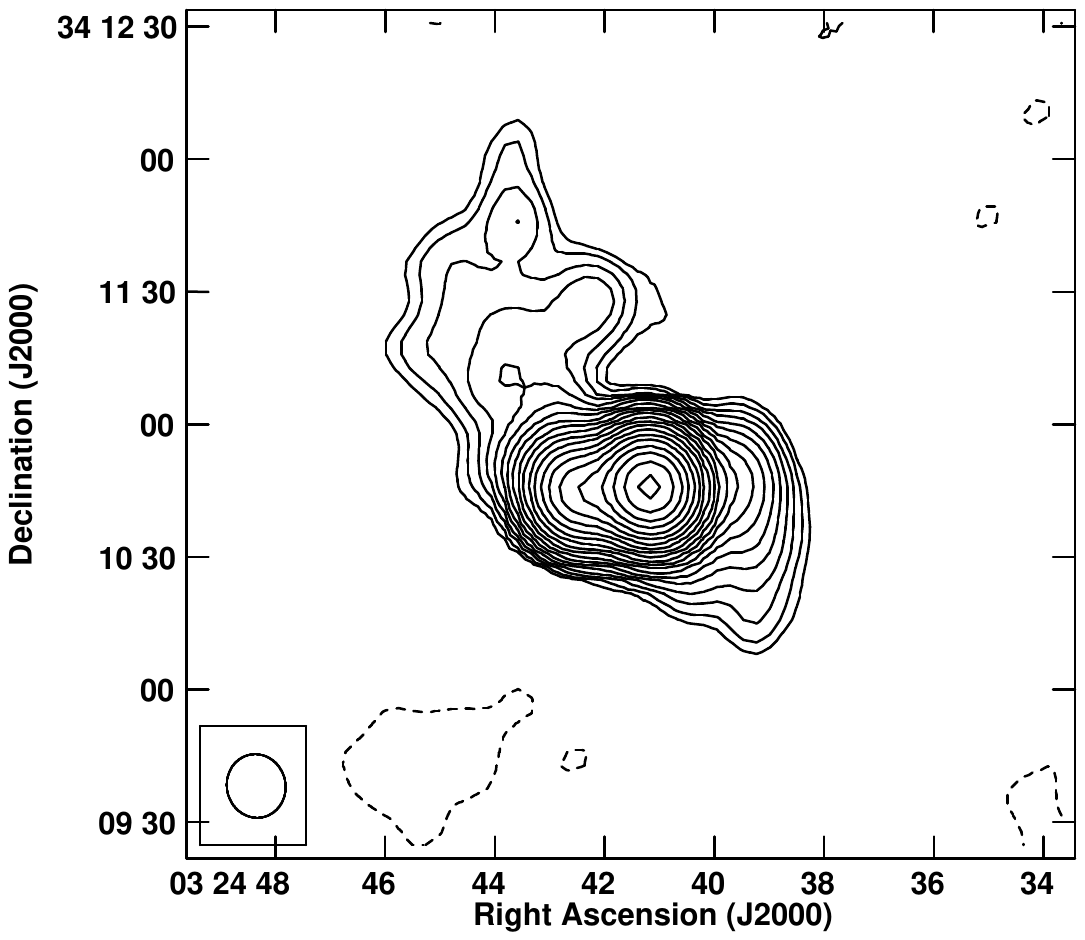}
\caption{Archival VLA C-array 1.42 GHz image of 1H~0323+342 showing the Z-shaped structure. The contours are at $3.807\times (\pm 0.12, 0.17, 0.25, 0.35, 0.5, 0.7, 1, 1.4, 2, 2.8, 4, 5.6, 8, 11.25, 16, 22.5, 32, 45, 64, 90)$~mJy~beam$^{-1}$. The synthesized beam shown in the bottom left corner is of size $14.44''\times 13.27''$ at a position angle of 12.4 degrees. }
\label{fig:radio}
\end{figure*}

We present the radio image of 1H~0323+342 from the VLA data archive\footnote{\url{https://www.vla.nrao.edu/cgi-bin/nvas-pos.pl}} at 1.42 GHz in Figure~\ref{fig:radio}.

\section{Acknowledgement}
The authors thank the anonymous referee for reviewing this manuscript and providing useful suggestions for improvement.
The authors also acknowledge the use of observations from \emph{Fermi}-LAT, NuSTAR, and \emph{Swift} telescopes. The data used in this work were
obtained from \emph{Fermi} Science Support Center, and High Energy Astrophysics Science Archive Research Center (HEASARC), a service of the Goddard Space Flight Center and the Smithsonian Astrophysical Observatory. AT and PK acknowledge the support of the Department of Atomic Energy, Government of India, under the project 12-R\&D-TFR-5.02-0700.

\section{Data Availability}
The data utilized in this study are publicly accessible and can be obtained from the archives
at https://heasarc.gsfc.nasa.gov/ and https://
Fermi.gsfc.nasa.gov/.
\bibliographystyle{elsarticle-num-names} 
\bibliography{ref}

\begin{thebibliography}{81}
\expandafter\ifx\csname natexlab\endcsname\relax\def\natexlab#1{#1}\fi
\providecommand{\url}[1]{\texttt{#1}}
\providecommand{\href}[2]{#2}
\providecommand{\path}[1]{#1}
\providecommand{\DOIprefix}{doi:}
\providecommand{\ArXivprefix}{arXiv:}
\providecommand{\URLprefix}{URL: }
\providecommand{\Pubmedprefix}{pmid:}
\providecommand{\doi}[1]{\href{http://dx.doi.org/#1}{\path{#1}}}
\providecommand{\Pubmed}[1]{\href{pmid:#1}{\path{#1}}}
\providecommand{\bibinfo}[2]{#2}
\ifx\xfnm\relax \def\xfnm[#1]{\unskip,\space#1}\fi
\bibitem[{{Urry} and {Padovani}(1995)}]{1995PASP..107..803U}
\bibinfo{author}{C.~M. {Urry}}, \bibinfo{author}{P.~{Padovani}},
\newblock \bibinfo{title}{{Unified Schemes for Radio-Loud Active Galactic
  Nuclei}},
\newblock \bibinfo{journal}{\pasp} \bibinfo{volume}{107} (\bibinfo{year}{1995})
  \bibinfo{pages}{803}. \DOIprefix\doi{10.1086/133630}.
  \href{http://arxiv.org/abs/astro-ph/9506063}{{\tt arXiv:astro-ph/9506063}}.
\bibitem[{{Osterbrock} and {Ferland}(2006)}]{2006agna.book.....O}
\bibinfo{author}{D.~E. {Osterbrock}}, \bibinfo{author}{G.~J. {Ferland}},
  \bibinfo{title}{{Astrophysics of gaseous nebulae and active galactic
  nuclei}}, \bibinfo{year}{2006}.
\bibitem[{{Caccianiga} and {Severgnini}(2011)}]{2011MNRAS.415.1928C}
\bibinfo{author}{A.~{Caccianiga}}, \bibinfo{author}{P.~{Severgnini}},
\newblock \bibinfo{title}{{The relationship between [O
  III]{\ensuremath{\lambda}}5007 {\r{A}} equivalent width and obscuration in
  active galactic nuclei}},
\newblock \bibinfo{journal}{\mnras} \bibinfo{volume}{415}
  (\bibinfo{year}{2011}) \bibinfo{pages}{1928--1934}.
  \DOIprefix\doi{10.1111/j.1365-2966.2011.18838.x}.
  \href{http://arxiv.org/abs/1104.1348}{{\tt arXiv:1104.1348}}.
\bibitem[{{Baldwin} et~al.(1981){Baldwin}, {Phillips}, and
  {Terlevich}}]{1981PASP...93....5B}
\bibinfo{author}{J.~A. {Baldwin}}, \bibinfo{author}{M.~M. {Phillips}},
  \bibinfo{author}{R.~{Terlevich}},
\newblock \bibinfo{title}{{Classification parameters for the emission-line
  spectra of extragalactic objects.}},
\newblock \bibinfo{journal}{\pasp} \bibinfo{volume}{93} (\bibinfo{year}{1981})
  \bibinfo{pages}{5--19}. \DOIprefix\doi{10.1086/130766}.
\bibitem[{{Goodrich}(1989)}]{1989ApJ...342..224G}
\bibinfo{author}{R.~W. {Goodrich}},
\newblock \bibinfo{title}{{Spectropolarimetry of 'narrow-line' Seyfert 1
  galaxies}},
\newblock \bibinfo{journal}{\apj} \bibinfo{volume}{342} (\bibinfo{year}{1989})
  \bibinfo{pages}{224--234}. \DOIprefix\doi{10.1086/167586}.
\bibitem[{{Osterbrock} and {Pogge}(1985)}]{1985ApJ...297..166O}
\bibinfo{author}{D.~E. {Osterbrock}}, \bibinfo{author}{R.~W. {Pogge}},
\newblock \bibinfo{title}{{The spectra of narrow-line Seyfert 1 galaxies}},
\newblock \bibinfo{journal}{\apj} \bibinfo{volume}{297} (\bibinfo{year}{1985})
  \bibinfo{pages}{166--176}. \DOIprefix\doi{10.1086/163513}.
\bibitem[{{Mathur}(2000)}]{2000MNRAS.314L..17M}
\bibinfo{author}{S.~{Mathur}},
\newblock \bibinfo{title}{{Narrow-line Seyfert 1 galaxies and the evolution of
  galaxies and active galaxies}},
\newblock \bibinfo{journal}{\mnras} \bibinfo{volume}{314}
  (\bibinfo{year}{2000}) \bibinfo{pages}{L17--L20}.
  \DOIprefix\doi{10.1046/j.1365-8711.2000.03530.x}.
  \href{http://arxiv.org/abs/astro-ph/0003111}{{\tt arXiv:astro-ph/0003111}}.
\bibitem[{{Xu} et~al.(2012){Xu}, {Komossa}, {Zhou}, {Lu}, {Li}, {Grupe},
  {Wang}, and {Yuan}}]{2012AJ....143...83X}
\bibinfo{author}{D.~{Xu}}, \bibinfo{author}{S.~{Komossa}},
  \bibinfo{author}{H.~{Zhou}}, \bibinfo{author}{H.~{Lu}},
  \bibinfo{author}{C.~{Li}}, \bibinfo{author}{D.~{Grupe}},
  \bibinfo{author}{J.~{Wang}}, \bibinfo{author}{W.~{Yuan}},
\newblock \bibinfo{title}{{Correlation Analysis of a Large Sample of
  Narrow-line Seyfert 1 Galaxies: Linking Central Engine and Host Properties}},
\newblock \bibinfo{journal}{\aj} \bibinfo{volume}{143} (\bibinfo{year}{2012})
  \bibinfo{pages}{83}. \DOIprefix\doi{10.1088/0004-6256/143/4/83}.
  \href{http://arxiv.org/abs/1201.2810}{{\tt arXiv:1201.2810}}.
\bibitem[{{Paliya} et~al.(2024){Paliya}, {Stalin}, {Dom{\'\i}nguez}, and
  {Saikia}}]{2024MNRAS.527.7055P}
\bibinfo{author}{V.~S. {Paliya}}, \bibinfo{author}{C.~S. {Stalin}},
  \bibinfo{author}{A.~{Dom{\'\i}nguez}}, \bibinfo{author}{D.~J. {Saikia}},
\newblock \bibinfo{title}{{Narrow-line Seyfert 1 galaxies in Sloan Digital Sky
  Survey: a new optical spectroscopic catalogue}},
\newblock \bibinfo{journal}{\mnras} \bibinfo{volume}{527}
  (\bibinfo{year}{2024}) \bibinfo{pages}{7055--7069}.
  \DOIprefix\doi{10.1093/mnras/stad3650}.
  \href{http://arxiv.org/abs/2311.13818}{{\tt arXiv:2311.13818}}.
\bibitem[{{Boller} et~al.(1996){Boller}, {Brandt}, and
  {Fink}}]{1996A&A...305...53B}
\bibinfo{author}{T.~{Boller}}, \bibinfo{author}{W.~N. {Brandt}},
  \bibinfo{author}{H.~{Fink}},
\newblock \bibinfo{title}{{Soft X-ray properties of narrow-line Seyfert 1
  galaxies.}},
\newblock \bibinfo{journal}{\aap} \bibinfo{volume}{305} (\bibinfo{year}{1996})
  \bibinfo{pages}{53}. \href{http://arxiv.org/abs/arXiv:astro-ph/9504093}{{\tt
  arXiv:arXiv:astro-ph/9504093}}.
\bibitem[{{Parker} et~al.(2019){Parker}, {Longinotti}, {Schartel}, {Grupe},
  {Komossa}, {Kriss}, {Fabian}, {Gallo}, {Harrison}, {Jiang}, {Kara},
  {Krongold}, {Matzeu}, {Pinto}, and {Santos-Lle{\'o}}}]{2019MNRAS.490..683P}
\bibinfo{author}{M.~L. {Parker}}, \bibinfo{author}{A.~L. {Longinotti}},
  \bibinfo{author}{N.~{Schartel}}, \bibinfo{author}{D.~{Grupe}},
  \bibinfo{author}{S.~{Komossa}}, \bibinfo{author}{G.~{Kriss}},
  \bibinfo{author}{A.~C. {Fabian}}, \bibinfo{author}{L.~{Gallo}},
  \bibinfo{author}{F.~A. {Harrison}}, \bibinfo{author}{J.~{Jiang}},
  \bibinfo{author}{E.~{Kara}}, \bibinfo{author}{Y.~{Krongold}},
  \bibinfo{author}{G.~A. {Matzeu}}, \bibinfo{author}{C.~{Pinto}},
  \bibinfo{author}{M.~{Santos-Lle{\'o}}},
\newblock \bibinfo{title}{{The nuclear environment of the NLS1 Mrk 335:
  Obscuration of the X-ray line emission by a variable outflow}},
\newblock \bibinfo{journal}{\mnras} \bibinfo{volume}{490}
  (\bibinfo{year}{2019}) \bibinfo{pages}{683--697}.
  \DOIprefix\doi{10.1093/mnras/stz2566}.
  \href{http://arxiv.org/abs/1909.04924}{{\tt arXiv:1909.04924}}.
\bibitem[{{Berton} et~al.(2016){Berton}, {Caccianiga}, {Foschini}, {Peterson},
  {Mathur}, {Terreran}, {Ciroi}, {Congiu}, {Cracco}, {Frezzato}, {La Mura}, and
  {Rafanelli}}]{2016A&A...591A..98B}
\bibinfo{author}{M.~{Berton}}, \bibinfo{author}{A.~{Caccianiga}},
  \bibinfo{author}{L.~{Foschini}}, \bibinfo{author}{B.~M. {Peterson}},
  \bibinfo{author}{S.~{Mathur}}, \bibinfo{author}{G.~{Terreran}},
  \bibinfo{author}{S.~{Ciroi}}, \bibinfo{author}{E.~{Congiu}},
  \bibinfo{author}{V.~{Cracco}}, \bibinfo{author}{M.~{Frezzato}},
  \bibinfo{author}{G.~{La Mura}}, \bibinfo{author}{P.~{Rafanelli}},
\newblock \bibinfo{title}{{Compact steep-spectrum sources as the parent
  population of flat-spectrum radio-loud narrow-line Seyfert 1 galaxies}},
\newblock \bibinfo{journal}{\aap} \bibinfo{volume}{591} (\bibinfo{year}{2016})
  \bibinfo{pages}{A98}. \DOIprefix\doi{10.1051/0004-6361/201628171}.
  \href{http://arxiv.org/abs/1601.06165}{{\tt arXiv:1601.06165}}.
\bibitem[{{Singh} and {Chand}(2018)}]{2018MNRAS.480.1796S}
\bibinfo{author}{V.~{Singh}}, \bibinfo{author}{H.~{Chand}},
\newblock \bibinfo{title}{{Investigating kpc-scale radio emission properties of
  narrow-line Seyfert 1 galaxies}},
\newblock \bibinfo{journal}{\mnras} \bibinfo{volume}{480}
  (\bibinfo{year}{2018}) \bibinfo{pages}{1796--1818}.
  \DOIprefix\doi{10.1093/mnras/sty1818}.
  \href{http://arxiv.org/abs/1807.01945}{{\tt arXiv:1807.01945}}.
\bibitem[{{Yuan} et~al.(2008){Yuan}, {Zhou}, {Komossa}, {Dong}, {Wang}, {Lu},
  and {Bai}}]{2008ApJ...685..801Y}
\bibinfo{author}{W.~{Yuan}}, \bibinfo{author}{H.~Y. {Zhou}},
  \bibinfo{author}{S.~{Komossa}}, \bibinfo{author}{X.~B. {Dong}},
  \bibinfo{author}{T.~G. {Wang}}, \bibinfo{author}{H.~L. {Lu}},
  \bibinfo{author}{J.~M. {Bai}},
\newblock \bibinfo{title}{{A Population of Radio-Loud Narrow-Line Seyfert 1
  Galaxies with Blazar-Like Properties?}},
\newblock \bibinfo{journal}{\apj} \bibinfo{volume}{685} (\bibinfo{year}{2008})
  \bibinfo{pages}{801--827}. \DOIprefix\doi{10.1086/591046}.
  \href{http://arxiv.org/abs/0806.3755}{{\tt arXiv:0806.3755}}.
\bibitem[{{Congiu} et~al.(2017){Congiu}, {Berton}, {Giroletti}, {Antonucci},
  {Caccianiga}, {Kharb}, {Lister}, {Foschini}, {Ciroi}, {Cracco}, {Frezzato},
  {J{\"a}rvel{\"a}}, {La Mura}, {Richards}, and {Rafanelli}}]{Congiu2017}
\bibinfo{author}{E.~{Congiu}}, \bibinfo{author}{M.~{Berton}},
  \bibinfo{author}{M.~{Giroletti}}, \bibinfo{author}{R.~{Antonucci}},
  \bibinfo{author}{A.~{Caccianiga}}, \bibinfo{author}{P.~{Kharb}},
  \bibinfo{author}{M.~L. {Lister}}, \bibinfo{author}{L.~{Foschini}},
  \bibinfo{author}{S.~{Ciroi}}, \bibinfo{author}{V.~{Cracco}},
  \bibinfo{author}{M.~{Frezzato}}, \bibinfo{author}{E.~{J{\"a}rvel{\"a}}},
  \bibinfo{author}{G.~{La Mura}}, \bibinfo{author}{J.~L. {Richards}},
  \bibinfo{author}{P.~{Rafanelli}},
\newblock \bibinfo{title}{{Kiloparsec-scale emission in the narrow-line Seyfert
  1 galaxy Mrk 783}},
\newblock \bibinfo{journal}{\aap} \bibinfo{volume}{603} (\bibinfo{year}{2017})
  \bibinfo{pages}{A32}. \DOIprefix\doi{10.1051/0004-6361/201730616}.
  \href{http://arxiv.org/abs/1704.03881}{{\tt arXiv:1704.03881}}.
\bibitem[{{Rakshit} et~al.(2018){Rakshit}, {Stalin}, {Hota}, and
  {Konar}}]{2018ApJ...869..173R}
\bibinfo{author}{S.~{Rakshit}}, \bibinfo{author}{C.~S. {Stalin}},
  \bibinfo{author}{A.~{Hota}}, \bibinfo{author}{C.~{Konar}},
\newblock \bibinfo{title}{{Rare Finding of a 100 Kpc Large, Double-lobed Radio
  Galaxy Hosted in the Narrow-line Seyfert 1 Galaxy SDSS J103024.95+551622.7}},
\newblock \bibinfo{journal}{\apj} \bibinfo{volume}{869} (\bibinfo{year}{2018})
  \bibinfo{pages}{173}. \DOIprefix\doi{10.3847/1538-4357/aaefe8}.
  \href{http://arxiv.org/abs/1811.02147}{{\tt arXiv:1811.02147}}.
\bibitem[{{Chen} et~al.(2024){Chen}, {Kharb}, {Silpa}, {Nandi}, {Berton},
  {J{\"a}rvel{\"a}}, {Laor}, {Behar}, {Foschini}, {Vietri}, {Gu}, {La Mura},
  {Crepaldi}, and {Zhou}}]{2024ApJ...963...32C}
\bibinfo{author}{S.~{Chen}}, \bibinfo{author}{P.~{Kharb}},
  \bibinfo{author}{S.~{Silpa}}, \bibinfo{author}{S.~{Nandi}},
  \bibinfo{author}{M.~{Berton}}, \bibinfo{author}{E.~{J{\"a}rvel{\"a}}},
  \bibinfo{author}{A.~{Laor}}, \bibinfo{author}{E.~{Behar}},
  \bibinfo{author}{L.~{Foschini}}, \bibinfo{author}{A.~{Vietri}},
  \bibinfo{author}{M.~{Gu}}, \bibinfo{author}{G.~{La Mura}},
  \bibinfo{author}{L.~{Crepaldi}}, \bibinfo{author}{M.~{Zhou}},
\newblock \bibinfo{title}{{A Large Jet Narrow-line Seyfert 1 Galaxy:
  Observations from Parsec to 100 kpc Scales}},
\newblock \bibinfo{journal}{\apj} \bibinfo{volume}{963} (\bibinfo{year}{2024})
  \bibinfo{pages}{32}. \DOIprefix\doi{10.3847/1538-4357/ad182a}.
  \href{http://arxiv.org/abs/2312.13351}{{\tt arXiv:2312.13351}}.
\bibitem[{{Umayal} et~al.(2025){Umayal}, {Paliya}, {Saikia}, {Stalin},
  {Muneer}, and {Gopinathan}}]{2025ApJ...995..125U}
\bibinfo{author}{S.~{Umayal}}, \bibinfo{author}{V.~S. {Paliya}},
  \bibinfo{author}{D.~J. {Saikia}}, \bibinfo{author}{C.~S. {Stalin}},
  \bibinfo{author}{S.~{Muneer}}, \bibinfo{author}{M.~{Gopinathan}},
\newblock \bibinfo{title}{{Identification of Large-scale (>100 kpc) Radio Jets
  in Narrow-line Seyfert 1 Galaxies}},
\newblock \bibinfo{journal}{\apj} \bibinfo{volume}{995} (\bibinfo{year}{2025})
  \bibinfo{pages}{125}. \DOIprefix\doi{10.3847/1538-4357/ae19e7}.
  \href{http://arxiv.org/abs/2510.25678}{{\tt arXiv:2510.25678}}.
\bibitem[{{Abdo} et~al.(2009){Abdo}, {Ackermann}, {Ajello}, {Baldini},
  {Ballet}, {Barbiellini}, {Bastieri}, {Bechtol}, {Bellazzini}, {Berenji},
  {Bloom}, {Bonamente}, {Borgland}, {Bregeon}, {Brez}, {Brigida}, {Bruel},
  {Burnett}, {Caliandro}, {Cameron}, {Caraveo}, {Casandjian}, {Cecchi}, {{\c
  C}elik}, {Chekhtman}, {Cheung}, {Chiang}, {Ciprini}, {Claus}, {Cohen-Tanugi},
  {Conrad}, {Cutini}, {Dermer}, {de Palma}, {Silva}, {Drell}, {Dubois},
  {Dumora}, {Farnier}, {Favuzzi}, {Fegan}, {Focke}, {Foschini}, {Frailis},
  {Fukazawa}, {Fusco}, {Gargano}, {Gehrels}, {Germani}, {Giebels}, {Giglietto},
  {Giordano}, {Giroletti}, {Glanzman}, {Godfrey}, {Grenier}, {Grove},
  {Guillemot}, {Guiriec}, {Hayashida}, {Hays}, {Horan}, {Hughes},
  {J{\'o}hannesson}, {Johnson}, {Johnson}, {Kadler}, {Kamae}, {Katagiri},
  {Kataoka}, {Kerr}, {Kn{\"o}dlseder}, {Kuss}, {Lande}, {Latronico}, {Longo},
  {Loparco}, {Lott}, {Lovellette}, {Lubrano}, {Makeev}, {Mazziotta},
  {McConville}, {McEnery}, {Meurer}, {Michelson}, {Mitthumsiri}, {Mizuno},
  {Monte}, {Monzani}, {Morselli}, {Moskalenko}, {Murgia}, {Nolan}, {Norris},
  {Nuss}, {Ohsugi}, {Omodei}, {Orlando}, {Ormes}, {Pelassa}, {Pepe}, {Persic},
  {Pesce-Rollins}, {Piron}, {Porter}, {Rain{\`o}}, {Rando}, {Razzano},
  {Rochester}, {Rodriguez}, {Ryde}, {Sadrozinski}, {Sambruna}, {Sander}, {Saz
  Parkinson}, {Scargle}, {Sgr{\`o}}, {Smith}, {Spandre}, {Spinelli},
  {Strickman}, {Suson}, {Tagliaferri}, {Takahashi}, {Takahashi}, {Tanaka},
  {Thayer}, {Thayer}, {Thompson}, {Tibaldo}, {Tibolla}, {Torres}, {Tosti},
  {Tramacere}, {Uchiyama}, {Usher}, {Vasileiou}, {Vilchez}, {Vitale}, {Waite},
  {Wang}, {Winer}, {Wood}, {Ylinen}, {Ziegler}, {Fermi/LAT Collaboration},
  {Ghisellini}, {Maraschi}, and {Tavecchio}}]{2009ApJ...707L.142A}
\bibinfo{author}{A.~A. {Abdo}}, \bibinfo{author}{M.~{Ackermann}},
  \bibinfo{author}{M.~{Ajello}}, \bibinfo{author}{L.~{Baldini}},
  \bibinfo{author}{J.~{Ballet}}, \bibinfo{author}{G.~{Barbiellini}},
  \bibinfo{author}{D.~{Bastieri}}, \bibinfo{author}{K.~{Bechtol}},
  \bibinfo{author}{R.~{Bellazzini}}, \bibinfo{author}{B.~{Berenji}},
  \bibinfo{author}{E.~D. {Bloom}}, \bibinfo{author}{E.~{Bonamente}},
  \bibinfo{author}{A.~W. {Borgland}}, \bibinfo{author}{J.~{Bregeon}},
  \bibinfo{author}{A.~{Brez}}, \bibinfo{author}{M.~{Brigida}},
  \bibinfo{author}{P.~{Bruel}}, \bibinfo{author}{T.~H. {Burnett}},
  \bibinfo{author}{G.~A. {Caliandro}}, \bibinfo{author}{R.~A. {Cameron}},
  \bibinfo{author}{P.~A. {Caraveo}}, \bibinfo{author}{J.~M. {Casandjian}},
  \bibinfo{author}{C.~{Cecchi}}, \bibinfo{author}{{\"O}.~{{\c C}elik}},
  \bibinfo{author}{A.~{Chekhtman}}, \bibinfo{author}{C.~C. {Cheung}},
  \bibinfo{author}{J.~{Chiang}}, \bibinfo{author}{S.~{Ciprini}},
  \bibinfo{author}{R.~{Claus}}, \bibinfo{author}{J.~{Cohen-Tanugi}},
  \bibinfo{author}{J.~{Conrad}}, \bibinfo{author}{S.~{Cutini}},
  \bibinfo{author}{C.~D. {Dermer}}, \bibinfo{author}{F.~{de Palma}},
  \bibinfo{author}{E.~d.~C.~e. {Silva}}, \bibinfo{author}{P.~S. {Drell}},
  \bibinfo{author}{R.~{Dubois}}, \bibinfo{author}{D.~{Dumora}},
  \bibinfo{author}{C.~{Farnier}}, \bibinfo{author}{C.~{Favuzzi}},
  \bibinfo{author}{S.~J. {Fegan}}, \bibinfo{author}{W.~B. {Focke}},
  \bibinfo{author}{L.~{Foschini}}, \bibinfo{author}{M.~{Frailis}},
  \bibinfo{author}{Y.~{Fukazawa}}, \bibinfo{author}{P.~{Fusco}},
  \bibinfo{author}{F.~{Gargano}}, \bibinfo{author}{N.~{Gehrels}},
  \bibinfo{author}{S.~{Germani}}, \bibinfo{author}{B.~{Giebels}},
  \bibinfo{author}{N.~{Giglietto}}, \bibinfo{author}{F.~{Giordano}},
  \bibinfo{author}{M.~{Giroletti}}, \bibinfo{author}{T.~{Glanzman}},
  \bibinfo{author}{G.~{Godfrey}}, \bibinfo{author}{I.~A. {Grenier}},
  \bibinfo{author}{J.~E. {Grove}}, \bibinfo{author}{L.~{Guillemot}},
  \bibinfo{author}{S.~{Guiriec}}, \bibinfo{author}{M.~{Hayashida}},
  \bibinfo{author}{E.~{Hays}}, \bibinfo{author}{D.~{Horan}},
  \bibinfo{author}{R.~E. {Hughes}}, \bibinfo{author}{G.~{J{\'o}hannesson}},
  \bibinfo{author}{A.~S. {Johnson}}, \bibinfo{author}{W.~N. {Johnson}},
  \bibinfo{author}{M.~{Kadler}}, \bibinfo{author}{T.~{Kamae}},
  \bibinfo{author}{H.~{Katagiri}}, \bibinfo{author}{J.~{Kataoka}},
  \bibinfo{author}{M.~{Kerr}}, \bibinfo{author}{J.~{Kn{\"o}dlseder}},
  \bibinfo{author}{M.~{Kuss}}, \bibinfo{author}{J.~{Lande}},
  \bibinfo{author}{L.~{Latronico}}, \bibinfo{author}{F.~{Longo}},
  \bibinfo{author}{F.~{Loparco}}, \bibinfo{author}{B.~{Lott}},
  \bibinfo{author}{M.~N. {Lovellette}}, \bibinfo{author}{P.~{Lubrano}},
  \bibinfo{author}{A.~{Makeev}}, \bibinfo{author}{M.~N. {Mazziotta}},
  \bibinfo{author}{W.~{McConville}}, \bibinfo{author}{J.~E. {McEnery}},
  \bibinfo{author}{C.~{Meurer}}, \bibinfo{author}{P.~F. {Michelson}},
  \bibinfo{author}{W.~{Mitthumsiri}}, \bibinfo{author}{T.~{Mizuno}},
  \bibinfo{author}{C.~{Monte}}, \bibinfo{author}{M.~E. {Monzani}},
  \bibinfo{author}{A.~{Morselli}}, \bibinfo{author}{I.~V. {Moskalenko}},
  \bibinfo{author}{S.~{Murgia}}, \bibinfo{author}{P.~L. {Nolan}},
  \bibinfo{author}{J.~P. {Norris}}, \bibinfo{author}{E.~{Nuss}},
  \bibinfo{author}{T.~{Ohsugi}}, \bibinfo{author}{N.~{Omodei}},
  \bibinfo{author}{E.~{Orlando}}, \bibinfo{author}{J.~F. {Ormes}},
  \bibinfo{author}{V.~{Pelassa}}, \bibinfo{author}{M.~{Pepe}},
  \bibinfo{author}{M.~{Persic}}, \bibinfo{author}{M.~{Pesce-Rollins}},
  \bibinfo{author}{F.~{Piron}}, \bibinfo{author}{T.~A. {Porter}},
  \bibinfo{author}{S.~{Rain{\`o}}}, \bibinfo{author}{R.~{Rando}},
  \bibinfo{author}{M.~{Razzano}}, \bibinfo{author}{L.~S. {Rochester}},
  \bibinfo{author}{A.~Y. {Rodriguez}}, \bibinfo{author}{F.~{Ryde}},
  \bibinfo{author}{H.~F.-W. {Sadrozinski}}, \bibinfo{author}{R.~{Sambruna}},
  \bibinfo{author}{A.~{Sander}}, \bibinfo{author}{P.~M. {Saz Parkinson}},
  \bibinfo{author}{J.~D. {Scargle}}, \bibinfo{author}{C.~{Sgr{\`o}}},
  \bibinfo{author}{P.~D. {Smith}}, \bibinfo{author}{G.~{Spandre}},
  \bibinfo{author}{P.~{Spinelli}}, \bibinfo{author}{M.~S. {Strickman}},
  \bibinfo{author}{D.~J. {Suson}}, \bibinfo{author}{G.~{Tagliaferri}},
  \bibinfo{author}{H.~{Takahashi}}, \bibinfo{author}{T.~{Takahashi}},
  \bibinfo{author}{T.~{Tanaka}}, \bibinfo{author}{J.~B. {Thayer}},
  \bibinfo{author}{J.~G. {Thayer}}, \bibinfo{author}{D.~J. {Thompson}},
  \bibinfo{author}{L.~{Tibaldo}}, \bibinfo{author}{O.~{Tibolla}},
  \bibinfo{author}{D.~F. {Torres}}, \bibinfo{author}{G.~{Tosti}},
  \bibinfo{author}{A.~{Tramacere}}, \bibinfo{author}{Y.~{Uchiyama}},
  \bibinfo{author}{T.~L. {Usher}}, \bibinfo{author}{V.~{Vasileiou}},
  \bibinfo{author}{N.~{Vilchez}}, \bibinfo{author}{V.~{Vitale}},
  \bibinfo{author}{A.~P. {Waite}}, \bibinfo{author}{P.~{Wang}},
  \bibinfo{author}{B.~L. {Winer}}, \bibinfo{author}{K.~S. {Wood}},
  \bibinfo{author}{T.~{Ylinen}}, \bibinfo{author}{M.~{Ziegler}},
  \bibinfo{author}{{Fermi/LAT Collaboration}},
  \bibinfo{author}{G.~{Ghisellini}}, \bibinfo{author}{L.~{Maraschi}},
  \bibinfo{author}{F.~{Tavecchio}},
\newblock \bibinfo{title}{{Radio-Loud Narrow-Line Seyfert 1 as a New Class of
  Gamma-Ray Active Galactic Nuclei}},
\newblock \bibinfo{journal}{\apjl} \bibinfo{volume}{707} (\bibinfo{year}{2009})
  \bibinfo{pages}{L142--L147}. \DOIprefix\doi{10.1088/0004-637X/707/2/L142}.
  \href{http://arxiv.org/abs/0911.3485}{{\tt arXiv:0911.3485}}.
\bibitem[{{Foschini}(2011)}]{2011nlsg.confE..24F}
\bibinfo{author}{L.~{Foschini}},
\newblock \bibinfo{title}{{Evidence of powerful relativistic jets in
  narrow-line Seyfert 1 galaxies}},
\newblock in: \bibinfo{booktitle}{Foschini, L. 2011, in Proceedings of Science,
  vol. 126, Narrow-Line Seyfert 1 Galaxies and their Place in the Universe, ed.
  L. Foschini, M. Colpi, L. Gallo et al.}, \bibinfo{year}{2011}.
  \href{http://arxiv.org/abs/1105.0772}{{\tt arXiv:1105.0772}}.
\bibitem[{{Paliya} et~al.(2018){Paliya}, {Ajello}, {Rakshit}, {Mandal},
  {Stalin}, {Kaur}, and {Hartmann}}]{2018ApJ...853L...2P}
\bibinfo{author}{V.~S. {Paliya}}, \bibinfo{author}{M.~{Ajello}},
  \bibinfo{author}{S.~{Rakshit}}, \bibinfo{author}{A.~K. {Mandal}},
  \bibinfo{author}{C.~S. {Stalin}}, \bibinfo{author}{A.~{Kaur}},
  \bibinfo{author}{D.~{Hartmann}},
\newblock \bibinfo{title}{{Gamma-Ray-emitting Narrow-line Seyfert 1 Galaxies in
  the Sloan Digital Sky Survey}},
\newblock \bibinfo{journal}{\apjl} \bibinfo{volume}{853} (\bibinfo{year}{2018})
  \bibinfo{pages}{L2}. \DOIprefix\doi{10.3847/2041-8213/aaa5ab}.
  \href{http://arxiv.org/abs/1801.01905}{{\tt arXiv:1801.01905}}.
\bibitem[{{Mao} and {Yi}(2021)}]{2021ApJS..255...10M}
\bibinfo{author}{L.~{Mao}}, \bibinfo{author}{T.~{Yi}},
\newblock \bibinfo{title}{{A Search for Rapid Mid-infrared Variability in
  Gamma-Ray-emitting Narrow-line Seyfert 1 Galaxies}},
\newblock \bibinfo{journal}{\apjs} \bibinfo{volume}{255} (\bibinfo{year}{2021})
  \bibinfo{pages}{10}. \DOIprefix\doi{10.3847/1538-4365/abfd3b}.
\bibitem[{{Wajima} et~al.(2014){Wajima}, {Fujisawa}, {Hayashida}, {Isobe},
  {Ishida}, and {Yonekura}}]{2014ApJ...781...75W}
\bibinfo{author}{K.~{Wajima}}, \bibinfo{author}{K.~{Fujisawa}},
  \bibinfo{author}{M.~{Hayashida}}, \bibinfo{author}{N.~{Isobe}},
  \bibinfo{author}{T.~{Ishida}}, \bibinfo{author}{Y.~{Yonekura}},
\newblock \bibinfo{title}{{Short-term Radio Variability and Parsec-scale
  Structure in a Gamma-Ray Narrow-line Seyfert 1 Galaxy 1H 0323+342}},
\newblock \bibinfo{journal}{\apj} \bibinfo{volume}{781} (\bibinfo{year}{2014})
  \bibinfo{pages}{75}. \DOIprefix\doi{10.1088/0004-637X/781/2/75}.
  \href{http://arxiv.org/abs/1312.3118}{{\tt arXiv:1312.3118}}.
\bibitem[{{Shao} et~al.(2025){Shao}, {Edwards}, {Stevens}, {Gu}, {Galvin}, and
  {Huynh}}]{2025MNRAS.536.1344S}
\bibinfo{author}{X.~{Shao}}, \bibinfo{author}{P.~G. {Edwards}},
  \bibinfo{author}{J.~{Stevens}}, \bibinfo{author}{M.~{Gu}},
  \bibinfo{author}{T.~J. {Galvin}}, \bibinfo{author}{M.~T. {Huynh}},
\newblock \bibinfo{title}{{The spectral behaviour and variability of
  narrow-line Seyfert 1 galaxies with Australia Telescope Compact Array
  observations}},
\newblock \bibinfo{journal}{\mnras} \bibinfo{volume}{536}
  (\bibinfo{year}{2025}) \bibinfo{pages}{1344--1356}.
  \DOIprefix\doi{10.1093/mnras/stae2662}.
  \href{http://arxiv.org/abs/2412.05933}{{\tt arXiv:2412.05933}}.
\bibitem[{{Foschini} et~al.(2011){Foschini}, {Ghisellini}, {Kovalev}, {Lister},
  {D'Ammando}, {Thompson}, {Tramacere}, {Angelakis}, {Donato}, {Falcone},
  {Fuhrmann}, {Hauser}, {Kovalev}, {Mannheim}, {Maraschi}, {Max-Moerbeck},
  {Nestoras}, {Pavlidou}, {Pearson}, {Pushkarev}, {Readhead}, {Richards},
  {Stevenson}, {Tagliaferri}, {Tibolla}, {Tavecchio}, and
  {Wagner}}]{2011MNRAS.413.1671F}
\bibinfo{author}{L.~{Foschini}}, \bibinfo{author}{G.~{Ghisellini}},
  \bibinfo{author}{Y.~Y. {Kovalev}}, \bibinfo{author}{M.~L. {Lister}},
  \bibinfo{author}{F.~{D'Ammando}}, \bibinfo{author}{D.~J. {Thompson}},
  \bibinfo{author}{A.~{Tramacere}}, \bibinfo{author}{E.~{Angelakis}},
  \bibinfo{author}{D.~{Donato}}, \bibinfo{author}{A.~{Falcone}},
  \bibinfo{author}{L.~{Fuhrmann}}, \bibinfo{author}{M.~{Hauser}},
  \bibinfo{author}{Y.~A. {Kovalev}}, \bibinfo{author}{K.~{Mannheim}},
  \bibinfo{author}{L.~{Maraschi}}, \bibinfo{author}{W.~{Max-Moerbeck}},
  \bibinfo{author}{I.~{Nestoras}}, \bibinfo{author}{V.~{Pavlidou}},
  \bibinfo{author}{T.~J. {Pearson}}, \bibinfo{author}{A.~B. {Pushkarev}},
  \bibinfo{author}{A.~C.~S. {Readhead}}, \bibinfo{author}{J.~L. {Richards}},
  \bibinfo{author}{M.~A. {Stevenson}}, \bibinfo{author}{G.~{Tagliaferri}},
  \bibinfo{author}{O.~{Tibolla}}, \bibinfo{author}{F.~{Tavecchio}},
  \bibinfo{author}{S.~{Wagner}},
\newblock \bibinfo{title}{{The first gamma-ray outburst of a narrow-line
  Seyfert 1 galaxy: the case of PMN J0948+0022 in 2010 July}},
\newblock \bibinfo{journal}{\mnras} \bibinfo{volume}{413}
  (\bibinfo{year}{2011}) \bibinfo{pages}{1671--1677}.
  \DOIprefix\doi{10.1111/j.1365-2966.2011.18240.x}.
  \href{http://arxiv.org/abs/1010.4434}{{\tt arXiv:1010.4434}}.
\bibitem[{{Paliya} et~al.(2019){Paliya}, {Parker}, {Jiang}, {Fabian},
  {Brenneman}, {Ajello}, and {Hartmann}}]{2019ApJ...872..169P}
\bibinfo{author}{V.~S. {Paliya}}, \bibinfo{author}{M.~L. {Parker}},
  \bibinfo{author}{J.~{Jiang}}, \bibinfo{author}{A.~C. {Fabian}},
  \bibinfo{author}{L.~{Brenneman}}, \bibinfo{author}{M.~{Ajello}},
  \bibinfo{author}{D.~{Hartmann}},
\newblock \bibinfo{title}{{General Physical Properties of Gamma-Ray-emitting
  Narrow-line Seyfert 1 Galaxies}},
\newblock \bibinfo{journal}{\apj} \bibinfo{volume}{872} (\bibinfo{year}{2019})
  \bibinfo{pages}{169}. \DOIprefix\doi{10.3847/1538-4357/ab01ce}.
  \href{http://arxiv.org/abs/1901.07613}{{\tt arXiv:1901.07613}}.
\bibitem[{{Bhattacharyya} et~al.(2014){Bhattacharyya}, {Bhatt}, {Bhatt}, and
  {Singh}}]{2014MNRAS.440..106B}
\bibinfo{author}{S.~{Bhattacharyya}}, \bibinfo{author}{H.~{Bhatt}},
  \bibinfo{author}{N.~{Bhatt}}, \bibinfo{author}{K.~K. {Singh}},
\newblock \bibinfo{title}{{The XMM-Newton view of the radio-loud narrow-line
  Seyfert 1 galaxy PMN J0948+0022}},
\newblock \bibinfo{journal}{\mnras} \bibinfo{volume}{440}
  (\bibinfo{year}{2014}) \bibinfo{pages}{106--118}.
  \DOIprefix\doi{10.1093/mnras/stu239}.
  \href{http://arxiv.org/abs/1301.1150}{{\tt arXiv:1301.1150}}.
\bibitem[{{Mundo} et~al.(2020){Mundo}, {Kara}, {Cackett}, {Fabian}, {Jiang},
  {Mushotzky}, {Parker}, {Pinto}, {Reynolds}, and
  {Zoghbi}}]{2020MNRAS.496.2922M}
\bibinfo{author}{S.~A. {Mundo}}, \bibinfo{author}{E.~{Kara}},
  \bibinfo{author}{E.~M. {Cackett}}, \bibinfo{author}{A.~C. {Fabian}},
  \bibinfo{author}{J.~{Jiang}}, \bibinfo{author}{R.~F. {Mushotzky}},
  \bibinfo{author}{M.~L. {Parker}}, \bibinfo{author}{C.~{Pinto}},
  \bibinfo{author}{C.~S. {Reynolds}}, \bibinfo{author}{A.~{Zoghbi}},
\newblock \bibinfo{title}{{The origin of X-ray emission in the gamma-ray
  emitting narrow-line Seyfert 1 1H 0323+342}},
\newblock \bibinfo{journal}{\mnras} \bibinfo{volume}{496}
  (\bibinfo{year}{2020}) \bibinfo{pages}{2922--2931}.
  \DOIprefix\doi{10.1093/mnras/staa1744}.
  \href{http://arxiv.org/abs/2006.07537}{{\tt arXiv:2006.07537}}.
\bibitem[{{D'Ammando} et~al.(2015){D'Ammando}, {Orienti}, {Finke}, {Raiteri},
  {Hovatta}, {Larsson}, {Max-Moerbeck}, {Perkins}, {Readhead}, {Richards},
  {Beilicke}, {Benbow}, {Berger}, {Bird}, {Bugaev}, {Cardenzana}, {Cerruti},
  {Chen}, {Ciupik}, {Dickinson}, {Eisch}, {Errando}, {Falcone}, {Finley},
  {Fleischhack}, {Fortin}, {Fortson}, {Furniss}, {Gerard}, {Gillanders},
  {Griffiths}, {Grube}, {Gyuk}, {H{\r{a}}kansson}, {Holder}, {Humensky}, {Kar},
  {Kertzman}, {Khassen}, {Kieda}, {Krennrich}, {Kumar}, {Lang}, {Maier},
  {McCann}, {Meagher}, {Moriarty}, {Mukherjee}, {Nieto}, {de Bhr{\'o}ithe},
  {Ong}, {Otte}, {Pohl}, {Popkow}, {Prokoph}, {Pueschel}, {Quinn}, {Ragan},
  {Reynolds}, {Richards}, {Roache}, {Rousselle}, {Santander}, {Sembroski},
  {Smith}, {Staszak}, {Telezhinsky}, {Tucci}, {Tyler}, {Varlotta}, {Vassiliev},
  {Wakely}, {Weinstein}, {Welsing}, {Williams}, and
  {Zitzer}}]{2015MNRAS.446.2456D}
\bibinfo{author}{F.~{D'Ammando}}, \bibinfo{author}{M.~{Orienti}},
  \bibinfo{author}{J.~{Finke}}, \bibinfo{author}{C.~M. {Raiteri}},
  \bibinfo{author}{T.~{Hovatta}}, \bibinfo{author}{J.~{Larsson}},
  \bibinfo{author}{W.~{Max-Moerbeck}}, \bibinfo{author}{J.~{Perkins}},
  \bibinfo{author}{A.~C.~S. {Readhead}}, \bibinfo{author}{J.~L. {Richards}},
  \bibinfo{author}{M.~{Beilicke}}, \bibinfo{author}{W.~{Benbow}},
  \bibinfo{author}{K.~{Berger}}, \bibinfo{author}{R.~{Bird}},
  \bibinfo{author}{V.~{Bugaev}}, \bibinfo{author}{J.~V. {Cardenzana}},
  \bibinfo{author}{M.~{Cerruti}}, \bibinfo{author}{X.~{Chen}},
  \bibinfo{author}{L.~{Ciupik}}, \bibinfo{author}{H.~J. {Dickinson}},
  \bibinfo{author}{J.~D. {Eisch}}, \bibinfo{author}{M.~{Errando}},
  \bibinfo{author}{A.~{Falcone}}, \bibinfo{author}{J.~P. {Finley}},
  \bibinfo{author}{H.~{Fleischhack}}, \bibinfo{author}{P.~{Fortin}},
  \bibinfo{author}{L.~{Fortson}}, \bibinfo{author}{A.~{Furniss}},
  \bibinfo{author}{L.~{Gerard}}, \bibinfo{author}{G.~H. {Gillanders}},
  \bibinfo{author}{S.~T. {Griffiths}}, \bibinfo{author}{J.~{Grube}},
  \bibinfo{author}{G.~{Gyuk}}, \bibinfo{author}{N.~{H{\r{a}}kansson}},
  \bibinfo{author}{J.~{Holder}}, \bibinfo{author}{T.~B. {Humensky}},
  \bibinfo{author}{P.~{Kar}}, \bibinfo{author}{M.~{Kertzman}},
  \bibinfo{author}{Y.~{Khassen}}, \bibinfo{author}{D.~{Kieda}},
  \bibinfo{author}{F.~{Krennrich}}, \bibinfo{author}{S.~{Kumar}},
  \bibinfo{author}{M.~J. {Lang}}, \bibinfo{author}{G.~{Maier}},
  \bibinfo{author}{A.~{McCann}}, \bibinfo{author}{K.~{Meagher}},
  \bibinfo{author}{P.~{Moriarty}}, \bibinfo{author}{R.~{Mukherjee}},
  \bibinfo{author}{D.~{Nieto}}, \bibinfo{author}{A.~O. {de Bhr{\'o}ithe}},
  \bibinfo{author}{R.~A. {Ong}}, \bibinfo{author}{A.~N. {Otte}},
  \bibinfo{author}{M.~{Pohl}}, \bibinfo{author}{A.~{Popkow}},
  \bibinfo{author}{H.~{Prokoph}}, \bibinfo{author}{E.~{Pueschel}},
  \bibinfo{author}{J.~{Quinn}}, \bibinfo{author}{K.~{Ragan}},
  \bibinfo{author}{P.~T. {Reynolds}}, \bibinfo{author}{G.~T. {Richards}},
  \bibinfo{author}{E.~{Roache}}, \bibinfo{author}{J.~{Rousselle}},
  \bibinfo{author}{M.~{Santander}}, \bibinfo{author}{G.~H. {Sembroski}},
  \bibinfo{author}{A.~W. {Smith}}, \bibinfo{author}{D.~{Staszak}},
  \bibinfo{author}{I.~{Telezhinsky}}, \bibinfo{author}{J.~V. {Tucci}},
  \bibinfo{author}{J.~{Tyler}}, \bibinfo{author}{A.~{Varlotta}},
  \bibinfo{author}{V.~V. {Vassiliev}}, \bibinfo{author}{S.~P. {Wakely}},
  \bibinfo{author}{A.~{Weinstein}}, \bibinfo{author}{R.~{Welsing}},
  \bibinfo{author}{D.~A. {Williams}}, \bibinfo{author}{B.~{Zitzer}},
\newblock \bibinfo{title}{{The most powerful flaring activity from the NLSy1
  PMN J0948+0022}},
\newblock \bibinfo{journal}{\mnras} \bibinfo{volume}{446}
  (\bibinfo{year}{2015}) \bibinfo{pages}{2456--2467}.
  \DOIprefix\doi{10.1093/mnras/stu2251}.
  \href{http://arxiv.org/abs/1410.7144}{{\tt arXiv:1410.7144}}.
\bibitem[{{Paliya} et~al.(2015){Paliya}, {Stalin}, and
  {Ravikumar}}]{2015AJ....149...41P}
\bibinfo{author}{V.~S. {Paliya}}, \bibinfo{author}{C.~S. {Stalin}},
  \bibinfo{author}{C.~D. {Ravikumar}},
\newblock \bibinfo{title}{{Fermi Monitoring of Radio-Loud Narrow-Line Seyfert 1
  Galaxies}},
\newblock \bibinfo{journal}{\aj} \bibinfo{volume}{149} (\bibinfo{year}{2015})
  \bibinfo{pages}{41}. \DOIprefix\doi{10.1088/0004-6256/149/2/41}.
  \href{http://arxiv.org/abs/1410.0755}{{\tt arXiv:1410.0755}}.
\bibitem[{{Gokus} et~al.(2021){Gokus}, {Paliya}, {Wagner}, {Buson},
  {D'Ammando}, {Edwards}, {Kadler}, {Meyer}, {Ojha}, {Stevens}, and
  {Wilms}}]{2021A&A...649A..77G}
\bibinfo{author}{A.~{Gokus}}, \bibinfo{author}{V.~S. {Paliya}},
  \bibinfo{author}{S.~M. {Wagner}}, \bibinfo{author}{S.~{Buson}},
  \bibinfo{author}{F.~{D'Ammando}}, \bibinfo{author}{P.~G. {Edwards}},
  \bibinfo{author}{M.~{Kadler}}, \bibinfo{author}{M.~{Meyer}},
  \bibinfo{author}{R.~{Ojha}}, \bibinfo{author}{J.~{Stevens}},
  \bibinfo{author}{J.~{Wilms}},
\newblock \bibinfo{title}{{The first GeV flare of the radio-loud narrow-line
  Seyfert 1 galaxy PKS 2004-447}},
\newblock \bibinfo{journal}{\aap} \bibinfo{volume}{649} (\bibinfo{year}{2021})
  \bibinfo{pages}{A77}. \DOIprefix\doi{10.1051/0004-6361/202039378}.
  \href{http://arxiv.org/abs/2102.11633}{{\tt arXiv:2102.11633}}.
\bibitem[{{Paliya} et~al.(2013){Paliya}, {Stalin}, {Shukla}, and
  {Sahayanathan}}]{2013ApJ...768...52P}
\bibinfo{author}{V.~S. {Paliya}}, \bibinfo{author}{C.~S. {Stalin}},
  \bibinfo{author}{A.~{Shukla}}, \bibinfo{author}{S.~{Sahayanathan}},
\newblock \bibinfo{title}{{The Nature of {$\gamma$}-Ray Loud Narrow-line
  Seyfert I Galaxies PKS 1502+036 and PKS 2004-447}},
\newblock \bibinfo{journal}{\apj} \bibinfo{volume}{768} (\bibinfo{year}{2013})
  \bibinfo{pages}{52}. \DOIprefix\doi{10.1088/0004-637X/768/1/52}.
  \href{http://arxiv.org/abs/1303.3443}{{\tt arXiv:1303.3443}}.
\bibitem[{{Foschini} et~al.(2015){Foschini}, {Berton}, {Caccianiga}, {Ciroi},
  {Cracco}, {Peterson}, {Angelakis}, {Braito}, {Fuhrmann}, {Gallo}, {Grupe},
  {J{\"a}rvel{\"a}}, {Kaufmann}, {Komossa}, {Kovalev}, {L{\"a}hteenm{\"a}ki},
  {Lisakov}, {Lister}, {Mathur}, {Richards}, {Romano}, {Sievers},
  {Tagliaferri}, {Tammi}, {Tibolla}, {Tornikoski}, {Vercellone}, {La Mura},
  {Maraschi}, and {Rafanelli}}]{2015ANA...575A..13F}
\bibinfo{author}{L.~{Foschini}}, \bibinfo{author}{M.~{Berton}},
  \bibinfo{author}{A.~{Caccianiga}}, \bibinfo{author}{S.~{Ciroi}},
  \bibinfo{author}{V.~{Cracco}}, \bibinfo{author}{B.~M. {Peterson}},
  \bibinfo{author}{E.~{Angelakis}}, \bibinfo{author}{V.~{Braito}},
  \bibinfo{author}{L.~{Fuhrmann}}, \bibinfo{author}{L.~{Gallo}},
  \bibinfo{author}{D.~{Grupe}}, \bibinfo{author}{E.~{J{\"a}rvel{\"a}}},
  \bibinfo{author}{S.~{Kaufmann}}, \bibinfo{author}{S.~{Komossa}},
  \bibinfo{author}{Y.~Y. {Kovalev}},
  \bibinfo{author}{A.~{L{\"a}hteenm{\"a}ki}}, \bibinfo{author}{M.~M.
  {Lisakov}}, \bibinfo{author}{M.~L. {Lister}}, \bibinfo{author}{S.~{Mathur}},
  \bibinfo{author}{J.~L. {Richards}}, \bibinfo{author}{P.~{Romano}},
  \bibinfo{author}{A.~{Sievers}}, \bibinfo{author}{G.~{Tagliaferri}},
  \bibinfo{author}{J.~{Tammi}}, \bibinfo{author}{O.~{Tibolla}},
  \bibinfo{author}{M.~{Tornikoski}}, \bibinfo{author}{S.~{Vercellone}},
  \bibinfo{author}{G.~{La Mura}}, \bibinfo{author}{L.~{Maraschi}},
  \bibinfo{author}{P.~{Rafanelli}},
\newblock \bibinfo{title}{{Properties of flat-spectrum radio-loud narrow-line
  Seyfert 1 galaxies}},
\newblock \bibinfo{journal}{\aap} \bibinfo{volume}{575} (\bibinfo{year}{2015})
  \bibinfo{pages}{A13}. \DOIprefix\doi{10.1051/0004-6361/201424972}.
  \href{http://arxiv.org/abs/1409.3716}{{\tt arXiv:1409.3716}}.
\bibitem[{{Paliya} et~al.(2020){Paliya}, {P{\'e}rez}, {Garc{\'\i}a-Benito},
  {Ajello}, {Prada}, {Alberdi}, {Suh}, {Chandra}, {Dom{\'\i}nguez}, {Marchesi},
  {Di Matteo}, {Hartmann}, and {Chiaberge}}]{2020ApJ...892..133P}
\bibinfo{author}{V.~S. {Paliya}}, \bibinfo{author}{E.~{P{\'e}rez}},
  \bibinfo{author}{R.~{Garc{\'\i}a-Benito}}, \bibinfo{author}{M.~{Ajello}},
  \bibinfo{author}{F.~{Prada}}, \bibinfo{author}{A.~{Alberdi}},
  \bibinfo{author}{H.~{Suh}}, \bibinfo{author}{C.~H.~I. {Chandra}},
  \bibinfo{author}{A.~{Dom{\'\i}nguez}}, \bibinfo{author}{S.~{Marchesi}},
  \bibinfo{author}{T.~{Di Matteo}}, \bibinfo{author}{D.~{Hartmann}},
  \bibinfo{author}{M.~{Chiaberge}},
\newblock \bibinfo{title}{{TXS 2116-077: A Gamma-Ray Emitting Relativistic Jet
  Hosted in a Galaxy Merger}},
\newblock \bibinfo{journal}{\apj} \bibinfo{volume}{892} (\bibinfo{year}{2020})
  \bibinfo{pages}{133}. \DOIprefix\doi{10.3847/1538-4357/ab754f}.
  \href{http://arxiv.org/abs/2004.02703}{{\tt arXiv:2004.02703}}.
\bibitem[{{Ant{\'o}n} et~al.(2008){Ant{\'o}n}, {Browne}, and
  {March{\~a}}}]{2008A&A...490..583A}
\bibinfo{author}{S.~{Ant{\'o}n}}, \bibinfo{author}{I.~W.~A. {Browne}},
  \bibinfo{author}{M.~J. {March{\~a}}},
\newblock \bibinfo{title}{{The colour of the narrow line Sy1-blazar
  0324+3410}},
\newblock \bibinfo{journal}{\aap} \bibinfo{volume}{490} (\bibinfo{year}{2008})
  \bibinfo{pages}{583--587}. \DOIprefix\doi{10.1051/0004-6361:20078926}.
  \href{http://arxiv.org/abs/0907.2400}{{\tt arXiv:0907.2400}}.
\bibitem[{{Landt} et~al.(2017){Landt}, {Ward}, {Balokovi{\'c}}, {Kynoch},
  {Storchi-Bergmann}, {Boisson}, {Done}, {Schimoia}, and
  {Stern}}]{2017MNRAS.464.2565L}
\bibinfo{author}{H.~{Landt}}, \bibinfo{author}{M.~J. {Ward}},
  \bibinfo{author}{M.~{Balokovi{\'c}}}, \bibinfo{author}{D.~{Kynoch}},
  \bibinfo{author}{T.~{Storchi-Bergmann}}, \bibinfo{author}{C.~{Boisson}},
  \bibinfo{author}{C.~{Done}}, \bibinfo{author}{J.~{Schimoia}},
  \bibinfo{author}{D.~{Stern}},
\newblock \bibinfo{title}{{On the black hole mass of the {$\gamma$}-ray
  emitting narrow-line Seyfert 1 galaxy 1H 0323+342}},
\newblock \bibinfo{journal}{\mnras} \bibinfo{volume}{464}
  (\bibinfo{year}{2017}) \bibinfo{pages}{2565--2576}.
  \DOIprefix\doi{10.1093/mnras/stw2447}.
  \href{http://arxiv.org/abs/1609.08002}{{\tt arXiv:1609.08002}}.
\bibitem[{{Doi} et~al.(2020){Doi}, {Kino}, {Kawakatu}, and
  {Hada}}]{2020MNRAS.496.1757D}
\bibinfo{author}{A.~{Doi}}, \bibinfo{author}{M.~{Kino}},
  \bibinfo{author}{N.~{Kawakatu}}, \bibinfo{author}{K.~{Hada}},
\newblock \bibinfo{title}{{The radio-loud narrow-line Seyfert 1 galaxy 1H
  0323+342 in a galaxy merger}},
\newblock \bibinfo{journal}{\mnras} \bibinfo{volume}{496}
  (\bibinfo{year}{2020}) \bibinfo{pages}{1757--1765}.
  \DOIprefix\doi{10.1093/mnras/staa1525}.
  \href{http://arxiv.org/abs/2005.12510}{{\tt arXiv:2005.12510}}.
\bibitem[{{Kellermann} et~al.(2016){Kellermann}, {Condon}, {Kimball}, {Perley},
  and {Ivezi{\'c}}}]{Kellermann2016}
\bibinfo{author}{K.~I. {Kellermann}}, \bibinfo{author}{J.~J. {Condon}},
  \bibinfo{author}{A.~E. {Kimball}}, \bibinfo{author}{R.~A. {Perley}},
  \bibinfo{author}{{\v{Z}}.~{Ivezi{\'c}}},
\newblock \bibinfo{title}{{Radio-loud and Radio-quiet QSOs}},
\newblock \bibinfo{journal}{\apj} \bibinfo{volume}{831} (\bibinfo{year}{2016})
  \bibinfo{pages}{168}. \DOIprefix\doi{10.3847/0004-637X/831/2/168}.
  \href{http://arxiv.org/abs/1608.04586}{{\tt arXiv:1608.04586}}.
\bibitem[{{Lister} et~al.(2016){Lister}, {Aller}, {Aller}, {Homan},
  {Kellermann}, {Kovalev}, {Pushkarev}, {Richards}, {Ros}, and
  {Savolainen}}]{Lister2016}
\bibinfo{author}{M.~L. {Lister}}, \bibinfo{author}{M.~F. {Aller}},
  \bibinfo{author}{H.~D. {Aller}}, \bibinfo{author}{D.~C. {Homan}},
  \bibinfo{author}{K.~I. {Kellermann}}, \bibinfo{author}{Y.~Y. {Kovalev}},
  \bibinfo{author}{A.~B. {Pushkarev}}, \bibinfo{author}{J.~L. {Richards}},
  \bibinfo{author}{E.~{Ros}}, \bibinfo{author}{T.~{Savolainen}},
\newblock \bibinfo{title}{{MOJAVE: XIII. Parsec-scale AGN Jet Kinematics
  Analysis Based on 19 years of VLBA Observations at 15 GHz}},
\newblock \bibinfo{journal}{\aj} \bibinfo{volume}{152} (\bibinfo{year}{2016})
  \bibinfo{pages}{12}. \DOIprefix\doi{10.3847/0004-6256/152/1/12}.
  \href{http://arxiv.org/abs/1603.03882}{{\tt arXiv:1603.03882}}.
\bibitem[{{Paliya} et~al.(2014){Paliya}, {Sahayanathan}, {Parker}, {Fabian},
  {Stalin}, {Anjum}, and {Pandey}}]{2014ApJ...789..143P}
\bibinfo{author}{V.~S. {Paliya}}, \bibinfo{author}{S.~{Sahayanathan}},
  \bibinfo{author}{M.~L. {Parker}}, \bibinfo{author}{A.~C. {Fabian}},
  \bibinfo{author}{C.~S. {Stalin}}, \bibinfo{author}{A.~{Anjum}},
  \bibinfo{author}{S.~B. {Pandey}},
\newblock \bibinfo{title}{{The Peculiar Radio-loud Narrow Line Seyfert 1 Galaxy
  1H 0323+342}},
\newblock \bibinfo{journal}{\apj} \bibinfo{volume}{789} (\bibinfo{year}{2014})
  \bibinfo{pages}{143}. \DOIprefix\doi{10.1088/0004-637X/789/2/143}.
  \href{http://arxiv.org/abs/1405.0715}{{\tt arXiv:1405.0715}}.
\bibitem[{{Carpenter} and {Ojha}(2013)}]{2013ATel.5344....1C}
\bibinfo{author}{B.~{Carpenter}}, \bibinfo{author}{R.~{Ojha}},
\newblock \bibinfo{title}{{Fermi LAT Detection of a GeV Flare from the
  Radio-Loud Narrow-Line Sy1 1H 0323+342}},
\newblock \bibinfo{journal}{The Astronomer's Telegram} \bibinfo{volume}{5344}
  (\bibinfo{year}{2013}) \bibinfo{pages}{1}.
\bibitem[{{Itoh} et~al.(2014){Itoh}, {Tanaka}, {Akitaya}, {Uemura}, {Fukazawa},
  {Inoue}, {Doi}, {Arai}, {Hanayama}, {Hashimoto}, {Hayashi}, {Izumiura},
  {Kanda}, {Kawabata}, {Kawaguchi}, {Kawai}, {Kinugasa}, {Kuroda}, {Miyaji},
  {Moritani}, {Morokuma}, {Murata}, {Nagayama}, {Oasa}, {Ohshima}, {Ohsugi},
  {Saito}, {Sakata}, {Sasada}, {Sekiguchi}, {Takagi}, {Takahashi}, {Takaki},
  {Ui}, {Watanabe}, {Yamanaka}, {Yamashita}, and
  {Yoshida}}]{2014PASJ...66..108I}
\bibinfo{author}{R.~{Itoh}}, \bibinfo{author}{Y.~T. {Tanaka}},
  \bibinfo{author}{H.~{Akitaya}}, \bibinfo{author}{M.~{Uemura}},
  \bibinfo{author}{Y.~{Fukazawa}}, \bibinfo{author}{Y.~{Inoue}},
  \bibinfo{author}{A.~{Doi}}, \bibinfo{author}{A.~{Arai}},
  \bibinfo{author}{H.~{Hanayama}}, \bibinfo{author}{O.~{Hashimoto}},
  \bibinfo{author}{M.~{Hayashi}}, \bibinfo{author}{H.~{Izumiura}},
  \bibinfo{author}{Y.~{Kanda}}, \bibinfo{author}{K.~S. {Kawabata}},
  \bibinfo{author}{K.~{Kawaguchi}}, \bibinfo{author}{N.~{Kawai}},
  \bibinfo{author}{K.~{Kinugasa}}, \bibinfo{author}{D.~{Kuroda}},
  \bibinfo{author}{T.~{Miyaji}}, \bibinfo{author}{Y.~{Moritani}},
  \bibinfo{author}{T.~{Morokuma}}, \bibinfo{author}{K.~L. {Murata}},
  \bibinfo{author}{T.~{Nagayama}}, \bibinfo{author}{Y.~{Oasa}},
  \bibinfo{author}{T.~{Ohshima}}, \bibinfo{author}{T.~{Ohsugi}},
  \bibinfo{author}{Y.~{Saito}}, \bibinfo{author}{S.~{Sakata}},
  \bibinfo{author}{M.~{Sasada}}, \bibinfo{author}{K.~{Sekiguchi}},
  \bibinfo{author}{Y.~{Takagi}}, \bibinfo{author}{J.~{Takahashi}},
  \bibinfo{author}{K.~{Takaki}}, \bibinfo{author}{T.~{Ui}},
  \bibinfo{author}{M.~{Watanabe}}, \bibinfo{author}{M.~{Yamanaka}},
  \bibinfo{author}{S.~{Yamashita}}, \bibinfo{author}{M.~{Yoshida}},
\newblock \bibinfo{title}{{Variable optical polarization during high state in
  {\ensuremath{\gamma}}-ray loud, narrow-line Seyfert 1 galaxy 1H 0323+342}},
\newblock \bibinfo{journal}{\pasj} \bibinfo{volume}{66} (\bibinfo{year}{2014})
  \bibinfo{pages}{108}. \DOIprefix\doi{10.1093/pasj/psu095}.
  \href{http://arxiv.org/abs/1405.3731}{{\tt arXiv:1405.3731}}.
\bibitem[{{Rosa} et~al.(2025){Rosa}, {Foschini}, and
  {Ciroi}}]{2025A&A...698A.160R}
\bibinfo{author}{V.~{Rosa}}, \bibinfo{author}{L.~{Foschini}},
  \bibinfo{author}{S.~{Ciroi}},
\newblock \bibinfo{title}{{Accretion and ejection at work in the Narrow Line
  Seyfert 1 galaxy 1H 0323+342: A case of intermittent activity?}},
\newblock \bibinfo{journal}{\aap} \bibinfo{volume}{698} (\bibinfo{year}{2025})
  \bibinfo{pages}{A160}. \DOIprefix\doi{10.1051/0004-6361/202453046}.
  \href{http://arxiv.org/abs/2504.20551}{{\tt arXiv:2504.20551}}.
\bibitem[{{Longo} et~al.(2025){Longo}, {Holzmann Airasca}, and {La
  Mura}}]{2025ATel17407....1L}
\bibinfo{author}{F.~{Longo}}, \bibinfo{author}{A.~{Holzmann Airasca}},
  \bibinfo{author}{G.~{La Mura}},
\newblock \bibinfo{title}{{Fermi-LAT detection of renewed gamma-ray activity
  from the Radio-Loud Narrow-Line Seyfert 1 1H 0323+342}},
\newblock \bibinfo{journal}{The Astronomer's Telegram} \bibinfo{volume}{17407}
  (\bibinfo{year}{2025}) \bibinfo{pages}{1}.
\bibitem[{{D'Ammando} et~al.(2025){D'Ammando}, {Longo}, {Holzmann Airasca}, and
  {La Mura}}]{2025ATel17411....1D}
\bibinfo{author}{F.~{D'Ammando}}, \bibinfo{author}{F.~{Longo}},
  \bibinfo{author}{A.~{Holzmann Airasca}}, \bibinfo{author}{G.~{La Mura}},
\newblock \bibinfo{title}{{Swift follow-up of the gamma-ray flaring NLSy1 1H
  0323+342}},
\newblock \bibinfo{journal}{The Astronomer's Telegram} \bibinfo{volume}{17411}
  (\bibinfo{year}{2025}) \bibinfo{pages}{1}.
\bibitem[{{D'Ammando} et~al.(2013){D'Ammando}, {Orienti}, {Finke}, {Raiteri},
  {Angelakis}, {Fuhrmann}, {Giroletti}, {Hovatta}, {Karamanavis},
  {Max-Moerbeck}, {Myserlis}, {Readhead}, and {Richards}}]{2013MNRAS.436..191D}
\bibinfo{author}{F.~{D'Ammando}}, \bibinfo{author}{M.~{Orienti}},
  \bibinfo{author}{J.~{Finke}}, \bibinfo{author}{C.~M. {Raiteri}},
  \bibinfo{author}{E.~{Angelakis}}, \bibinfo{author}{L.~{Fuhrmann}},
  \bibinfo{author}{M.~{Giroletti}}, \bibinfo{author}{T.~{Hovatta}},
  \bibinfo{author}{V.~{Karamanavis}}, \bibinfo{author}{W.~{Max-Moerbeck}},
  \bibinfo{author}{I.~{Myserlis}}, \bibinfo{author}{A.~C.~S. {Readhead}},
  \bibinfo{author}{J.~L. {Richards}},
\newblock \bibinfo{title}{{Multifrequency studies of the narrow-line Seyfert 1
  galaxy SBS 0846+513}},
\newblock \bibinfo{journal}{\mnras} \bibinfo{volume}{436}
  (\bibinfo{year}{2013}) \bibinfo{pages}{191--201}.
  \DOIprefix\doi{10.1093/mnras/stt1560}.
  \href{http://arxiv.org/abs/1308.3709}{{\tt arXiv:1308.3709}}.
\bibitem[{{Paliya} and {Stalin}(2016)}]{2016ApJ...820...52P}
\bibinfo{author}{V.~S. {Paliya}}, \bibinfo{author}{C.~S. {Stalin}},
\newblock \bibinfo{title}{{The First GeV Outburst of the Radio-loud Narrow-line
  Seyfert 1 Galaxy PKS 1502+036}},
\newblock \bibinfo{journal}{\apj} \bibinfo{volume}{820} (\bibinfo{year}{2016})
  \bibinfo{pages}{52}. \DOIprefix\doi{10.3847/0004-637X/820/1/52}.
  \href{http://arxiv.org/abs/1603.01543}{{\tt arXiv:1603.01543}}.
\bibitem[{{Planck Collaboration} et~al.(2016){Planck Collaboration}, {Ade},
  {Aghanim}, {Arnaud}, {Ashdown}, {Aumont}, {Baccigalupi}, {Banday},
  {Barreiro}, {Bartlett}, {Bartolo}, {Battaner}, {Battye}, {Benabed},
  {Beno{\^\i}t}, {Benoit-L{\'e}vy}, {Bernard}, {Bersanelli}, {Bielewicz},
  {Bock}, {Bonaldi}, {Bonavera}, {Bond}, {Borrill}, {Bouchet}, {Boulanger},
  {Bucher}, {Burigana}, {Butler}, {Calabrese}, {Cardoso}, {Catalano},
  {Challinor}, {Chamballu}, {Chary}, {Chiang}, {Chluba}, {Christensen},
  {Church}, {Clements}, {Colombi}, {Colombo}, {Combet}, {Coulais}, {Crill},
  {Curto}, {Cuttaia}, {Danese}, {Davies}, {Davis}, {de Bernardis}, {de Rosa},
  {de Zotti}, {Delabrouille}, {D{\'e}sert}, {Di Valentino}, {Dickinson},
  {Diego}, {Dolag}, {Dole}, {Donzelli}, {Dor{\'e}}, {Douspis}, {Ducout},
  {Dunkley}, {Dupac}, {Efstathiou}, {Elsner}, {En{\ss}lin}, {Eriksen},
  {Farhang}, {Fergusson}, {Finelli}, {Forni}, {Frailis}, {Fraisse},
  {Franceschi}, {Frejsel}, {Galeotta}, {Galli}, {Ganga}, {Gauthier}, {Gerbino},
  {Ghosh}, {Giard}, {Giraud-H{\'e}raud}, {Giusarma}, {Gjerl{\o}w},
  {Gonz{\'a}lez-Nuevo}, {G{\'o}rski}, {Gratton}, {Gregorio}, {Gruppuso},
  {Gudmundsson}, {Hamann}, {Hansen}, {Hanson}, {Harrison}, {Helou},
  {Henrot-Versill{\'e}}, {Hern{\'a}ndez-Monteagudo}, {Herranz}, {Hildebrandt},
  {Hivon}, {Hobson}, {Holmes}, {Hornstrup}, {Hovest}, {Huang}, {Huffenberger},
  {Hurier}, {Jaffe}, {Jaffe}, {Jones}, {Juvela}, {Keih{\"a}nen}, {Keskitalo},
  {Kisner}, {Kneissl}, {Knoche}, {Knox}, {Kunz}, {Kurki-Suonio}, {Lagache},
  {L{\"a}hteenm{\"a}ki}, {Lamarre}, {Lasenby}, {Lattanzi}, {Lawrence}, {Leahy},
  {Leonardi}, {Lesgourgues}, {Levrier}, {Lewis}, {Liguori}, {Lilje},
  {Linden-V{\o}rnle}, {L{\'o}pez-Caniego}, {Lubin}, {Mac{\'\i}as-P{\'e}rez},
  {Maggio}, {Maino}, {Mandolesi}, {Mangilli}, {Marchini}, {Maris}, {Martin},
  {Martinelli}, {Mart{\'\i}nez-Gonz{\'a}lez}, {Masi}, {Matarrese}, {McGehee},
  {Meinhold}, {Melchiorri}, {Melin}, {Mendes}, {Mennella}, {Migliaccio},
  {Millea}, {Mitra}, {Miville-Desch{\^e}nes}, {Moneti}, {Montier}, {Morgante},
  {Mortlock}, {Moss}, {Munshi}, {Murphy}, {Naselsky}, {Nati}, {Natoli},
  {Netterfield}, {N{\o}rgaard-Nielsen}, {Noviello}, {Novikov}, {Novikov},
  {Oxborrow}, {Paci}, {Pagano}, {Pajot}, {Paladini}, {Paoletti}, {Partridge},
  {Pasian}, {Patanchon}, {Pearson}, {Perdereau}, {Perotto}, {Perrotta},
  {Pettorino}, {Piacentini}, {Piat}, {Pierpaoli}, {Pietrobon}, {Plaszczynski},
  {Pointecouteau}, {Polenta}, {Popa}, {Pratt}, and
  {Pr{\'e}zeau}}]{2016A&A...594A..13P}
\bibinfo{author}{{Planck Collaboration}}, \bibinfo{author}{P.~A.~R. {Ade}},
  \bibinfo{author}{N.~{Aghanim}}, \bibinfo{author}{M.~{Arnaud}},
  \bibinfo{author}{M.~{Ashdown}}, \bibinfo{author}{J.~{Aumont}},
  \bibinfo{author}{C.~{Baccigalupi}}, \bibinfo{author}{A.~J. {Banday}},
  \bibinfo{author}{R.~B. {Barreiro}}, \bibinfo{author}{J.~G. {Bartlett}},
  \bibinfo{author}{N.~{Bartolo}}, \bibinfo{author}{E.~{Battaner}},
  \bibinfo{author}{R.~{Battye}}, \bibinfo{author}{K.~{Benabed}},
  \bibinfo{author}{A.~{Beno{\^\i}t}}, \bibinfo{author}{A.~{Benoit-L{\'e}vy}},
  \bibinfo{author}{J.-P. {Bernard}}, \bibinfo{author}{M.~{Bersanelli}},
  \bibinfo{author}{P.~{Bielewicz}}, \bibinfo{author}{J.~J. {Bock}},
  \bibinfo{author}{A.~{Bonaldi}}, \bibinfo{author}{L.~{Bonavera}},
  \bibinfo{author}{J.~R. {Bond}}, \bibinfo{author}{J.~{Borrill}},
  \bibinfo{author}{F.~R. {Bouchet}}, \bibinfo{author}{F.~{Boulanger}},
  \bibinfo{author}{M.~{Bucher}}, \bibinfo{author}{C.~{Burigana}},
  \bibinfo{author}{R.~C. {Butler}}, \bibinfo{author}{E.~{Calabrese}},
  \bibinfo{author}{J.-F. {Cardoso}}, \bibinfo{author}{A.~{Catalano}},
  \bibinfo{author}{A.~{Challinor}}, \bibinfo{author}{A.~{Chamballu}},
  \bibinfo{author}{R.-R. {Chary}}, \bibinfo{author}{H.~C. {Chiang}},
  \bibinfo{author}{J.~{Chluba}}, \bibinfo{author}{P.~R. {Christensen}},
  \bibinfo{author}{S.~{Church}}, \bibinfo{author}{D.~L. {Clements}},
  \bibinfo{author}{S.~{Colombi}}, \bibinfo{author}{L.~P.~L. {Colombo}},
  \bibinfo{author}{C.~{Combet}}, \bibinfo{author}{A.~{Coulais}},
  \bibinfo{author}{B.~P. {Crill}}, \bibinfo{author}{A.~{Curto}},
  \bibinfo{author}{F.~{Cuttaia}}, \bibinfo{author}{L.~{Danese}},
  \bibinfo{author}{R.~D. {Davies}}, \bibinfo{author}{R.~J. {Davis}},
  \bibinfo{author}{P.~{de Bernardis}}, \bibinfo{author}{A.~{de Rosa}},
  \bibinfo{author}{G.~{de Zotti}}, \bibinfo{author}{J.~{Delabrouille}},
  \bibinfo{author}{F.-X. {D{\'e}sert}}, \bibinfo{author}{E.~{Di Valentino}},
  \bibinfo{author}{C.~{Dickinson}}, \bibinfo{author}{J.~M. {Diego}},
  \bibinfo{author}{K.~{Dolag}}, \bibinfo{author}{H.~{Dole}},
  \bibinfo{author}{S.~{Donzelli}}, \bibinfo{author}{O.~{Dor{\'e}}},
  \bibinfo{author}{M.~{Douspis}}, \bibinfo{author}{A.~{Ducout}},
  \bibinfo{author}{J.~{Dunkley}}, \bibinfo{author}{X.~{Dupac}},
  \bibinfo{author}{G.~{Efstathiou}}, \bibinfo{author}{F.~{Elsner}},
  \bibinfo{author}{T.~A. {En{\ss}lin}}, \bibinfo{author}{H.~K. {Eriksen}},
  \bibinfo{author}{M.~{Farhang}}, \bibinfo{author}{J.~{Fergusson}},
  \bibinfo{author}{F.~{Finelli}}, \bibinfo{author}{O.~{Forni}},
  \bibinfo{author}{M.~{Frailis}}, \bibinfo{author}{A.~A. {Fraisse}},
  \bibinfo{author}{E.~{Franceschi}}, \bibinfo{author}{A.~{Frejsel}},
  \bibinfo{author}{S.~{Galeotta}}, \bibinfo{author}{S.~{Galli}},
  \bibinfo{author}{K.~{Ganga}}, \bibinfo{author}{C.~{Gauthier}},
  \bibinfo{author}{M.~{Gerbino}}, \bibinfo{author}{T.~{Ghosh}},
  \bibinfo{author}{M.~{Giard}}, \bibinfo{author}{Y.~{Giraud-H{\'e}raud}},
  \bibinfo{author}{E.~{Giusarma}}, \bibinfo{author}{E.~{Gjerl{\o}w}},
  \bibinfo{author}{J.~{Gonz{\'a}lez-Nuevo}}, \bibinfo{author}{K.~M.
  {G{\'o}rski}}, \bibinfo{author}{S.~{Gratton}},
  \bibinfo{author}{A.~{Gregorio}}, \bibinfo{author}{A.~{Gruppuso}},
  \bibinfo{author}{J.~E. {Gudmundsson}}, \bibinfo{author}{J.~{Hamann}},
  \bibinfo{author}{F.~K. {Hansen}}, \bibinfo{author}{D.~{Hanson}},
  \bibinfo{author}{D.~L. {Harrison}}, \bibinfo{author}{G.~{Helou}},
  \bibinfo{author}{S.~{Henrot-Versill{\'e}}},
  \bibinfo{author}{C.~{Hern{\'a}ndez-Monteagudo}},
  \bibinfo{author}{D.~{Herranz}}, \bibinfo{author}{S.~R. {Hildebrandt}},
  \bibinfo{author}{E.~{Hivon}}, \bibinfo{author}{M.~{Hobson}},
  \bibinfo{author}{W.~A. {Holmes}}, \bibinfo{author}{A.~{Hornstrup}},
  \bibinfo{author}{W.~{Hovest}}, \bibinfo{author}{Z.~{Huang}},
  \bibinfo{author}{K.~M. {Huffenberger}}, \bibinfo{author}{G.~{Hurier}},
  \bibinfo{author}{A.~H. {Jaffe}}, \bibinfo{author}{T.~R. {Jaffe}},
  \bibinfo{author}{W.~C. {Jones}}, \bibinfo{author}{M.~{Juvela}},
  \bibinfo{author}{E.~{Keih{\"a}nen}}, \bibinfo{author}{R.~{Keskitalo}},
  \bibinfo{author}{T.~S. {Kisner}}, \bibinfo{author}{R.~{Kneissl}},
  \bibinfo{author}{J.~{Knoche}}, \bibinfo{author}{L.~{Knox}},
  \bibinfo{author}{M.~{Kunz}}, \bibinfo{author}{H.~{Kurki-Suonio}},
  \bibinfo{author}{G.~{Lagache}}, \bibinfo{author}{A.~{L{\"a}hteenm{\"a}ki}},
  \bibinfo{author}{J.-M. {Lamarre}}, \bibinfo{author}{A.~{Lasenby}},
  \bibinfo{author}{M.~{Lattanzi}}, \bibinfo{author}{C.~R. {Lawrence}},
  \bibinfo{author}{J.~P. {Leahy}}, \bibinfo{author}{R.~{Leonardi}},
  \bibinfo{author}{J.~{Lesgourgues}}, \bibinfo{author}{F.~{Levrier}},
  \bibinfo{author}{A.~{Lewis}}, \bibinfo{author}{M.~{Liguori}},
  \bibinfo{author}{P.~B. {Lilje}}, \bibinfo{author}{M.~{Linden-V{\o}rnle}},
  \bibinfo{author}{M.~{L{\'o}pez-Caniego}}, \bibinfo{author}{P.~M. {Lubin}},
  \bibinfo{author}{J.~F. {Mac{\'\i}as-P{\'e}rez}},
  \bibinfo{author}{G.~{Maggio}}, \bibinfo{author}{D.~{Maino}},
  \bibinfo{author}{N.~{Mandolesi}}, \bibinfo{author}{A.~{Mangilli}},
  \bibinfo{author}{A.~{Marchini}}, \bibinfo{author}{M.~{Maris}},
  \bibinfo{author}{P.~G. {Martin}}, \bibinfo{author}{M.~{Martinelli}},
  \bibinfo{author}{E.~{Mart{\'\i}nez-Gonz{\'a}lez}},
  \bibinfo{author}{S.~{Masi}}, \bibinfo{author}{S.~{Matarrese}},
  \bibinfo{author}{P.~{McGehee}}, \bibinfo{author}{P.~R. {Meinhold}},
  \bibinfo{author}{A.~{Melchiorri}}, \bibinfo{author}{J.-B. {Melin}},
  \bibinfo{author}{L.~{Mendes}}, \bibinfo{author}{A.~{Mennella}},
  \bibinfo{author}{M.~{Migliaccio}}, \bibinfo{author}{M.~{Millea}},
  \bibinfo{author}{S.~{Mitra}}, \bibinfo{author}{M.-A.
  {Miville-Desch{\^e}nes}}, \bibinfo{author}{A.~{Moneti}},
  \bibinfo{author}{L.~{Montier}}, \bibinfo{author}{G.~{Morgante}},
  \bibinfo{author}{D.~{Mortlock}}, \bibinfo{author}{A.~{Moss}},
  \bibinfo{author}{D.~{Munshi}}, \bibinfo{author}{J.~A. {Murphy}},
  \bibinfo{author}{P.~{Naselsky}}, \bibinfo{author}{F.~{Nati}},
  \bibinfo{author}{P.~{Natoli}}, \bibinfo{author}{C.~B. {Netterfield}},
  \bibinfo{author}{H.~U. {N{\o}rgaard-Nielsen}},
  \bibinfo{author}{F.~{Noviello}}, \bibinfo{author}{D.~{Novikov}},
  \bibinfo{author}{I.~{Novikov}}, \bibinfo{author}{C.~A. {Oxborrow}},
  \bibinfo{author}{F.~{Paci}}, \bibinfo{author}{L.~{Pagano}},
  \bibinfo{author}{F.~{Pajot}}, \bibinfo{author}{R.~{Paladini}},
  \bibinfo{author}{D.~{Paoletti}}, \bibinfo{author}{B.~{Partridge}},
  \bibinfo{author}{F.~{Pasian}}, \bibinfo{author}{G.~{Patanchon}},
  \bibinfo{author}{T.~J. {Pearson}}, \bibinfo{author}{O.~{Perdereau}},
  \bibinfo{author}{L.~{Perotto}}, \bibinfo{author}{F.~{Perrotta}},
  \bibinfo{author}{V.~{Pettorino}}, \bibinfo{author}{F.~{Piacentini}},
  \bibinfo{author}{M.~{Piat}}, \bibinfo{author}{E.~{Pierpaoli}},
  \bibinfo{author}{D.~{Pietrobon}}, \bibinfo{author}{S.~{Plaszczynski}},
  \bibinfo{author}{E.~{Pointecouteau}}, \bibinfo{author}{G.~{Polenta}},
  \bibinfo{author}{L.~{Popa}}, \bibinfo{author}{G.~W. {Pratt}},
  \bibinfo{author}{G.~{Pr{\'e}zeau}},
\newblock \bibinfo{title}{{Planck 2015 results. XIII. Cosmological
  parameters}},
\newblock \bibinfo{journal}{\aap} \bibinfo{volume}{594} (\bibinfo{year}{2016})
  \bibinfo{pages}{A13}. \DOIprefix\doi{10.1051/0004-6361/201525830}.
  \href{http://arxiv.org/abs/1502.01589}{{\tt arXiv:1502.01589}}.
\bibitem[{{Wood} et~al.(2017){Wood}, {Caputo}, {Charles}, {Di Mauro}, {Magill},
  {Perkins}, and {Fermi-LAT Collaboration}}]{2017ICRC...35..824W}
\bibinfo{author}{M.~{Wood}}, \bibinfo{author}{R.~{Caputo}},
  \bibinfo{author}{E.~{Charles}}, \bibinfo{author}{M.~{Di Mauro}},
  \bibinfo{author}{J.~{Magill}}, \bibinfo{author}{J.~S. {Perkins}},
  \bibinfo{author}{{Fermi-LAT Collaboration}},
\newblock \bibinfo{title}{{Fermipy: An open-source Python package for analysis
  of Fermi-LAT Data}},
\newblock in: \bibinfo{booktitle}{35th International Cosmic Ray Conference
  (ICRC2017)}, volume \bibinfo{volume}{301} of
  \textit{\bibinfo{series}{International Cosmic Ray Conference}},
  \bibinfo{year}{2017}, p. \bibinfo{pages}{824}.
  \DOIprefix\doi{10.22323/1.301.0824}.
  \href{http://arxiv.org/abs/1707.09551}{{\tt arXiv:1707.09551}}.
\bibitem[{{Abdollahi} et~al.(2022){Abdollahi}, {Acero}, {Baldini}, {Ballet},
  {Bastieri}, {Bellazzini}, {Berenji}, {Berretta}, {Bissaldi}, {Blandford},
  {Bloom}, {Bonino}, {Brill}, {Britto}, {Bruel}, {Burnett}, {Buson}, {Cameron},
  {Caputo}, {Caraveo}, {Castro}, {Chaty}, {Cheung}, {Chiaro}, {Cibrario},
  {Ciprini}, {Coronado-Bl{\'a}zquez}, {Crnogorcevic}, {Cutini}, {D'Ammando},
  {De Gaetano}, {Digel}, {Di Lalla}, {Dirirsa}, {Di Venere}, {Dom{\'\i}nguez},
  {Fallah Ramazani}, {Fegan}, {Ferrara}, {Fiori}, {Fleischhack}, {Franckowiak},
  {Fukazawa}, {Funk}, {Fusco}, {Galanti}, {Gammaldi}, {Gargano}, {Garrappa},
  {Gasparrini}, {Giacchino}, {Giglietto}, {Giordano}, {Giroletti}, {Glanzman},
  {Green}, {Grenier}, {Grondin}, {Guillemot}, {Guiriec}, {Gustafsson},
  {Harding}, {Hays}, {Hewitt}, {Horan}, {Hou}, {J{\'o}hannesson}, {Karwin},
  {Kayanoki}, {Kerr}, {Kuss}, {Landriu}, {Larsson}, {Latronico},
  {Lemoine-Goumard}, {Li}, {Liodakis}, {Longo}, {Loparco}, {Lott}, {Lubrano},
  {Maldera}, {Malyshev}, {Manfreda}, {Mart{\'\i}-Devesa}, {Mazziotta}, {Mereu},
  {Meyer}, {Michelson}, {Mirabal}, {Mitthumsiri}, {Mizuno}, {Moiseev},
  {Monzani}, {Morselli}, {Moskalenko}, {Negro}, {Nuss}, {Omodei}, {Orienti},
  {Orlando}, {Paneque}, {Pei}, {Perkins}, {Persic}, {Pesce-Rollins},
  {Petrosian}, {Pillera}, {Poon}, {Porter}, {Principe}, {Rain{\`o}}, {Rando},
  {Rani}, {Razzano}, {Razzaque}, {Reimer}, {Reimer}, {Reposeur},
  {S{\'a}nchez-Conde}, {Saz Parkinson}, {Scotton}, {Serini}, {Sgr{\`o}},
  {Siskind}, {Smith}, {Spandre}, {Spinelli}, {Sueoka}, {Suson}, {Tajima},
  {Tak}, {Thayer}, {Thompson}, {Torres}, {Troja}, {Valverde}, {Wood}, and
  {Zaharijas}}]{2022ApJS..260...53A}
\bibinfo{author}{S.~{Abdollahi}}, \bibinfo{author}{F.~{Acero}},
  \bibinfo{author}{L.~{Baldini}}, \bibinfo{author}{J.~{Ballet}},
  \bibinfo{author}{D.~{Bastieri}}, \bibinfo{author}{R.~{Bellazzini}},
  \bibinfo{author}{B.~{Berenji}}, \bibinfo{author}{A.~{Berretta}},
  \bibinfo{author}{E.~{Bissaldi}}, \bibinfo{author}{R.~D. {Blandford}},
  \bibinfo{author}{E.~{Bloom}}, \bibinfo{author}{R.~{Bonino}},
  \bibinfo{author}{A.~{Brill}}, \bibinfo{author}{R.~J. {Britto}},
  \bibinfo{author}{P.~{Bruel}}, \bibinfo{author}{T.~H. {Burnett}},
  \bibinfo{author}{S.~{Buson}}, \bibinfo{author}{R.~A. {Cameron}},
  \bibinfo{author}{R.~{Caputo}}, \bibinfo{author}{P.~A. {Caraveo}},
  \bibinfo{author}{D.~{Castro}}, \bibinfo{author}{S.~{Chaty}},
  \bibinfo{author}{C.~C. {Cheung}}, \bibinfo{author}{G.~{Chiaro}},
  \bibinfo{author}{N.~{Cibrario}}, \bibinfo{author}{S.~{Ciprini}},
  \bibinfo{author}{J.~{Coronado-Bl{\'a}zquez}},
  \bibinfo{author}{M.~{Crnogorcevic}}, \bibinfo{author}{S.~{Cutini}},
  \bibinfo{author}{F.~{D'Ammando}}, \bibinfo{author}{S.~{De Gaetano}},
  \bibinfo{author}{S.~W. {Digel}}, \bibinfo{author}{N.~{Di Lalla}},
  \bibinfo{author}{F.~{Dirirsa}}, \bibinfo{author}{L.~{Di Venere}},
  \bibinfo{author}{A.~{Dom{\'\i}nguez}}, \bibinfo{author}{V.~{Fallah
  Ramazani}}, \bibinfo{author}{S.~J. {Fegan}}, \bibinfo{author}{E.~C.
  {Ferrara}}, \bibinfo{author}{A.~{Fiori}}, \bibinfo{author}{H.~{Fleischhack}},
  \bibinfo{author}{A.~{Franckowiak}}, \bibinfo{author}{Y.~{Fukazawa}},
  \bibinfo{author}{S.~{Funk}}, \bibinfo{author}{P.~{Fusco}},
  \bibinfo{author}{G.~{Galanti}}, \bibinfo{author}{V.~{Gammaldi}},
  \bibinfo{author}{F.~{Gargano}}, \bibinfo{author}{S.~{Garrappa}},
  \bibinfo{author}{D.~{Gasparrini}}, \bibinfo{author}{F.~{Giacchino}},
  \bibinfo{author}{N.~{Giglietto}}, \bibinfo{author}{F.~{Giordano}},
  \bibinfo{author}{M.~{Giroletti}}, \bibinfo{author}{T.~{Glanzman}},
  \bibinfo{author}{D.~{Green}}, \bibinfo{author}{I.~A. {Grenier}},
  \bibinfo{author}{M.~H. {Grondin}}, \bibinfo{author}{L.~{Guillemot}},
  \bibinfo{author}{S.~{Guiriec}}, \bibinfo{author}{M.~{Gustafsson}},
  \bibinfo{author}{A.~K. {Harding}}, \bibinfo{author}{E.~{Hays}},
  \bibinfo{author}{J.~W. {Hewitt}}, \bibinfo{author}{D.~{Horan}},
  \bibinfo{author}{X.~{Hou}}, \bibinfo{author}{G.~{J{\'o}hannesson}},
  \bibinfo{author}{C.~{Karwin}}, \bibinfo{author}{T.~{Kayanoki}},
  \bibinfo{author}{M.~{Kerr}}, \bibinfo{author}{M.~{Kuss}},
  \bibinfo{author}{D.~{Landriu}}, \bibinfo{author}{S.~{Larsson}},
  \bibinfo{author}{L.~{Latronico}}, \bibinfo{author}{M.~{Lemoine-Goumard}},
  \bibinfo{author}{J.~{Li}}, \bibinfo{author}{I.~{Liodakis}},
  \bibinfo{author}{F.~{Longo}}, \bibinfo{author}{F.~{Loparco}},
  \bibinfo{author}{B.~{Lott}}, \bibinfo{author}{P.~{Lubrano}},
  \bibinfo{author}{S.~{Maldera}}, \bibinfo{author}{D.~{Malyshev}},
  \bibinfo{author}{A.~{Manfreda}}, \bibinfo{author}{G.~{Mart{\'\i}-Devesa}},
  \bibinfo{author}{M.~N. {Mazziotta}}, \bibinfo{author}{I.~{Mereu}},
  \bibinfo{author}{M.~{Meyer}}, \bibinfo{author}{P.~F. {Michelson}},
  \bibinfo{author}{N.~{Mirabal}}, \bibinfo{author}{W.~{Mitthumsiri}},
  \bibinfo{author}{T.~{Mizuno}}, \bibinfo{author}{A.~A. {Moiseev}},
  \bibinfo{author}{M.~E. {Monzani}}, \bibinfo{author}{A.~{Morselli}},
  \bibinfo{author}{I.~V. {Moskalenko}}, \bibinfo{author}{M.~{Negro}},
  \bibinfo{author}{E.~{Nuss}}, \bibinfo{author}{N.~{Omodei}},
  \bibinfo{author}{M.~{Orienti}}, \bibinfo{author}{E.~{Orlando}},
  \bibinfo{author}{D.~{Paneque}}, \bibinfo{author}{Z.~{Pei}},
  \bibinfo{author}{J.~S. {Perkins}}, \bibinfo{author}{M.~{Persic}},
  \bibinfo{author}{M.~{Pesce-Rollins}}, \bibinfo{author}{V.~{Petrosian}},
  \bibinfo{author}{R.~{Pillera}}, \bibinfo{author}{H.~{Poon}},
  \bibinfo{author}{T.~A. {Porter}}, \bibinfo{author}{G.~{Principe}},
  \bibinfo{author}{S.~{Rain{\`o}}}, \bibinfo{author}{R.~{Rando}},
  \bibinfo{author}{B.~{Rani}}, \bibinfo{author}{M.~{Razzano}},
  \bibinfo{author}{S.~{Razzaque}}, \bibinfo{author}{A.~{Reimer}},
  \bibinfo{author}{O.~{Reimer}}, \bibinfo{author}{T.~{Reposeur}},
  \bibinfo{author}{M.~{S{\'a}nchez-Conde}}, \bibinfo{author}{P.~M. {Saz
  Parkinson}}, \bibinfo{author}{L.~{Scotton}}, \bibinfo{author}{D.~{Serini}},
  \bibinfo{author}{C.~{Sgr{\`o}}}, \bibinfo{author}{E.~J. {Siskind}},
  \bibinfo{author}{D.~A. {Smith}}, \bibinfo{author}{G.~{Spandre}},
  \bibinfo{author}{P.~{Spinelli}}, \bibinfo{author}{K.~{Sueoka}},
  \bibinfo{author}{D.~J. {Suson}}, \bibinfo{author}{H.~{Tajima}},
  \bibinfo{author}{D.~{Tak}}, \bibinfo{author}{J.~B. {Thayer}},
  \bibinfo{author}{D.~J. {Thompson}}, \bibinfo{author}{D.~F. {Torres}},
  \bibinfo{author}{E.~{Troja}}, \bibinfo{author}{J.~{Valverde}},
  \bibinfo{author}{K.~{Wood}}, \bibinfo{author}{G.~{Zaharijas}},
\newblock \bibinfo{title}{{Incremental Fermi Large Area Telescope Fourth Source
  Catalog}},
\newblock \bibinfo{journal}{\apjs} \bibinfo{volume}{260} (\bibinfo{year}{2022})
  \bibinfo{pages}{53}. \DOIprefix\doi{10.3847/1538-4365/ac6751}.
  \href{http://arxiv.org/abs/2201.11184}{{\tt arXiv:2201.11184}}.
\bibitem[{{Ballet} et~al.(2023){Ballet}, {Bruel}, {Burnett}, {Lott}, and {The
  Fermi-LAT collaboration}}]{2023arXiv230712546B}
\bibinfo{author}{J.~{Ballet}}, \bibinfo{author}{P.~{Bruel}},
  \bibinfo{author}{T.~H. {Burnett}}, \bibinfo{author}{B.~{Lott}},
  \bibinfo{author}{{The Fermi-LAT collaboration}},
\newblock \bibinfo{title}{{Fermi Large Area Telescope Fourth Source Catalog
  Data Release 4 (4FGL-DR4)}},
\newblock \bibinfo{journal}{arXiv e-prints}  (\bibinfo{year}{2023})
  \bibinfo{pages}{arXiv:2307.12546}. \DOIprefix\doi{10.48550/arXiv.2307.12546}.
  \href{http://arxiv.org/abs/2307.12546}{{\tt arXiv:2307.12546}}.
\bibitem[{{Arnaud}(1996)}]{1996ASPC..101...17A}
\bibinfo{author}{K.~A. {Arnaud}},
\newblock \bibinfo{title}{{XSPEC: The First Ten Years}},
\newblock in: \bibinfo{editor}{G.~H. {Jacoby}}, \bibinfo{editor}{J.~{Barnes}}
  (Eds.), \bibinfo{booktitle}{Astronomical Data Analysis Software and Systems
  V}, volume \bibinfo{volume}{101} of \textit{\bibinfo{series}{Astronomical
  Society of the Pacific Conference Series}}, \bibinfo{year}{1996},
  p.~\bibinfo{pages}{17}.
\bibitem[{{Evans} et~al.(2009){Evans}, {Beardmore}, {Page}, {Osborne},
  {O'Brien}, {Willingale}, {Starling}, {Burrows}, {Godet}, {Vetere}, {Racusin},
  {Goad}, {Wiersema}, {Angelini}, {Capalbi}, {Chincarini}, {Gehrels}, {Kennea},
  {Margutti}, {Morris}, {Mountford}, {Pagani}, {Perri}, {Romano}, and
  {Tanvir}}]{2009MNRAS.397.1177E}
\bibinfo{author}{P.~A. {Evans}}, \bibinfo{author}{A.~P. {Beardmore}},
  \bibinfo{author}{K.~L. {Page}}, \bibinfo{author}{J.~P. {Osborne}},
  \bibinfo{author}{P.~T. {O'Brien}}, \bibinfo{author}{R.~{Willingale}},
  \bibinfo{author}{R.~L.~C. {Starling}}, \bibinfo{author}{D.~N. {Burrows}},
  \bibinfo{author}{O.~{Godet}}, \bibinfo{author}{L.~{Vetere}},
  \bibinfo{author}{J.~{Racusin}}, \bibinfo{author}{M.~R. {Goad}},
  \bibinfo{author}{K.~{Wiersema}}, \bibinfo{author}{L.~{Angelini}},
  \bibinfo{author}{M.~{Capalbi}}, \bibinfo{author}{G.~{Chincarini}},
  \bibinfo{author}{N.~{Gehrels}}, \bibinfo{author}{J.~A. {Kennea}},
  \bibinfo{author}{R.~{Margutti}}, \bibinfo{author}{D.~C. {Morris}},
  \bibinfo{author}{C.~J. {Mountford}}, \bibinfo{author}{C.~{Pagani}},
  \bibinfo{author}{M.~{Perri}}, \bibinfo{author}{P.~{Romano}},
  \bibinfo{author}{N.~{Tanvir}},
\newblock \bibinfo{title}{{Methods and results of an automatic analysis of a
  complete sample of Swift-XRT observations of GRBs}},
\newblock \bibinfo{journal}{\mnras} \bibinfo{volume}{397}
  (\bibinfo{year}{2009}) \bibinfo{pages}{1177--1201}.
  \DOIprefix\doi{10.1111/j.1365-2966.2009.14913.x}.
  \href{http://arxiv.org/abs/0812.3662}{{\tt arXiv:0812.3662}}.
\bibitem[{{Kalberla} et~al.(2005){Kalberla}, {Burton}, {Hartmann}, {Arnal},
  {Bajaja}, {Morras}, and {P{\"o}ppel}}]{Kalberla_2005}
\bibinfo{author}{P.~M.~W. {Kalberla}}, \bibinfo{author}{W.~B. {Burton}},
  \bibinfo{author}{D.~{Hartmann}}, \bibinfo{author}{E.~M. {Arnal}},
  \bibinfo{author}{E.~{Bajaja}}, \bibinfo{author}{R.~{Morras}},
  \bibinfo{author}{W.~G.~L. {P{\"o}ppel}},
\newblock \bibinfo{title}{{The Leiden/Argentine/Bonn (LAB) Survey of Galactic
  HI. Final data release of the combined LDS and IAR surveys with improved
  stray-radiation corrections}},
\newblock \bibinfo{journal}{\aap} \bibinfo{volume}{440} (\bibinfo{year}{2005})
  \bibinfo{pages}{775--782}. \DOIprefix\doi{10.1051/0004-6361:20041864}.
  \href{http://arxiv.org/abs/astro-ph/0504140}{{\tt arXiv:astro-ph/0504140}}.
\bibitem[{{Schlafly} and {Finkbeiner}(2011)}]{Schlafly_2011}
\bibinfo{author}{E.~F. {Schlafly}}, \bibinfo{author}{D.~P. {Finkbeiner}},
\newblock \bibinfo{title}{{Measuring Reddening with Sloan Digital Sky Survey
  Stellar Spectra and Recalibrating SFD}},
\newblock \bibinfo{journal}{\apj} \bibinfo{volume}{737} (\bibinfo{year}{2011})
  \bibinfo{pages}{103}. \DOIprefix\doi{10.1088/0004-637X/737/2/103}.
  \href{http://arxiv.org/abs/1012.4804}{{\tt arXiv:1012.4804}}.
\bibitem[{{Breeveld} et~al.(2011){Breeveld}, {Landsman}, {Holland}, {Roming},
  {Kuin}, and {Page}}]{2011AIPC.1358..373B}
\bibinfo{author}{A.~A. {Breeveld}}, \bibinfo{author}{W.~{Landsman}},
  \bibinfo{author}{S.~T. {Holland}}, \bibinfo{author}{P.~{Roming}},
  \bibinfo{author}{N.~P.~M. {Kuin}}, \bibinfo{author}{M.~J. {Page}},
\newblock \bibinfo{title}{{An Updated Ultraviolet Calibration for the
  Swift/UVOT}},
\newblock in: \bibinfo{editor}{J.~E. {McEnery}}, \bibinfo{editor}{J.~L.
  {Racusin}}, \bibinfo{editor}{N.~{Gehrels}} (Eds.),
  \bibinfo{booktitle}{American Institute of Physics Conference Series}, volume
  \bibinfo{volume}{1358} of \textit{\bibinfo{series}{American Institute of
  Physics Conference Series}}, \bibinfo{year}{2011}, pp.
  \bibinfo{pages}{373--376}. \DOIprefix\doi{10.1063/1.3621807}.
  \href{http://arxiv.org/abs/1102.4717}{{\tt arXiv:1102.4717}}.
\bibitem[{{Ackermann} et~al.(2016){Ackermann}, {Anantua}, {Asano}, {Baldini},
  {Barbiellini}, {Bastieri}, {Becerra Gonzalez}, {Bellazzini}, {Bissaldi},
  {Blandford}, {Bloom}, {Bonino}, {Bottacini}, {Bruel}, {Buehler}, {Caliandro},
  {Cameron}, {Caragiulo}, {Caraveo}, {Cavazzuti}, {Cecchi}, {Cheung}, {Chiang},
  {Chiaro}, {Ciprini}, {Cohen-Tanugi}, {Costanza}, {Cutini}, {D'Ammando}, {de
  Palma}, {Desiante}, {Digel}, {Di Lalla}, {Di Mauro}, {Di Venere}, {Drell},
  {Favuzzi}, {Fegan}, {Ferrara}, {Fukazawa}, {Funk}, {Fusco}, {Gargano},
  {Gasparrini}, {Giglietto}, {Giordano}, {Giroletti}, {Grenier}, {Guillemot},
  {Guiriec}, {Hayashida}, {Hays}, {Horan}, {J{\'o}hannesson}, {Kensei},
  {Kocevski}, {Kuss}, {La Mura}, {Larsson}, {Latronico}, {Li}, {Longo},
  {Loparco}, {Lott}, {Lovellette}, {Lubrano}, {Madejski}, {Magill}, {Maldera},
  {Manfreda}, {Mayer}, {Mazziotta}, {Michelson}, {Mirabal}, {Mizuno},
  {Monzani}, {Morselli}, {Moskalenko}, {Nalewajko}, {Negro}, {Nuss}, {Ohsugi},
  {Orlando}, {Paneque}, {Perkins}, {Pesce-Rollins}, {Piron}, {Pivato},
  {Porter}, {Principe}, {Rando}, {Razzano}, {Razzaque}, {Reimer}, {Scargle},
  {Sgr{\`o}}, {Sikora}, {Simone}, {Siskind}, {Spada}, {Spinelli}, {Stawarz},
  {Thayer}, {Thompson}, {Torres}, {Troja}, {Uchiyama}, {Yuan}, and
  {Zimmer}}]{2016ApJ...824L..20A}
\bibinfo{author}{M.~{Ackermann}}, \bibinfo{author}{R.~{Anantua}},
  \bibinfo{author}{K.~{Asano}}, \bibinfo{author}{L.~{Baldini}},
  \bibinfo{author}{G.~{Barbiellini}}, \bibinfo{author}{D.~{Bastieri}},
  \bibinfo{author}{J.~{Becerra Gonzalez}}, \bibinfo{author}{R.~{Bellazzini}},
  \bibinfo{author}{E.~{Bissaldi}}, \bibinfo{author}{R.~D. {Blandford}},
  \bibinfo{author}{E.~D. {Bloom}}, \bibinfo{author}{R.~{Bonino}},
  \bibinfo{author}{E.~{Bottacini}}, \bibinfo{author}{P.~{Bruel}},
  \bibinfo{author}{R.~{Buehler}}, \bibinfo{author}{G.~A. {Caliandro}},
  \bibinfo{author}{R.~A. {Cameron}}, \bibinfo{author}{M.~{Caragiulo}},
  \bibinfo{author}{P.~A. {Caraveo}}, \bibinfo{author}{E.~{Cavazzuti}},
  \bibinfo{author}{C.~{Cecchi}}, \bibinfo{author}{C.~C. {Cheung}},
  \bibinfo{author}{J.~{Chiang}}, \bibinfo{author}{G.~{Chiaro}},
  \bibinfo{author}{S.~{Ciprini}}, \bibinfo{author}{J.~{Cohen-Tanugi}},
  \bibinfo{author}{F.~{Costanza}}, \bibinfo{author}{S.~{Cutini}},
  \bibinfo{author}{F.~{D'Ammando}}, \bibinfo{author}{F.~{de Palma}},
  \bibinfo{author}{R.~{Desiante}}, \bibinfo{author}{S.~W. {Digel}},
  \bibinfo{author}{N.~{Di Lalla}}, \bibinfo{author}{M.~{Di Mauro}},
  \bibinfo{author}{L.~{Di Venere}}, \bibinfo{author}{P.~S. {Drell}},
  \bibinfo{author}{C.~{Favuzzi}}, \bibinfo{author}{S.~J. {Fegan}},
  \bibinfo{author}{E.~C. {Ferrara}}, \bibinfo{author}{Y.~{Fukazawa}},
  \bibinfo{author}{S.~{Funk}}, \bibinfo{author}{P.~{Fusco}},
  \bibinfo{author}{F.~{Gargano}}, \bibinfo{author}{D.~{Gasparrini}},
  \bibinfo{author}{N.~{Giglietto}}, \bibinfo{author}{F.~{Giordano}},
  \bibinfo{author}{M.~{Giroletti}}, \bibinfo{author}{I.~A. {Grenier}},
  \bibinfo{author}{L.~{Guillemot}}, \bibinfo{author}{S.~{Guiriec}},
  \bibinfo{author}{M.~{Hayashida}}, \bibinfo{author}{E.~{Hays}},
  \bibinfo{author}{D.~{Horan}}, \bibinfo{author}{G.~{J{\'o}hannesson}},
  \bibinfo{author}{S.~{Kensei}}, \bibinfo{author}{D.~{Kocevski}},
  \bibinfo{author}{M.~{Kuss}}, \bibinfo{author}{G.~{La Mura}},
  \bibinfo{author}{S.~{Larsson}}, \bibinfo{author}{L.~{Latronico}},
  \bibinfo{author}{J.~{Li}}, \bibinfo{author}{F.~{Longo}},
  \bibinfo{author}{F.~{Loparco}}, \bibinfo{author}{B.~{Lott}},
  \bibinfo{author}{M.~N. {Lovellette}}, \bibinfo{author}{P.~{Lubrano}},
  \bibinfo{author}{G.~M. {Madejski}}, \bibinfo{author}{J.~D. {Magill}},
  \bibinfo{author}{S.~{Maldera}}, \bibinfo{author}{A.~{Manfreda}},
  \bibinfo{author}{M.~{Mayer}}, \bibinfo{author}{M.~N. {Mazziotta}},
  \bibinfo{author}{P.~F. {Michelson}}, \bibinfo{author}{N.~{Mirabal}},
  \bibinfo{author}{T.~{Mizuno}}, \bibinfo{author}{M.~E. {Monzani}},
  \bibinfo{author}{A.~{Morselli}}, \bibinfo{author}{I.~V. {Moskalenko}},
  \bibinfo{author}{K.~{Nalewajko}}, \bibinfo{author}{M.~{Negro}},
  \bibinfo{author}{E.~{Nuss}}, \bibinfo{author}{T.~{Ohsugi}},
  \bibinfo{author}{E.~{Orlando}}, \bibinfo{author}{D.~{Paneque}},
  \bibinfo{author}{J.~S. {Perkins}}, \bibinfo{author}{M.~{Pesce-Rollins}},
  \bibinfo{author}{F.~{Piron}}, \bibinfo{author}{G.~{Pivato}},
  \bibinfo{author}{T.~A. {Porter}}, \bibinfo{author}{G.~{Principe}},
  \bibinfo{author}{R.~{Rando}}, \bibinfo{author}{M.~{Razzano}},
  \bibinfo{author}{S.~{Razzaque}}, \bibinfo{author}{A.~{Reimer}},
  \bibinfo{author}{J.~D. {Scargle}}, \bibinfo{author}{C.~{Sgr{\`o}}},
  \bibinfo{author}{M.~{Sikora}}, \bibinfo{author}{D.~{Simone}},
  \bibinfo{author}{E.~J. {Siskind}}, \bibinfo{author}{F.~{Spada}},
  \bibinfo{author}{P.~{Spinelli}}, \bibinfo{author}{L.~{Stawarz}},
  \bibinfo{author}{J.~B. {Thayer}}, \bibinfo{author}{D.~J. {Thompson}},
  \bibinfo{author}{D.~F. {Torres}}, \bibinfo{author}{E.~{Troja}},
  \bibinfo{author}{Y.~{Uchiyama}}, \bibinfo{author}{Y.~{Yuan}},
  \bibinfo{author}{S.~{Zimmer}},
\newblock \bibinfo{title}{{Minute-timescale >100 MeV {\ensuremath{\gamma}}-Ray
  Variability during the Giant Outburst of Quasar 3C 279 Observed by Fermi-LAT
  in 2015 June}},
\newblock \bibinfo{journal}{\apjl} \bibinfo{volume}{824} (\bibinfo{year}{2016})
  \bibinfo{pages}{L20}. \DOIprefix\doi{10.3847/2041-8205/824/2/L20}.
  \href{http://arxiv.org/abs/1605.05324}{{\tt arXiv:1605.05324}}.
\bibitem[{{Liu} and {Bai}(2006)}]{2006ApJ...653.1089L}
\bibinfo{author}{H.~T. {Liu}}, \bibinfo{author}{J.~M. {Bai}},
\newblock \bibinfo{title}{{Absorption of 10-200 GeV Gamma Rays by Radiation
  from Broad-Line Regions in Blazars}},
\newblock \bibinfo{journal}{\apj} \bibinfo{volume}{653} (\bibinfo{year}{2006})
  \bibinfo{pages}{1089--1097}. \DOIprefix\doi{10.1086/509097}.
  \href{http://arxiv.org/abs/0807.3135}{{\tt arXiv:0807.3135}}.
\bibitem[{{Costamante} et~al.(2018){Costamante}, {Cutini}, {Tosti}, {Antolini},
  and {Tramacere}}]{Costamante_2018}
\bibinfo{author}{L.~{Costamante}}, \bibinfo{author}{S.~{Cutini}},
  \bibinfo{author}{G.~{Tosti}}, \bibinfo{author}{E.~{Antolini}},
  \bibinfo{author}{A.~{Tramacere}},
\newblock \bibinfo{title}{{On the origin of gamma-rays in Fermi blazars:
  beyondthe broad-line region}},
\newblock \bibinfo{journal}{\mnras} \bibinfo{volume}{477}
  (\bibinfo{year}{2018}) \bibinfo{pages}{4749--4767}.
  \DOIprefix\doi{10.1093/mnras/sty887}.
  \href{http://arxiv.org/abs/1804.02408}{{\tt arXiv:1804.02408}}.
\bibitem[{{Finke} et~al.(2008){Finke}, {Dermer}, and
  {B{\"o}ttcher}}]{2008ApJ...686..181F}
\bibinfo{author}{J.~D. {Finke}}, \bibinfo{author}{C.~D. {Dermer}},
  \bibinfo{author}{M.~{B{\"o}ttcher}},
\newblock \bibinfo{title}{{Synchrotron Self-Compton Analysis of TeV
  X-Ray-Selected BL Lacertae Objects}},
\newblock \bibinfo{journal}{\apj} \bibinfo{volume}{686} (\bibinfo{year}{2008})
  \bibinfo{pages}{181--194}. \DOIprefix\doi{10.1086/590900}.
  \href{http://arxiv.org/abs/0802.1529}{{\tt arXiv:0802.1529}}.
\bibitem[{{Ghisellini} and {Tavecchio}(2009)}]{2009MNRAS.397..985G}
\bibinfo{author}{G.~{Ghisellini}}, \bibinfo{author}{F.~{Tavecchio}},
\newblock \bibinfo{title}{{Canonical high-power blazars}},
\newblock \bibinfo{journal}{\mnras} \bibinfo{volume}{397}
  (\bibinfo{year}{2009}) \bibinfo{pages}{985--1002}.
  \DOIprefix\doi{10.1111/j.1365-2966.2009.15007.x}.
  \href{http://arxiv.org/abs/0902.0793}{{\tt arXiv:0902.0793}}.
\bibitem[{{Dermer} et~al.(2009){Dermer}, {Finke}, {Krug}, and
  {B{\"o}ttcher}}]{2009ApJ...692...32D}
\bibinfo{author}{C.~D. {Dermer}}, \bibinfo{author}{J.~D. {Finke}},
  \bibinfo{author}{H.~{Krug}}, \bibinfo{author}{M.~{B{\"o}ttcher}},
\newblock \bibinfo{title}{{Gamma-Ray Studies of Blazars: Synchro-Compton
  Analysis of Flat Spectrum Radio Quasars}},
\newblock \bibinfo{journal}{\apj} \bibinfo{volume}{692} (\bibinfo{year}{2009})
  \bibinfo{pages}{32--46}. \DOIprefix\doi{10.1088/0004-637X/692/1/32}.
  \href{http://arxiv.org/abs/0808.3185}{{\tt arXiv:0808.3185}}.
\bibitem[{{Shakura} and {Sunyaev}(1973)}]{1973A&A....24..337S}
\bibinfo{author}{N.~I. {Shakura}}, \bibinfo{author}{R.~A. {Sunyaev}},
\newblock \bibinfo{title}{{Black holes in binary systems. Observational
  appearance.}},
\newblock \bibinfo{journal}{\aap} \bibinfo{volume}{24} (\bibinfo{year}{1973})
  \bibinfo{pages}{337--355}.
\bibitem[{{Sikora} et~al.(2002){Sikora}, {B{\l}a{\.z}ejowski}, {Moderski}, and
  {Madejski}}]{2002ApJ...577...78S}
\bibinfo{author}{M.~{Sikora}}, \bibinfo{author}{M.~{B{\l}a{\.z}ejowski}},
  \bibinfo{author}{R.~{Moderski}}, \bibinfo{author}{G.~M. {Madejski}},
\newblock \bibinfo{title}{{On the Nature of MeV Blazars}},
\newblock \bibinfo{journal}{\apj} \bibinfo{volume}{577} (\bibinfo{year}{2002})
  \bibinfo{pages}{78--84}. \DOIprefix\doi{10.1086/342164}.
  \href{http://arxiv.org/abs/astro-ph/0205527}{{\tt arXiv:astro-ph/0205527}}.
\bibitem[{{B{\l}a{\.z}ejowski} et~al.(2000){B{\l}a{\.z}ejowski}, {Sikora},
  {Moderski}, and {Madejski}}]{2000ApJ...545..107B}
\bibinfo{author}{M.~{B{\l}a{\.z}ejowski}}, \bibinfo{author}{M.~{Sikora}},
  \bibinfo{author}{R.~{Moderski}}, \bibinfo{author}{G.~M. {Madejski}},
\newblock \bibinfo{title}{{Comptonization of Infrared Radiation from Hot Dust
  by Relativistic Jets in Quasars}},
\newblock \bibinfo{journal}{\apj} \bibinfo{volume}{545} (\bibinfo{year}{2000})
  \bibinfo{pages}{107--116}. \DOIprefix\doi{10.1086/317791}.
  \href{http://arxiv.org/abs/astro-ph/0008154}{{\tt arXiv:astro-ph/0008154}}.
\bibitem[{{Dermer} and {Menon}(2009)}]{2009herb.book.....D}
\bibinfo{author}{C.~D. {Dermer}}, \bibinfo{author}{G.~{Menon}},
  \bibinfo{title}{{High Energy Radiation from Black Holes: Gamma Rays, Cosmic
  Rays, and Neutrinos}}, \bibinfo{year}{2009}.
\bibitem[{{Celotti} and {Ghisellini}(2008)}]{2008MNRAS.385..283C}
\bibinfo{author}{A.~{Celotti}}, \bibinfo{author}{G.~{Ghisellini}},
\newblock \bibinfo{title}{{The power of blazar jets}},
\newblock \bibinfo{journal}{\mnras} \bibinfo{volume}{385}
  (\bibinfo{year}{2008}) \bibinfo{pages}{283--300}.
  \DOIprefix\doi{10.1111/j.1365-2966.2007.12758.x}.
  \href{http://arxiv.org/abs/0711.4112}{{\tt arXiv:0711.4112}}.
\bibitem[{{Rakshit} et~al.(2017){Rakshit}, {Stalin}, {Chand}, and
  {Zhang}}]{2017ApJS..229...39R}
\bibinfo{author}{S.~{Rakshit}}, \bibinfo{author}{C.~S. {Stalin}},
  \bibinfo{author}{H.~{Chand}}, \bibinfo{author}{X.-G. {Zhang}},
\newblock \bibinfo{title}{{A Catalog of Narrow Line Seyfert 1 Galaxies from the
  Sloan Digital Sky Survey Data Release 12}},
\newblock \bibinfo{journal}{\apjs} \bibinfo{volume}{229} (\bibinfo{year}{2017})
  \bibinfo{pages}{39}. \DOIprefix\doi{10.3847/1538-4365/aa6971}.
  \href{http://arxiv.org/abs/1704.07700}{{\tt arXiv:1704.07700}}.
\bibitem[{{Ghisellini} and {Tavecchio}(2015)}]{2015MNRAS.448.1060G}
\bibinfo{author}{G.~{Ghisellini}}, \bibinfo{author}{F.~{Tavecchio}},
\newblock \bibinfo{title}{{Fermi/LAT broad emission line blazars}},
\newblock \bibinfo{journal}{\mnras} \bibinfo{volume}{448}
  (\bibinfo{year}{2015}) \bibinfo{pages}{1060--1077}.
  \DOIprefix\doi{10.1093/mnras/stv055}.
  \href{http://arxiv.org/abs/1501.03504}{{\tt arXiv:1501.03504}}.
\bibitem[{{Wang} et~al.(2016){Wang}, {Du}, {Hu}, {Bai}, {Wang}, {Yi}, {Wang},
  {Zhang}, {Xin}, {Lun}, {Chang}, and {Fan}}]{2016ApJ...824..149W}
\bibinfo{author}{F.~{Wang}}, \bibinfo{author}{P.~{Du}},
  \bibinfo{author}{C.~{Hu}}, \bibinfo{author}{J.-M. {Bai}},
  \bibinfo{author}{C.-J. {Wang}}, \bibinfo{author}{W.-M. {Yi}},
  \bibinfo{author}{J.-G. {Wang}}, \bibinfo{author}{J.-J. {Zhang}},
  \bibinfo{author}{Y.-X. {Xin}}, \bibinfo{author}{B.-L. {Lun}},
  \bibinfo{author}{L.~{Chang}}, \bibinfo{author}{Y.-F. {Fan}},
\newblock \bibinfo{title}{{Reverberation Mapping of the Gamma-Ray Loud
  Narrow-line Seyfert 1 Galaxy 1H 0323+342}},
\newblock \bibinfo{journal}{\apj} \bibinfo{volume}{824} (\bibinfo{year}{2016})
  \bibinfo{pages}{149}. \DOIprefix\doi{10.3847/0004-637X/824/2/149}.
\bibitem[{{Pan} et~al.(2018){Pan}, {Yuan}, {Yao}, {Komossa}, and
  {Jin}}]{2018ApJ...866...69P}
\bibinfo{author}{H.-W. {Pan}}, \bibinfo{author}{W.~{Yuan}},
  \bibinfo{author}{S.~{Yao}}, \bibinfo{author}{S.~{Komossa}},
  \bibinfo{author}{C.~{Jin}},
\newblock \bibinfo{title}{{Independent Estimation of Black Hole Mass for the
  {\ensuremath{\gamma}}-ray-detected Archetypal Narrow-line Seyfert 1 Galaxy 1H
  0323+342 from X-Ray Variability}},
\newblock \bibinfo{journal}{\apj} \bibinfo{volume}{866} (\bibinfo{year}{2018})
  \bibinfo{pages}{69}. \DOIprefix\doi{10.3847/1538-4357/aadd4a}.
  \href{http://arxiv.org/abs/1807.01002}{{\tt arXiv:1807.01002}}.
\bibitem[{{Paliya} et~al.(2017){Paliya}, {Marcotulli}, {Ajello}, {Joshi},
  {Sahayanathan}, {Rao}, and {Hartmann}}]{2017ApJ...851...33P}
\bibinfo{author}{V.~S. {Paliya}}, \bibinfo{author}{L.~{Marcotulli}},
  \bibinfo{author}{M.~{Ajello}}, \bibinfo{author}{M.~{Joshi}},
  \bibinfo{author}{S.~{Sahayanathan}}, \bibinfo{author}{A.~R. {Rao}},
  \bibinfo{author}{D.~{Hartmann}},
\newblock \bibinfo{title}{{General Physical Properties of CGRaBS Blazars}},
\newblock \bibinfo{journal}{\apj} \bibinfo{volume}{851} (\bibinfo{year}{2017})
  \bibinfo{pages}{33}. \DOIprefix\doi{10.3847/1538-4357/aa98e1}.
  \href{http://arxiv.org/abs/1711.01292}{{\tt arXiv:1711.01292}}.
\bibitem[{{Pandey} et~al.(2017){Pandey}, {Gupta}, and
  {Wiita}}]{2017ApJ...841..123P}
\bibinfo{author}{A.~{Pandey}}, \bibinfo{author}{A.~C. {Gupta}},
  \bibinfo{author}{P.~J. {Wiita}},
\newblock \bibinfo{title}{{X-Ray Intraday Variability of Five TeV Blazars with
  NuSTAR}},
\newblock \bibinfo{journal}{\apj} \bibinfo{volume}{841} (\bibinfo{year}{2017})
  \bibinfo{pages}{123}. \DOIprefix\doi{10.3847/1538-4357/aa705e}.
  \href{http://arxiv.org/abs/1705.02719}{{\tt arXiv:1705.02719}}.
\bibitem[{{Bhatta} et~al.(2018){Bhatta}, {Mohorian}, and
  {Bilinsky}}]{2018A&A...619A..93B}
\bibinfo{author}{G.~{Bhatta}}, \bibinfo{author}{M.~{Mohorian}},
  \bibinfo{author}{I.~{Bilinsky}},
\newblock \bibinfo{title}{{Hard X-ray properties of NuSTAR blazars}},
\newblock \bibinfo{journal}{\aap} \bibinfo{volume}{619} (\bibinfo{year}{2018})
  \bibinfo{pages}{A93}. \DOIprefix\doi{10.1051/0004-6361/201833628}.
  \href{http://arxiv.org/abs/1710.09910}{{\tt arXiv:1710.09910}}.
\bibitem[{{Kapanadze} et~al.(2020){Kapanadze}, {Gurchumelia}, {Dorner},
  {Vercellone}, {Romano}, {Hughes}, {Aller}, {Aller}, and
  {Kharshiladze}}]{2020ApJS..247...27K}
\bibinfo{author}{B.~{Kapanadze}}, \bibinfo{author}{A.~{Gurchumelia}},
  \bibinfo{author}{D.~{Dorner}}, \bibinfo{author}{S.~{Vercellone}},
  \bibinfo{author}{P.~{Romano}}, \bibinfo{author}{P.~{Hughes}},
  \bibinfo{author}{M.~{Aller}}, \bibinfo{author}{H.~{Aller}},
  \bibinfo{author}{O.~{Kharshiladze}},
\newblock \bibinfo{title}{{Swift Observations of Mrk 421 in Selected Epochs.
  III. Extreme X-Ray Timing/Spectral Properties and Multiwavelength
  Lognormality during 2015 December-2018 April}},
\newblock \bibinfo{journal}{\apjs} \bibinfo{volume}{247} (\bibinfo{year}{2020})
  \bibinfo{pages}{27}. \DOIprefix\doi{10.3847/1538-4365/ab6322}.
  \href{http://arxiv.org/abs/2004.00676}{{\tt arXiv:2004.00676}}.
\bibitem[{{Stern} and {Poutanen}(2011)}]{2011MNRAS.417L..11S}
\bibinfo{author}{B.~E. {Stern}}, \bibinfo{author}{J.~{Poutanen}},
\newblock \bibinfo{title}{{Variation of the
  {\ensuremath{\gamma}}{\ensuremath{\gamma}} opacity by the He II Lyman
  continuum constrains the location of the {\ensuremath{\gamma}}-ray emission
  region in the blazar 3C 454.3}},
\newblock \bibinfo{journal}{\mnras} \bibinfo{volume}{417}
  (\bibinfo{year}{2011}) \bibinfo{pages}{L11--L15}.
  \DOIprefix\doi{10.1111/j.1745-3933.2011.01107.x}.
  \href{http://arxiv.org/abs/1105.2762}{{\tt arXiv:1105.2762}}.
\bibitem[{{Cerruti} et~al.(2013){Cerruti}, {Dermer}, {Lott}, {Boisson}, and
  {Zech}}]{2013ApJ...771L...4C}
\bibinfo{author}{M.~{Cerruti}}, \bibinfo{author}{C.~D. {Dermer}},
  \bibinfo{author}{B.~{Lott}}, \bibinfo{author}{C.~{Boisson}},
  \bibinfo{author}{A.~{Zech}},
\newblock \bibinfo{title}{{Gamma-Ray Blazars near Equipartition and the Origin
  of the GeV Spectral Break in 3C 454.3}},
\newblock \bibinfo{journal}{\apjl} \bibinfo{volume}{771} (\bibinfo{year}{2013})
  \bibinfo{pages}{L4}. \DOIprefix\doi{10.1088/2041-8205/771/1/L4}.
  \href{http://arxiv.org/abs/1305.4159}{{\tt arXiv:1305.4159}}.
\bibitem[{{Kang} et~al.(2021){Kang}, {Zheng}, {Wu}, {Chen}, and
  {Yin}}]{2021MNRAS.502.5875K}
\bibinfo{author}{S.-J. {Kang}}, \bibinfo{author}{Y.-G. {Zheng}},
  \bibinfo{author}{Q.~{Wu}}, \bibinfo{author}{L.~{Chen}},
  \bibinfo{author}{Y.~{Yin}},
\newblock \bibinfo{title}{{On the origin of GeV spectral break for Fermi
  blazars: 3C 454.3}},
\newblock \bibinfo{journal}{\mnras} \bibinfo{volume}{502}
  (\bibinfo{year}{2021}) \bibinfo{pages}{5875--5881}.
  \DOIprefix\doi{10.1093/mnras/stab489}.
  \href{http://arxiv.org/abs/2102.08962}{{\tt arXiv:2102.08962}}.
\bibitem[{{Kynoch} et~al.(2018){Kynoch}, {Landt}, {Ward}, {Done}, {Gardner},
  {Boisson}, {Arrieta-Lobo}, {Zech}, {Steenbrugge}, and {Pereira
  Santaella}}]{2018MNRAS.475..404K}
\bibinfo{author}{D.~{Kynoch}}, \bibinfo{author}{H.~{Landt}},
  \bibinfo{author}{M.~J. {Ward}}, \bibinfo{author}{C.~{Done}},
  \bibinfo{author}{E.~{Gardner}}, \bibinfo{author}{C.~{Boisson}},
  \bibinfo{author}{M.~{Arrieta-Lobo}}, \bibinfo{author}{A.~{Zech}},
  \bibinfo{author}{K.~{Steenbrugge}}, \bibinfo{author}{M.~{Pereira Santaella}},
\newblock \bibinfo{title}{{The relativistic jet of the
  {\ensuremath{\gamma}}-ray emitting narrow-line Seyfert 1 galaxy 1H
  0323+342}},
\newblock \bibinfo{journal}{\mnras} \bibinfo{volume}{475}
  (\bibinfo{year}{2018}) \bibinfo{pages}{404--423}.
  \DOIprefix\doi{10.1093/mnras/stx3161}.
  \href{http://arxiv.org/abs/1712.01799}{{\tt arXiv:1712.01799}}.
\bibitem[{{Wilkins} et~al.(2015){Wilkins}, {Gallo}, {Grupe}, {Bonson},
  {Komossa}, and {Fabian}}]{2015MNRAS.454.4440W}
\bibinfo{author}{D.~R. {Wilkins}}, \bibinfo{author}{L.~C. {Gallo}},
  \bibinfo{author}{D.~{Grupe}}, \bibinfo{author}{K.~{Bonson}},
  \bibinfo{author}{S.~{Komossa}}, \bibinfo{author}{A.~C. {Fabian}},
\newblock \bibinfo{title}{{Flaring from the supermassive black hole in Mrk 335
  studied with Swift and NuSTAR}},
\newblock \bibinfo{journal}{\mnras} \bibinfo{volume}{454}
  (\bibinfo{year}{2015}) \bibinfo{pages}{4440--4451}.
  \DOIprefix\doi{10.1093/mnras/stv2130}.
  \href{http://arxiv.org/abs/1510.07656}{{\tt arXiv:1510.07656}}.
\bibitem[{{Agarwal} and {Paliya}(2026)}]{2026ApJS..282...35A}
\bibinfo{author}{S.~{Agarwal}}, \bibinfo{author}{V.~S. {Paliya}},
\newblock \bibinfo{title}{{Hunting Very High-energy-emitting (>100 GeV)
  High-synchrotron-peaked Blazars}},
\newblock \bibinfo{journal}{\apjs} \bibinfo{volume}{282} (\bibinfo{year}{2026})
  \bibinfo{pages}{35}. \DOIprefix\doi{10.3847/1538-4365/ae27be}.
  \href{http://arxiv.org/abs/2510.06017}{{\tt arXiv:2510.06017}}.

\end{thebibliography}

\end{document}